\documentclass[fleqn,usenatbib]{mnras}

\usepackage{newtxtext,newtxmath}

\usepackage[T1]{fontenc}

\DeclareRobustCommand{\VAN}[3]{#2}
\let\VANthebibliography\thebibliography
\def\thebibliography{\DeclareRobustCommand{\VAN}[3]{##3}\VANthebibliography}

\usepackage{graphicx}	
\usepackage{tabularx}   
\usepackage{amsmath}	
\usepackage{multirow}   
\usepackage{pdflscape}  
\usepackage{booktabs}   
\usepackage{xcolor}     
\usepackage{caption}    
\usepackage{subcaption} 
\usepackage{fix-cm}     
\usepackage{threeparttable} 

\graphicspath{ {Figures/} }

\title[Stellar Populations in nearby, dusty ETGs]{Stellar Population Characterisations in nearby, dusty Early-Type Galaxies}

\author[R. W. Savage et al.]{
Ron W. Savage,$^{1}$\thanks{E-mail: ronwsavage@gmail.com}
Anne E. Sansom,$^{1}$\thanks{E-mail: aesansom@lancashire.ac.uk}
and David H. W. Glass$^{1} $
\\
$^{1}$Jeremiah Horrocks Institute, University of Lancashire, Preston, Lancashire, PR1 2HE, UK\\
}

\date{Accepted XXX. Received YYY; in original form ZZZ}

\pubyear{2024}

\begin{document}
\label{firstpage}
\pagerange{\pageref{firstpage}--\pageref{lastpage}}
\maketitle

\begin{abstract}
Dust in Early-Type galaxies (ETGs) may originate from internal or external sources. 
In this paper we study the stellar populations of particularly dusty ETGs to search for evidence of the dust's origin. Using the Southern African Large Telescope (SALT), we obtained long-slit optical spectra within the effective radius ($R_e$), along the major axis of 15 nearby ETGs, selected from the GAMA and Herschel-ATLAS surveys for their high levels of interstellar dust. Using full spectrum fitting and Lick index 
fitting we analysed their major axis kinematics and stellar population characteristics. We used stellar population models from the newly developed sMILES library and from the empirical MILES library. Kinematic results 
show that most of our sample of dusty ETGs are rotationally supported and there are no detectable kinematic discontinuities. 12 of our sample of 15 dusty ETGs show evidence of young/intermediate age stellar population components suggesting ongoing/recent star formation. Using simulations, we show that these recent ($\approx$1~Gyr) populations are not artefacts of the fitting process or data. As a check with a control sample we use stacked SDSS spectra and find
that dusty ETGs show a component with intermediate age, whereas non-dusty ETGs do not. Age, metallicity and $\alpha$-element abundance ratio increase with increasing central velocity dispersion in the SALT spectra, as seen in previous studies of ETGs, but with larger scatter in our sample. Given our stellar population findings, we discuss formation scenarios that might cause or rule out a high dust/molecular gas content.
\end{abstract}

\begin{keywords}
galaxies: elliptical and lenticular, cD, galaxies: evolution, galaxies: stellar content, galaxies: ISM
\end{keywords}



\section{Introduction}
\label{sec:introduction}  
Early-type galaxies (ETGs), including E and S0 galaxies, are considered to be end points of galaxy evolution. Their smooth elliptical or lenticular morphology and lack of spiral arms can result from gravitational interactions and mergers \citep{Burkert2003, Bournaud2005, Bournaud2007, Conselice2014, Martin2018}. In general ETGs also have low star formation rates, older stellar populations, and low levels of gas and dust in their interstellar medium (ISM) \citep{Young2011, Cortese2012}, however, a number of authors, e.g. \citet{Goudfrooij1995}, identified some ETGs that contained higher than expected levels of dust.

More recently, using far-infrared and submillimeter observations, e.g. from the Herschel Space Observatory\footnote{https://www.esa.int/Science\_Exploration/Space\_Science/Herschel}, a number of investigations, e.g. \citet{Rowlands2012} and \citet{Agius2013}, showed that the ISM of some ETGs contains significant amounts of dust. Using data from the Herschel Reference Survey (HRS) \citet{Smith2012} detected dust in 24$\%$ of the E galaxies and 62$\%$ of the S0 in their sample of 62 nearby ETGs, with the dust detected ETGs having a mean dust to stellar mass ratio (log($M_d$/$M_*$)) of $\approx$-4.3. This is $\approx$6 times higher than the mean for their overall sample of ETGs including dust non-detected, but is $\approx$50 times lower than in late type galaxies (LTGs) in the HRS. From a sample of 38 ETGs observed with Spitzer \citet{Martini2013} determined that approximately $\approx$60$\%$ of typical ETGs have $\geq$$10^{5}$ $M_{\sun}$ of interstellar dust and it is now widely accepted that the ISM of some ETGs contains large amounts of dust \citep{Agius2013, Agius2015, Davies2019}. It has also been confirmed that some ETGs have substantial molecular gas reservoirs, e.g. \citet{Davis2015} and \citet{Sansom2019}, leading to some level of ongoing star formation, e.g. \citet{Schawinski2007, Werle2020} and \citet{Lee2023}. 

Possible origins of dust in ETGs divide into sources either internal or external to the galaxy. Internal sources include formation during cooling of hot ISM and ejection of dust by evolved massive stars \citep{Griffith2019}. External sources include accretion or gas rich mergers \citep{Smith2012}. ISM which has been externally sourced can have different angular momentum to the receiving galaxy, which may result, for a period of a few dynamical times (i.e. about 100 Myr for central regions), in a kinematic misalignment between the ISM and stellar population \citep{Davis2011, Bassett2017, vandevoort2018, Glass2022}. 

Despite ongoing replenishment from massive evolved stars, dust has a relatively short lifetime in the ISM due to destruction through mechanisms such as astration, shocks and particle collision-driven sputtering, see for example \citet{Draine1979} and \citet{Schneider2024}. \citeauthor{Draine1979} computed a likely dust destruction timescale ($\tau_{dust}$) of 0.001 < $\tau_{dust}$ < 0.1~Gyr, while \citet{Clemens2010} used infra-red observations to calculate a dust grain destruction time of <0.046~Gyr. More recently, \citet{Michalowski2019} used far-IR and submillimeter measurements to calculate dust lifetimes for a sample that included 42 elliptical and lenticular galaxies, finding a longer $\tau_{dust}$ of 2.5$\pm$0.4~Gyr. This longer $\tau_{dust}$  
could suggest that the dust source is mainly internal, driven by fast grain growth in the ISM \citep{Michalowski2019}. However, dust lifetime would also be extended if it is replenished by external sources such as gas-rich minor mergers \citep{Kaviraj2012}. Order-of-magnitude variations in dust mass between ETGs of the same morphological type and stellar mass measured by \citet{Martini2013} led them to conclude that at least some of the dust in some ETGs originates from external sources. \citet{Kokusho2019} (their fig. 2) plotted dust mass versus stellar mass for the 260 ETGs in ATLAS$^{3D}$ revealing no correlation, which indicates that continuous mass loss from stars is not a significant source of dust mass in ETGs. 

As the evolutionary paths of ETGs are not yet fully understood, see for example \citet{Bournaud2007, Eales2015} and \citet{Cappellari2016}, there is not a consensus as to the relative contribution of internal vs. external sources of dust. High dust content, gas reservoirs and stellar populations of different ages or metallicities could be pointers towards interactions or merger events which have affected the galaxies’ evolutionary paths. Therefore, it is of interest to investigate the stellar population properties of dusty ETGs, to identify evidence of past events or any trends that might point towards the dust source.

There have been some optical surveys of ETGs, some of which have associated dust or gas measurements. The small sample of ETGs studied by 
\citet{Crocker2011} revealed that 55\% of galaxies with molecular gas detections and only 11\% of those without molecular gas, had young simple stellar population (SSP) equivalent ages. The larger ATLAS$^{3D}$ survey \citep{Cappellari2011} used infrared colours from the WISE survey \citep{Wright2010} and showed that there is warm circumstellar dust in their ETGs, even at old stellar population ages \citep{SimonianMartini2017}, with stellar population (SP) parameters determined by \citet{Mcdermid2015}. The ATLAS$^{3D}$ ETGs also have CO molecular gas observations \citep{Young2011} and AKARI satellite mid- and far-infrared data, from which \citet{Kokusho2017} (e.g. their fig. 12) showed that higher cold gas fractions are associated with younger mass-weighted mean SP ages. Other, similar trends have been found between the amounts of cool ISM and mean SP ages, for example in the study of 
\citet{Lesniewska2023}, who selected $\sim$2000 elliptical galaxies using r-band Sersic indices $>4$, from the GAMA sample, with data from the Herschel Space Observatory used to characterise their dust properties. \citet{Lesniewska2023} (their fig. 1) showed that specific dust mass descreases with luminosity-weighted average SP age in those galaxies, with a dust-removal timescale of 2.26 Gyr. 
\citet{Michalowski2024} showed that a similar timescale applies to neutral HI and molecular H$_2$ gas as well as dust in ETGs. Recently, for post starburst (PSB) galaxies, \citet{French2023} used optical observations from the MANGA survey (\citep{Bundy2015}) and submm CO observations with ALMA array and found that molecular gas mass decreases with PSB age since the starburst, over several hundred million years (\citet{French2023}, their fig. 11). 

Whist these general trends have been found between cold ISM contents and average SP ages in ETGs, their more detailed star formation histories have not been well studied in relation to their cold ISM content. In this current work we look at the breakdown with time of SP contributions for a small sample of very dusty ETGs, selected from GAMA and Herschel surveys, using new optical spectra from the SALT telescope.

Information on the physical properties of stellar populations and kinematics in an ETG is encoded within the absorption features in its composite optical spectrum \citep{Worthey1994, Conroy2013}, with the most frequently studied properties being stellar population age, metallicity and the abundance ratio of $\alpha$-elements to iron ([$\alpha$/Fe]), where $\alpha$-elements are O, Ne, Mg, Si, S, Ar, Ca and Ti \citep{Worthey2014}. However, extraction of these parameters is non-trivial due to degeneracies between the parameters e.g. age and metallicity \citep{Worthey1994B}. For ETGs there are already well known trends between the key stellar population parameters, i.e. mean population age, metallicity and [$\alpha$/Fe], and overall galaxy mass, aliased by central velocity dispersion (${\sigma_0}$), with more massive ETGs tending towards higher stellar metallicities, and earlier/shorter star formation timescales resulting in older stellar populations, see for example \citet{laBarbera2014}.

Star formation timescales in ETGs will influence [$\alpha$/Fe]. During the first $\approx$0.1~Gyr of a galaxy’s star formation history (SFH) massive stars have evolved and produced core-collapse supernovae, which enrich the interstellar medium with a wide range of elements, producing high [$\alpha$/Fe]. As lower mass stars evolve the proportion of Type 1a supernova increases, preferentially seeding the ISM with iron peak elements i.e. Fe, Cr, Co, Ni, Cu, and Mn, thus reducing the overall [$\alpha$/Fe] in subsequently formed stars. Therefore, shorter star formation timescales lead to higher [$\alpha$/Fe] in stars \citep{Worthey2014, Walcher2015}.

To study the stellar population properties in dusty ETGs found in the Herschel-ATLAS survey \citep{Eales2010} we used data from the Galaxy and Mass Assembly (GAMA) survey \citep{Driver2009} and obtained new, major axis, optical, long-slit spectroscopic observations for 15 of the dustiest ETGs, with the Southern African Large Telescope (SALT). These dusty ETGs are nearby (z<0.06), with the majority being morphological type S0 and the remainder type E. From these spectra we extracted results for luminosity weighted mean stellar population age (Age$_L$), metallicity ([M/H]$_L$) and $\alpha$-element abundance ([$\alpha$/Fe]$_L$) using full spectrum fitting \citep{Cappellari2017} and Lick index \citep{Worthey1994} SSP fitting. In the remainder of this paper, luminosity weighted results are denoted using subscript~L.

An SSP is a collection of stars that originated at the same time from the same molecular gas, hence all the stars have identical ages and chemical compositions. We used SSP template spectra from the well established Medium-resolution Isaac Newton Telescope library of empirical spectra (MILES)\footnote{http://research.iac.es/proyecto/miles/pages/ssp-models.php} \citep{Vazdekis2010, Falcon2011}. 
We also made use of a recently developed library of semi-empirical MILES stellar population models with variable [$\alpha$/Fe] abundance ratios (sMILES) \citep{Knowles2021, Knowles2023}. Both MILES and sMILES SSPs use overall metallicity ([M/H]) therefore metallicity results from our fittings are also in terms of [M/H].

In Section~\ref{sec:sample} we describe the criteria for sample selection and characteristics of the target dusty ETGs, and in Section~\ref{sec:observations} we describe the SALT observations and data reduction. In Section~\ref{sec:analysis} we describe our measurements of the stellar kinematics, stellar population properties and histories. Section~\ref{sec:discussion} gives a discussion of our results and our conclusions are summarised in Section~\ref{sec:conclusions}.
We assume flat $\Lambda$CDM cosmology, with $H_0$ = 70 km~s$^{-1}$ Mpc$^{-1}$, $\Omega_M$ = 0.3, and $\Omega_\Lambda$ = 0.7.

\section{Sample Selection}
\label{sec:sample}  

Target dusty ETGs were chosen from a sample of galaxies of all morphological types from the GAMA survey \citep{Driver2009} equatorial regions, as described initially by \citet{Agius2013} and updated by \citet{Glass2024}. The updated sample is based on GAMA II data \citep{Liske2015, Driver2022} and pre-release data from GAMA-KiDS-GalaxyZoo (Kelvin, private communication), see also \citet{Porter-Temple2022} and \citet{Holwerda2024}. An Initial Complete Sample was constructed with flow-corrected redshift limits of 0.002 < z < 0.06, r-band effective radii $\geq$1.2 arcsec, absolute Petrosian magnitudes ($M_r$) <-17.4 derived from Sloan Digital Sky Survey (SDSS, \citet{York2000}) photometry and appropriate GAMA II target observational and spectral quality flags. These GAMA data contain galaxies in lower-density (field and group) environments. Galaxies with evidence of strong active galactic nuclei (AGN) were then removed using multiple methods as described in detail in \citet{Glass2024}, resulting in a Parent Sample of 4458 galaxies. Galaxies with strong AGN were removed from the sample because of potential contamination of submillimeter spectra from cool dust emission, or non-thermal radiation, associated with the central AGN, making it difficult to extract dust properties by spectral fitting.

ETGs within the Parent Sample were identified using GAMA II morphological classifications \citep{Kelvin2014, Moffett2016}, based on SDSS imaging. However, the depth and angular resolution limits of SDSS images may allow LTGs with weak spiral features to be identified as ETGs. Additional LTGs with weak spiral features were identified and removed using pre-release morphology classification statistics from GAMA-KiDS-GalaxyZoo. These classifications are based on imaging from the Kilo-Degree Survey (KiDS, \citealt{deJong2013}), with significantly better depth and angular resolution than SDSS. The remaining ETGs with SDSS r-band ellipticities $\leq$0.7, where GAMA-KiDS-GalaxyZoo is most effective in ruling out weak spiral galaxies, were assigned final morphological classifications of ETGs as Elliptical or Lenticular, giving 608 E type and 461 S0 type galaxies (\citealt{Glass2024}; \textcolor{blue}{Glass et al. in preparation}).

A sub-sample of dusty ETGs was extracted from this ETG sample, based on 250$\mu$m emission (assumed to be from cool dust) at $>$4$\sigma$ detection within Herschel-ATLAS DR1 \citep{Valiante2016}, leading to a sample of 202 dusty ETGs, i.e. approximately 20\% of the morphologically classified ETG sample. 

From the sample of 202 dusty ETGs, 32 were targeted for IRAM 30m observations, which reveal molecular gas detections (\textcolor{blue}{Glass et al.} \textcolor{blue}{in preparation}). Of these 32 dusty ETGs, 14 were also observed using SALT. Hence we have a sample of dusty, molecular gas-rich ETGs, for which we can study their stellar populations. An additional dusty ETG, GAMA65075,  
had been observed with SALT previously  \citep{Vaghmare2018}   
and is included in this work.

\begin{figure}
    \includegraphics[width=\columnwidth, alt={WHAN diagram showing the equivalent width of the H-alpha line against the ratio of [NII] to H-alpha for our dusty ETG targets, with a background of the Initial Complete Sample.}]{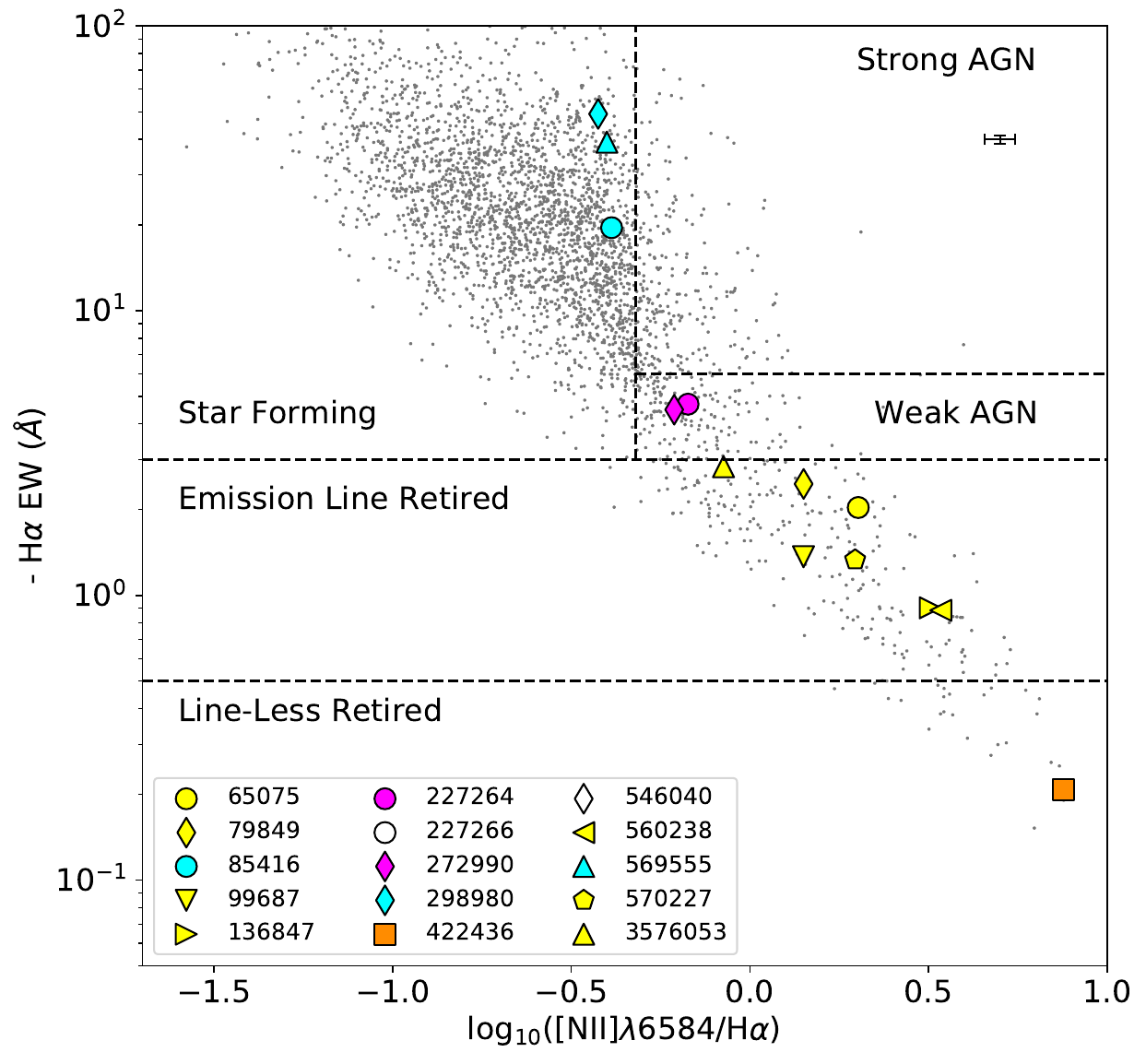}
\caption{
     H${\alpha}$ emission line equivalent width vs log$_{10}$([NII]$\lambda$6584/H${\alpha}$) emission line ratio WHAN diagram \citep{Cid-Fernandes2010, Cid-Fernandes2011}, with the selected dusty ETGs highlighted. For reference, the grey points show our Initial Complete Sample (see Section~\ref{sec:sample}) of galaxies with $\ge$3$\sigma$ detection of relevant emission lines. Emission line strengths and equivalent widths were sourced from GAMA II \textsc{GaussFitSimplev05}. The vertical demarcation between Star-forming and Strong AGN is set at log$_{10}$([NII]$\lambda$6584/H${\alpha}$)=-0.32 \citep{Glass2024}. Error bars, illustrated top right in the plot, are based on median uncertainties from GAMA II \textsc{GaussFitSimplev05} for the plotted Initial Complete Sample galaxies. GAMA catalogue numbers (CATAID) and symbols shown in the legend box are also used in Fig.~\ref{fig:fig_2}.
    }
    \label{fig:Fig_1}
\end{figure}

These 15 dusty ETGs, all from field and group environment \citep{Robotham2011}, form the sample for this paper. A list of their key physical properties is given in Table~\ref{tab:table_1} and their location on a plot of H$\alpha$ emission line equivalent width vs. log$_{10}$([NII]$\lambda$6584/H$_{\alpha}$) emission line strengths, WHAN diagram, \citep{Cid-Fernandes2010, Cid-Fernandes2011} is presented in Fig.~\ref{fig:Fig_1}. No galaxies with strong AGN are included in our SALT sample because of filtering during the sample build. Three galaxies, GAMA85416, 298980 and 569555 (cyan), are shown as star-forming. GAMA272990 and 227264 (magenta) are shown as having weak AGN. The remainder are emission-line retired (yellow) or line-less retired (orange), see Fig.~\ref{fig:Fig_1}. GAMA227266 and 546040 (white) are not plotted as they have no detected emission lines (line-less retired).

\begin{figure}
    \includegraphics[width=\columnwidth, alt={This two panel figure plots the sample of 15 dusty ETGs against a background of the galaxies in the Parent Samale. The upper panel shows dust mass versus stellar mass. The lower panel shows star formation rate versus stellar mass.}]{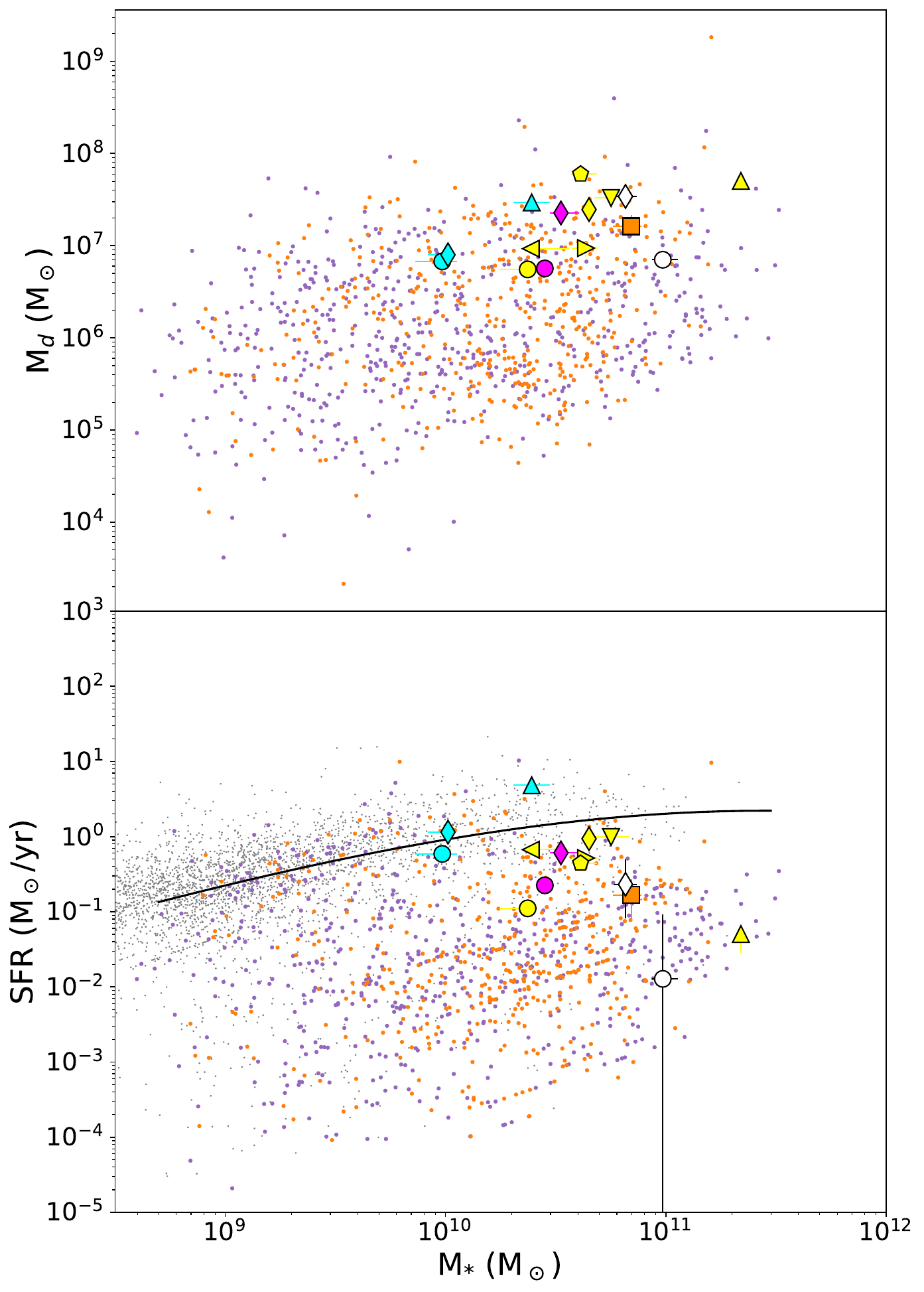}
    \caption{
    The top panel plots dust mass ($M_d$) vs. stellar mass ($M_*$) showing the relatively high dust content of the 15 selected ETGs compared with the 608 E type (purple dots) and 461 S0 type (orange dots) galaxies in the Parent Sample. The lower panel plots star formation rate (SFR) vs. $M_*$.  The grey cloud of dots shows the overall Parent Sample of 4458 galaxies and the black line represents the Galaxy Main Sequence according to Equation 5 of \citet{Saintonge2016}. Most of the 15 dusty ETGs fall below the Galaxy Main Sequence, but have relatively high SFRs compared with most ETGs in the Parent Sample. The symbols used are identified by GAMA catalog number in Fig.~\ref{fig:Fig_1} legend. Data were sourced from GAMA II \textsc{MagPhysv06} best\_fit parameters. Error bars show 1-sigma uncertainties calculated from GAMA II \textsc{MagPhysv06} 16th and 84th percentiles. 
    }
    \label{fig:fig_2}
\end{figure}

Using stellar and dust masses from spectral energy distribution fitting in GAMA II\footnote{https://www.gama-survey.org} \textsc{MagPhys06}, the selected dusty ETGs have log stellar mass ($M_*$) from 10.0 to 11.3 (\(M_\odot\)) and log dust mass ($M_d$) from 6.7 to 7.8 (\(M_\odot\)), see Table~\ref{tab:table_1}. GAMA II dust mass values are combined warm and cold dust derived from \textsc{MagPhys} \citep{daCunha2008}, giving dust mass ratios (DMR), i.e. $M_d/M_*$, from 7.3$\times10^{-5}$ to 1.3$\times10^{-3}$, with a mean of 5.4$\times10^{-4}$ vs. a mean DMR of 2.8$\times10^{-4}$ for the 606 E and 461 S0 type galaxies in the Parent Sample. We plot dust mass and star formation rate (SFR) vs. stellar mass in Fig.~\ref{fig:fig_2} to show the relatively high dust content of our 15 selected dusty ETGs (Fig.~\ref{fig:fig_2} upper panel) and their location mostly within the Green Valley below the star-forming main-sequence (Fig.~\ref{fig:fig_2} lower panel).

\begin{table*}
    \centering
    \caption{
    Physical properties of the 15 target dusty ETGs. Stellar mass, dust mass and their uncertainties were obtained from GAMA II \textsc{MagPhysv06} parameters \textit{mass\_stellar\_best\_fit} 
    and \textit{mass\_dust\_best\_fit}.
    Half-Light Radius, Ellipticity and major axis Position Angle were recovered from GAMA II \textsc{SersicCatSDSSv09} parameters GALRE\_r, GALELLIP\_r and GALPA\_r. 
    Redshift was recovered from GAMA II \textsc{DistancesFramesv14} parameter Z\_HELIO.
    Visual morphology was recovered from \citet{Glass2024}.}
	\label{tab:table_1}
    \begin{tabular}{c c c c c c c c c}
    \toprule
    GAMA & Redshift & Half-light & Ellipticity & Mass of & Mass of & Visual & Position & Dust mass \\
    ~ & ~ & Radius ($R_e$) & ($\varepsilon$) & stars ($M_*$) & dust ($M_d$) & Morphology & Angle & ratio ($M_d/M_*$)\\
    ~ & ~ & arcsec & ~ & $\times10^{10}$ ($M_{\sun}$) & $\times10^{6}$ ($M_{\sun}$) & ~ & deg & $\times10^{-4}$ \\ 
    \midrule
    65075 & 0.0055 & 30.8 & 0.336 & $2.36^{+0.05}_{-0.61}$ & $5.53^{+0.43}_{-0.66}$ & Lenticular & 69.8 & $2.34^{+0.19}_{-0.66}$ \\ \addlinespace[0.5em]
    79849 & 0.0452 & 4.7 & 0.423 & $4.49^{+0.00}_{-0.10}$ & $24.6 ^{+5.81}_{-1.68}$ & Lenticular & 4.7 & $5.49^{+1.30}_{-0.40}$ \\ \addlinespace[0.5em] 
    85416 & 0.0194 & 5.9 & 0.665 & $0.97^{+0.16}_{-0.24}$ & $6.75 ^{+0.67}_{-0.46}$ & Elliptical & 80.1 & $6.99^{+1.36}_{-1.80}$ \\ \addlinespace[0.5em]  
    99687 & 0.0480 & 6.6 & 0.624 & $5.65^{+1.22}_{-0.90}$ & $32.9 ^{+6.73}_{-4.95}$ & Lenticular & 124.6 & $5.82^{+1.73}_{-1.28}$ \\ \addlinespace[0.5em]  
    136847 & 0.0277 & 8.1 & 0.298 & $4.34^{+0.56}_{-0.59}$ & $9.413 ^{+1.49}_{-1.85}$ & Lenticular & 44.9 & $2.17^{+0.44}_{-0.52}$ \\ \addlinespace[0.5em]  
    227264 & 0.0249 & 6.8 & 0.449 & $2.83^{+0.00}_{-0.64}$ & $5.66 ^{+0.70}_{-0.93}$ & Lenticular & 83.0 & $2.00^{+0.25}_{-0.33}$ \\ \addlinespace[0.5em]  
    227266 & 0.0249 & 9.6 & 0.112 & $9.70^{+1.64}_{-1.05}$ & $7.040^{+1.53}_{-0.95}$ & Elliptical & 58.2 & $0.73^{+0.20}_{-0.13}$ \\ \addlinespace[0.5em]  
    272990 & 0.0411 & 3.8 & 0.254 & $3.34^{+0.69}_{-0.37}$ & $22.6 ^{+4.17}_{-2.35}$ & Elliptical & 2.9 & $6.77^{+1.87}_{-1.02}$ \\ \addlinespace[0.5em]  
    298980 & 0.0271 & 3.8 & 0.322 & $1.03^{+0.09}_{-0.19}$ & $7.96 ^{+1.11}_{-0.89}$ & Elliptical & 153.8 & $7.75^{+1.26}_{-1.70}$ \\ \addlinespace[0.5em]  
    422436 & 0.0259 & 8.3 & 0.584 & $6.98^{+0.79}_{-1.21}$ & $16.2 ^{+5.33}_{-3.80}$ & Lenticular & 141.7 & $2.33^{+0.81}_{-0.68}$ \\ \addlinespace[0.5em]  
    546040 & 0.0266 & 11.8 & 0.086 & $6.57^{+0.81}_{-0.72}$ & $34.2 ^{+5.34}_{-6.34}$ & Lenticular & 117.5 & $5.21^{+1.04}_{-1.12}$ \\ \addlinespace[0.5em]  
    560238 & 0.0213 & 8.6 & 0.295 & $2.44^{+1.20}_{-0.24}$ & $9.28 ^{+1.11}_{-1.17}$ & Lenticular & 24.7 & $3.80^{+1.19}_{-0.61}$ \\ \addlinespace[0.5em]  
    569555 & 0.0569 & 2.7 & 0.392 & $2.46^{+0.51}_{-0.42}$ & $29.5 ^{+6.43}_{-4.42}$ & Lenticular & 77.0 & $11.97^{+3.59}_{-2.73}$ \\ \addlinespace[0.5em]  
    570227 & 0.0433 & 5.6 & 0.578 & $4.11^{+0.76}_{-0.38}$ & $59.7 ^{+9.22}_{-10.36}$ & Lenticular & 7.3 & $14.54^{+3.51}_{-2.87}$ \\ \addlinespace[0.5em] 
    3576053 & 0.0520 & 6.1 & 0.179 & $21.93^{+0.96}_{-0.47}$ & $50.4 ^{+9.66}_{-10.07}$ & Elliptical & 8.0 & $2.30^{+0.45}_{-0.46}$ \\ \bottomrule
    \end{tabular}
\end{table*}

\section{Observations and Data Reductions}
\label{sec:observations}  
We obtained longslit optical spectra for 14 of our dusty ETGs between 2019 and 2021, using the Robert Stobie Spectrograph (RSS) \citep{Burgh2003}
on SALT \citep{Buckley2006}. All observations used an 8 arcmin long by 1 arcsec wide slit, aligned along the target ETGs' major axes, and a 900 lines/mm transmission grating (PG0900), giving a spectral range of 3777 to 6850\r{A}. The detector has three 2048×4096 pixel CCD chips separated by 1.5mm gaps, covering an 8 arcmin field of view with a pixel scale of 0.1267 arcsec/pixel (unbinned). 
We chose the grating angle to minimize Lick indices falling into the CCD chip gaps.
On-chip binning of 2×4 was used to improve signal to noise (S/N) ratio, giving an image size of 3072×1024 pixels, spectral sampling of $\approx$1\r{A}/pixel and spatial sampling of $\approx$0.5 arcsec/pixel. Median zenith V-band seeing at SALT was $\approx$1.5 arcsec\footnote{https://pysalt.salt.ac.za/proposal$\_$calls/current/ProposalCall.pdf} \citep{Catala2013}.
Observations for the fifteenth ETG, GAMA65075, taken in 2014 by \citet{Vaghmare2018} using the same instrument set-up, were downloaded from the SALT Data Archive\footnote{https://ssda.saao.ac.za}. A summary of the observations is given in Table~\ref{tab:table_2}.

\begin{table*}
    \centering
    \caption{Summary of longslit observations using SALT/RSS. GAMA65075 was observed by \citet{Vaghmare2018} and spectra from these observations were obtained from the SALT Data Archive. The last column indicates the number and S/N ratio of spatial bins used to extract spectra for the major axis profiles described in Section~\ref{sec:observations}. Notes: [1] date format used is YYYYMMDD. [2] average value for images in the observing block. [3] measured east from north. } 
    \label{tab:table_2}
    \begin{tabular}{c c c c c c c c c}
    \toprule
    GAMA Galaxy & SALT & Night Observed & Exposure & Airmass & Slit Angle & Number of apertures \\
    ~ & Semester & [Note 1] & (s) & [Note 2] & [Note 3] & @ S/N ratio\\ \midrule
    \multirow{2}{*}{65075} & \multirow{2}{*}{2014-1} & 20140511 & 900 x 2 & 1.21 & 70.6 &   \multirow{2}{*}{10 @ 30} \\
    & ~ & 20140613 & 900 x 2 & 1.19 & 70.6 & \\ \midrule    
    79849 & 2019-1 & 20190426 & 600 x 4 & 1.22 & 4.7 & 7 @ 25 \\ \midrule
    85416 & 2020-1 & 20200510 & 648 x 3 & 1.20 & 80.1 & 7 @ 30 \\ \midrule
    \multirow{2}{*}{99687} & \multirow{2}{*}{2020-1} & 20200514 & 844 x 3 & 1.22 & 124.6 &   \multirow{2}{*}{7 @ 25} \\
    & ~ & 20200514 & 843 x 3 & 1.28 & 124.6 & \\ \midrule
    136847 & 2019-2 & 20200124 & 500 x 4 & 1.27 & 44.9 &  6 @ 25 \\ \midrule
    227264 & 2020-1 & 20200524 & 646 x 4 & 1.33 & 83.0 &  7 @ 25 \\ \midrule
    \multirow{2}{*}{227266} & \multirow{2}{*}{2020-1} & 20200622 & 801 x 3 & 1.27 & 58.2 & \multirow{2}{*}{13 @ 25} \\
    & ~ & 20200724 & 802 x 3 & 1.3 & 58.2 \\ \midrule
    272990 & 2019-2 & 20200324 & 500 x 3 & 1.27 & 2.9 & 5 @ 30 \\ \midrule
    \multirow{2}{*}{298980} & \multirow{2}{*}{2019-1} & 20190426 & 698 x 3 & 1.22 & 153.8 &  \multirow{2}{*}{5 @ 25} \\
    & ~ & 20190426 & 698 x 3 & 1.3 & 153.8 \\ \midrule
    422436 & 2019-2 & 20191201 & 468 x 4 & 1.28 & 141.7 & 11 @ 30 \\ \midrule
    \multirow{3}{*}{546040} & \multirow{3}{*}{2020-2} & 20210222 & 924 x 2 & 1.18 & 117.5 & \multirow{3}{*}{10 @ 30} \\
    & ~ & 20210324 & 924 x 2 & 1.34 & 117.5 \\
    & ~ & 20210410 & 924 x 2 & 1.26 & 117.5 \\ \midrule
    560238 &  2020-2 & 20210222 & 671 x 3 & 1.20 &  24.7 & 10 @ 30 \\ \midrule
    569555 & 2019-1 & 20190511 & 501 x 4 & 1.19 & 77.0 & 4 @ 20 \\ \midrule
    \multirow{2}{*}{570227} & \multirow{2}{*}{2019-1} & 20190512 & 586 x 3 & 1.19 & 7.3 & \multirow{2}{*}{6 @ 20} \\
    & ~ & 20190522 & 586 x 3 & 1.2 & 7.3 \\ \midrule
    \multirow{3}{*}{3576053} & \multirow{3}{*}{2019-2} & 20191122 & 538 x 2 & 1.29 & 8.0 & \multirow{3}{*}{5 @ 20} \\
    & ~ & 20191206 & 450 x 2 & 1.29 & 8.0 \\
    & ~ & 20200114 & 678 x 2 & 1.27 & 8.0 \\ \bottomrule
    \end{tabular}
\end{table*}

SALT and RSS were designed to have improved throughput in the blue part of the spectrum \citep{Buckley2006, Buckley2008}. This spectral region ($\approx 3500 - 4500$ \AA) contains many age sensitive absorption features, including higher-order Balmer lines, which respond strongly to A and F type stars produced by star formation episodes within the past one or two Gyr. This instrument therefore facilitates detection of any intermediate age components that may be embedded in old stellar populations of ETGs.

Raw RSS data were processed through the SALT Data Pipeline \citep{Crawford2010} to give bias and gain corrected longslit spectra in FITS format. These data were further reduced using a sequence of \textsc{IRAF} tasks \citep{Tody1986} adapted from the SALT Long Slit Reduction Recipe\footnote{https://www.saao.ac.za/$\sim$brent/Kniazev$\_$longslit.pdf}, including wavelength calibration, background subtraction, flux calibration and geometric distortion correction. During wavelength calibration the spectra were binned to a fixed 1.25\r{A} per pixel scale. As the target galaxies were significantly smaller than the RSS field, night sky spectra extracted away from the galaxy light were used for background subtraction. Median combination of multiple exposures of each target was used to improve S/N ratio, create a variance image and remove cosmic ray artefacts. We used \textsc{IRAF} task \textit{aptrace} to trace the galaxy centre in the 2D longslit spectra. Spectral resolutions were determined as $\approx$4.5\r{A} at red wavelengths and $\approx$5.5\r{A} at blue wavelengths, by fitting a Gaussian curve to lines in arc lamp spectra, giving a mid-spectrum Resolving power of $\approx$1050. The variable pupil of SALT makes absolute flux calibration impractical, therefore we used reduced SALT standard star spectra for relative flux calibration, to provide an approximate correction of the spectrum continuum shape, accurate to within $\approx$10\%. We masked the chip gaps to avoid artefacts from the CCD chip edges being included in the reduced spectra.

Spectra from the RSS are affected by partial scattering of light within the instrumentation light path \citep{Katkov2019}, resulting in redistribution of light in both spatial and spectral directions in sufficient quantities to broaden spectral absorption features and reduce their contrast. When using full spectrum fitting this broadening will cause an overestimate of the calculated velocity dispersions ($\sigma$) and reduction in contrast will systematically bias the stellar population fitting. We corrected for scattered light using the method of \citet{Katkov2019}, where a scattered light profile is determined from a longslit spectrum of a point source, e.g. a standard star, observed using the same instrument set-up as the science observations. Scattered light was removed from the science observations by subtracting the scattered light profile from the observed longslit spectra of the target galaxies. To assess the effect of scattered light removal we measured $\sigma$, age and metallicity from spectra covering the central $R_e$/8 radius aperture of each target ETG, with and without scattered light. We used full spectrum fitting with pPXF described in Section~\ref{subsec:analysis_population}. Removal of scattered light reduced $\sigma$ results by amounts less than than their 1-sigma uncertainties and increased metallicity results by less than their  1-sigma uncertainties. Age results were generally unaffected. Therefore, in the remainder of this paper we analysed apertures extracted from the longslit spectra with scattered light removed.

As described in Section~\ref{sec:analysis} below we applied full spectrum fitting to extract results for Age$_L$, [M/H]$_L$ and [$\alpha$/Fe]$_L$ from the central $R_e$/8 radius aperture of each target ETG. These $R_e$/8 apertures have S/N ratios of typically 44 per wavelength bin, see Table~\ref{tab:table_3}. In addition, we developed Python code to extract spatially binned 1D apertures from the reduced, scattered-light removed longslit spectra, based on S/N ratio. \citet{Ge2018} measured the effect of S/N ratio on fitted values for age and metallicity when using pPXF for full spectrum fitting, showing that both bias and scatter in fitted age and metallicity results are proportional to (S/N)$^{-1}$. In practice we selected the S/N ratio to give at least 5 apertures for each target ETG, see the last column in Table~\ref{tab:table_2}, giving 20 < S/N ratio < 30 per 1.25\r{A} wavelength bin at $\approx$5100 – 5200\r{A}. We used these apertures to construct $\sigma$, Age$_L$ and [M/H]$_L$ profiles along the slit, i.e. radially along the major axis, by full spectrum fitting. There was insufficient S/N to measure [$\alpha$/Fe]$_L$ for the radial profiles.

The methodology for estimation of $\alpha$-element enhancement by full spectrum fitting is less well established than for age and metallicity, see for example \citet{Liu2020} and \citet{Pernet2024}. Therefore, we developed a methodology and supporting Python code to estimate $\alpha$-element enhancement by Lick index fitting of SSPs.


\section{Analysis and Results}
\label{sec:analysis}  
We used penalised pixel fitting \citep{Cappellari2004} with pPXF version 8.2.3\footnote{From https://pypi.org/project/ppxf/} \citep{Cappellari2017, Cappellari2023} to extract stellar kinematics parameters, line-of-sight velocity and velocity dispersion, and luminosity weighted stellar populations parameters, Age$_L$, [M/H]$_L$ and [$\alpha$/Fe]$_L$, from the spatially binned 1D spectral apertures by full spectrum fitting between 3800 and 6800\r{A}. pPXF fits a linear combination of template SSP spectra to the target spectrum using a pixel-to-pixel chi squared ($\chi^2$) minimization algorithm. Throughout we used the pPXF default value of \textit{bias}=None, to bias the higher order Gauss-Hermite moments (h3, h4) towards zero, giving a Gaussian $\sigma$. We used a multiplicative Legendre polynomial (order 10) in pPXF. Multiplicative polynomials compensate for inaccuracies in the relative flux calibration of our SALT spectra and for dust reddening, without needing to use a specific reddening curve \citep{Cappellari2017}. Order 10 allows the polynomial to fit our spectra over scales that enable fits to continuum variations while preserving the absorption line strengths. Additive polynomials were not used as they can modify the depth of absorption lines. We used pPXF keyword \textit{norm\_range} between 3800 and 6800\r{A} to obtain luminosity weighted stellar population parameters. 

We sourced template SSP spectra from the MILES Library of SSPs \citep{Vazdekis2010, Falcon2011}, built using empirical star spectra, BaSTI isochrones and Chabrier initial mass function (IMF) and with [$\alpha$/Fe] reflecting those of the local Solar neighbourhood pattern. We also sourced SSPs from a recently developed library of semi-empirical star spectra, sMILES\footnote{From the MILES website at \newline https://cloud.iac.es/index.php/s/WnxBSX6G3CnaTkD} \citep{Knowles2023}. The sMILES SSP library provides a grid of spectra with ages from 0.03 to 14.0~Gyr, total metallicity (denoted [M/H]$_{SSP}$ in \citeauthor{Knowles2023}) from -1.79 to 0.26, and [$\alpha$/Fe] from -0.2 to 0.6. As for the MILES SSPs we chose sMILES SSPs with a Chabrier IMF.
\label{sec:fig8ref}

We assumed that the galaxy physical centre was the point of maximum signal on the spatial axis of the longslit spectrum, determined by \textsc{IRAF} task \textit{apall} during geometric distortion correction. We fitted spectra from apertures covering the central $R_e$/8 radius of each target ETG and apertures extracted along the major axis based on S/N ratio, using MILES SSPs and sMILES SSPs with an [$\alpha$/Fe] enhancement of 0 (later referred to as sMILES 2D fits). From these fits we obtained results for V, $\sigma$, their 1-sigma uncertainties, Age$_L$ and [M/H]$_L$ (see Section~\ref{subsec:analysis_population}). We also fitted spectra from apertures covering the central $R_e$/8 radius of each target ETG using the full grid of sMILES SSPs (later referred to as sMILES 3D fits), to obtain results for V, ${\sigma}$, their 1-sigma uncertainties, Age$_L$, [M/H]$_L$ and [$\alpha$/Fe]$_L$ (see Section~\ref{subsec:analysis_alpha_ppxf}).

We also used Python code, previously developed by \citet{Knowles2023}, to 
find the best fitting SSP
for the central $R_e$/8 aperture of each target ETG, by Lick index fitting. 
From the best fitting SSP we took values for Age, metallicity and [$\alpha$/Fe] (see Section~\ref{subsec:analysis_alpha_lick}).

\subsection{Stellar kinematics and profiles}
\label{subsec:analysis_kinematics}  
\begin{figure}
    \centering
    \includegraphics[width=\columnwidth, alt={This plot shows that the three values of velocity dispersions measured by pPXF full spectrum fitting are within their 1-sigma uncertainties for each of the target 15 dusty ETGs.}]{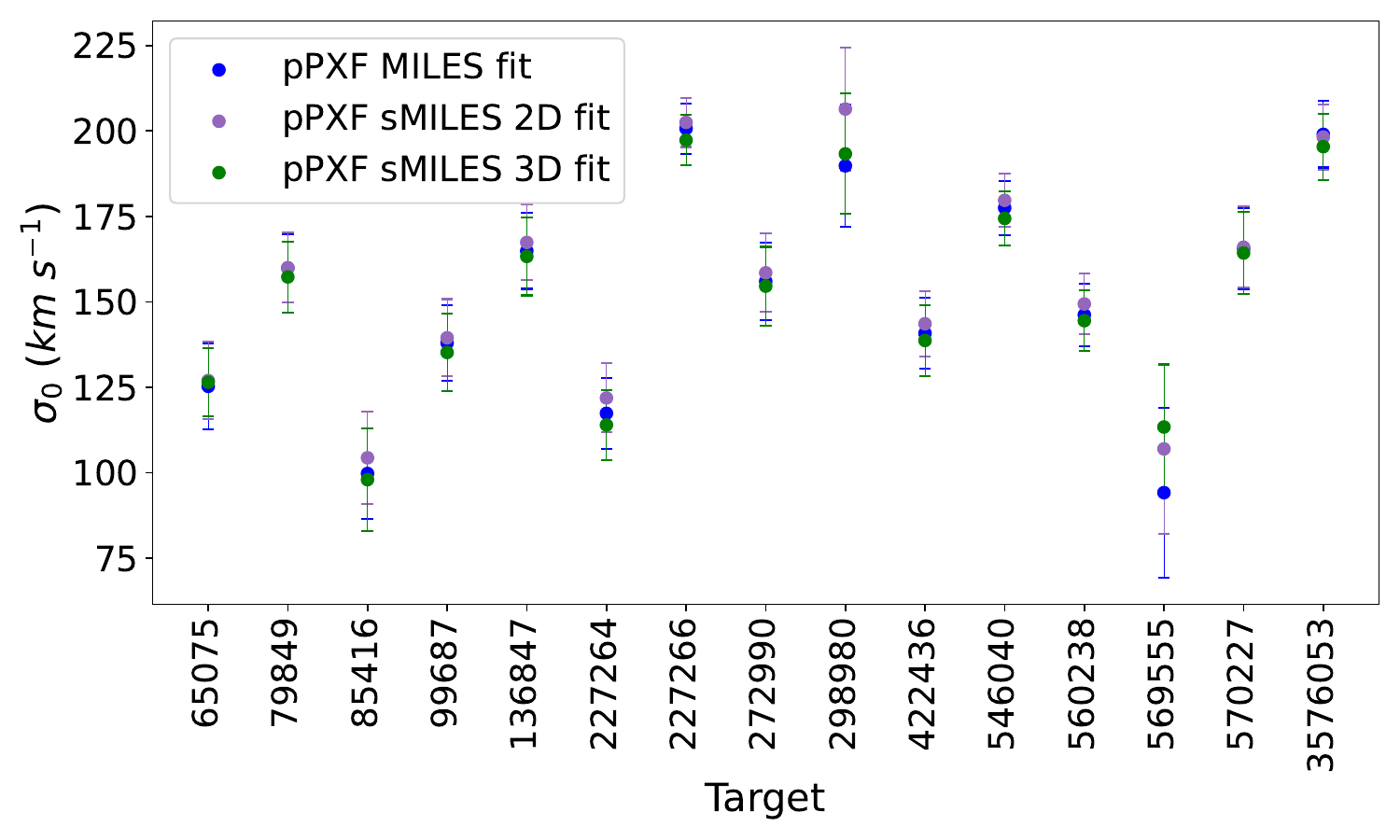}
    \caption{Plot of ${\sigma_0}$ for the central $R_e$/8 aperture of each target ETG, showing that $\sigma_0$ results from pPXF fittings using SSPs from the MILES and sMILES libraries are within 1-sigma uncertainties for each target ETG.}
    \label{fig:fig_3}
\end{figure}

The line-of-sight recession velocity (V) and central velocity dispersion ($\sigma_0$) were derived from the central $R_e$/8 aperture spectrum of each target, using pPXF \citep{Cappellari2017} with MILES SSPs. These stellar kinematics are presented in Table~\ref{tab:table_3}. 
In Fig.~\ref{fig:fig_3} we compare these $\sigma_0$ results with $\sigma_0$ results from fittings using sMILES SSPs with an [$\alpha$/Fe] enhancement of 0 (sMILES 2D fits), and $\sigma_0$ results from fittings using sMILES SSPs with the full grid of [$\alpha$/Fe] values (sMILES 3D fits). Fig.~\ref{fig:fig_3} shows the $\sigma_0$ results from MILES, sMILES 2D and sMILES 3D fittings are in good agreement within their 1-sigma uncertainties for all 15 targets, with an average standard deviation of 2.6~km~s$^{-1}$ from the mean of their three measured $\sigma_0$ values. These dusty ETGs cover a range of only $\approx 100 <\sigma_0 < 200$ km~s$^{-1}$, typical of intermediate mass ETGs.

\begin{table}
    \centering
    \caption{Kinematic results for the central $R_e$/8 apertures from full spectrum fitting using pPXF with MILES SSPs. V is line of sight recession velocity and $\sigma_0$ is line of sight velocity dispersion. Uncertainties shown are 1-sigma errors calculated by pPXF. The value given for maximum rotation velocity ($V_{max}$) is a lower limit, measured from the velocity profile described in Section~\ref{subsec:analysis_kinematics}, see also for example the top panel of Fig.~\ref{fig:fig_4}. S/N ratio is the mean per 1.25\r{A} wavelength bin in the central spectrum region between 4500-5500\r{A}.}
    \label{tab:table_3}
    \begin{tabular}{ccccc}
    \toprule
    GAMA  & V & $\sigma_0$ & $V_{max}$ & S/N Ratio  \\
    ~  & (km~s$^{-1}$) & (km~s$^{-1}$) & (km~s$^{-1}$) & ~  \\
    \midrule
    65075 & 1633 $\pm$ 6 & 125.3 $\pm$ 12.5 & 99 & 42.1  \\ \addlinespace[0.3em]
    79849 & 13229 $\pm$ 6 & 159.9 $\pm$ 10.0 & 180 & 56.9  \\ \addlinespace[0.3em]
    85416 & 5789 $\pm$ 13 & 99.8 $\pm$ 13.3 & 179 & 59.4  \\ \addlinespace[0.3em]
    99687 & 14040 $\pm$ 6 & 138.0 $\pm$ 11.1 & 170 & 45.2  \\ \addlinespace[0.3em]
    136847 & 8143 $\pm$ 6 & 164.9 $\pm$ 11.1 & 117 & 32.1  \\ \addlinespace[0.3em]
    227264 & 7400 $\pm$ 6 & 117.4 $\pm$ 10.3 & 96 & 42.5  \\ \addlinespace[0.3em]
    227266 & 7422 $\pm$ 4 & 200.7 $\pm$ 7.4 & 47 & 45.1  \\ \addlinespace[0.3em]
    272990 & 12077 $\pm$ 7 & 156.0 $\pm$ 11.4 & 68 & 56.7  \\ \addlinespace[0.3em]
    298980 & 8015 $\pm$ 12 & 189.8 $\pm$ 17.8 & 44 & 54.5  \\ \addlinespace[0.3em]
    422436 & 7639 $\pm$ 6 & 140.8 $\pm$ 11.4 & 187 & 30.9  \\ \addlinespace[0.3em]
    546040 & 7858 $\pm$ 4 & 177.5 $\pm$ 7.9 & 63 & 34.9   \\ \addlinespace[0.3em]
    560238 & 6325 $\pm$ 5 & 146.2 $\pm$ 9.1 & 70 & 43.9  \\ \addlinespace[0.3em]
    569555 & 16622 $\pm$ 20 & 94.2 $\pm$ 24.9 & 71 & 39.4  \\ \addlinespace[0.3em]
    570227 & 12733 $\pm$ 7 & 165.6 $\pm$ 11.8 & 69 & 32.0  \\ \addlinespace[0.3em]
    3576053 & 15157 $\pm$ 6 & 199.0 $\pm$ 9.7 & 152 & 47.8  \\ \addlinespace[0.3em]
\bottomrule
    \end{tabular}
\end{table}

\begin{figure}
	\includegraphics[width=\columnwidth, alt={This shows an example profiles from pPXF full spectrum fitting of apertures along the major axis. There are four panels with position on the x-axis. The upper panel shows rotational velocity, the second velocity dispersion, the third stellar population age and the lower panel shows stellar population metallicity. Each parameter was measured using MILES and sMILES SSPs.}]{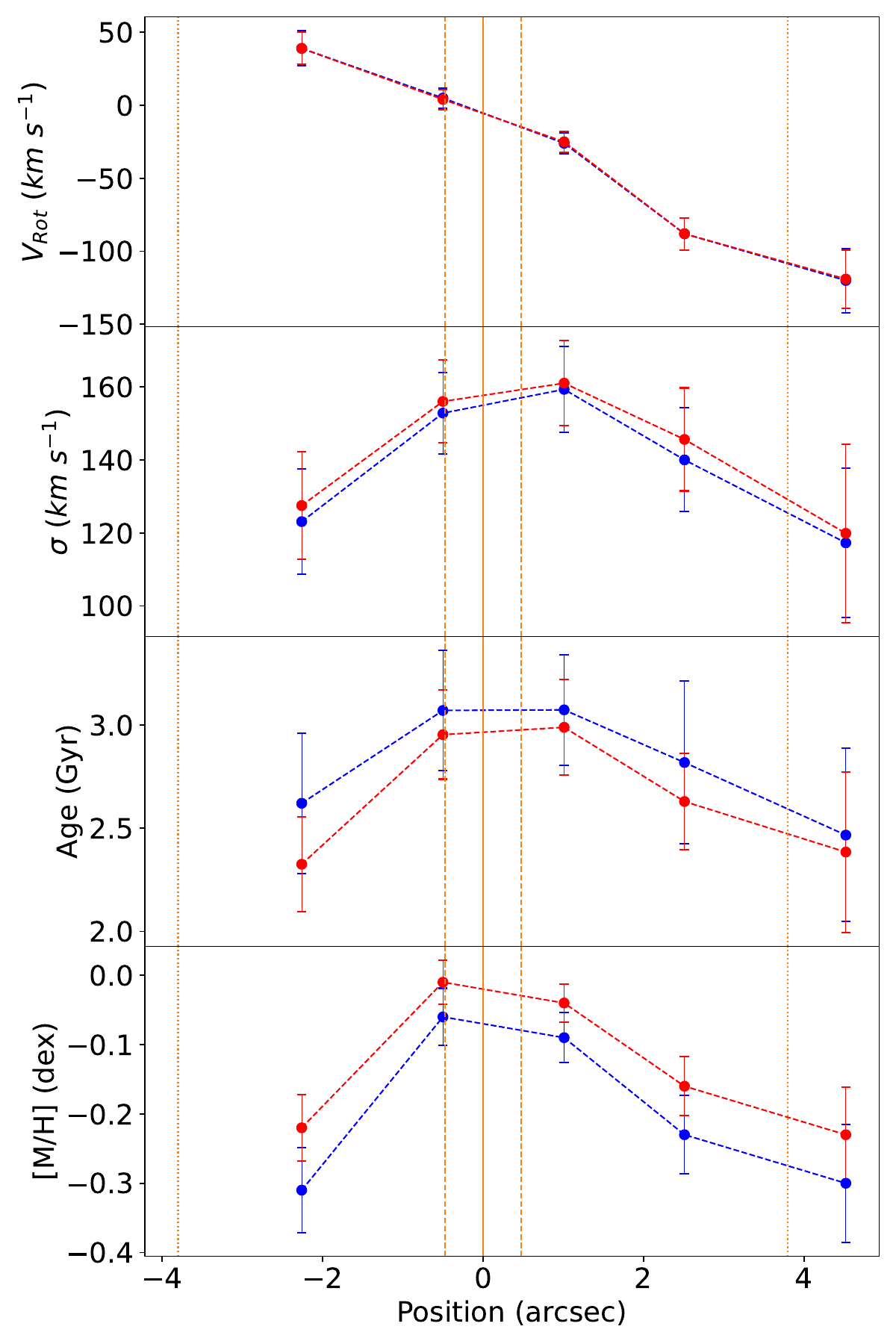}
    \caption{Example profiles of rotational velocity (V$_{Rot}$), line of sight velocity dispersion ($\sigma$), Age$_L$ and [M/H]$_L$ along the major axis of GAMA272990, with the x axis showing position relative to the galaxy centre. Results from fitting with MILES SSPs are shown in blue, results obtained using sMILES SSPs with [$\alpha$/Fe] = 0.0 are shown in red. Vertical orange lines show the spatial centre, $R_e$/8 and $R_e$. V$_{Rot}$ is the line of sight velocity for the aperture minus the central line of sight velocity. A complete set of major axis profile plots are available in the online Supplementary Material.}
    \label{fig:fig_4}
\end{figure}

We also extracted V, $\sigma$, 
Age$_L$ and [M/H]$_L$ from spatially binned apertures along the major axis using MILES and sMILES 2D fittings with pPXF. From these results we plotted profiles of rotational velocity (V$_{Rot}$), $\sigma$, Age$_L$ and [M/H]$_L$ along the major axis of each target ETG. An example is shown in Fig.~\ref{fig:fig_4} and the full set of profiles are available in online Supplementary Material. 
We find significant rotational support for all targets, with most lying close to the line of rotationally supported oblate spheroids. \citet{Bassett2017} found similar results for dusty ETGs. The kinematic profiles do not show any evidence of kinematic discontinuities, however, the S/N is not very high (see last column in Table~\ref{tab:table_2}). For the rest of the paper we concentrate mainly on results from central values (within $R_e$/8).

\subsection{Average age and metallicity by full spectrum fitting}
\label{subsec:analysis_population}  
As described above we extracted stellar population Age$_L$ and [M/H]$_L$ 
using MILES and sMILES 2D fittings. We used linear regularization \citep{Press1992}, controlled using the pPXF \textit{regul} keyword as discussed in \citet{Cappellari2017, Cappellari2023}, to smooth the solution for the stellar population parameters. The solution calculated by pPXF is defined by the weight of each individual template spectrum making up the solution and linear regularization constrains the weights between adjacent spectra on the age-metallicity grid to deliver a smoothed solution. We tested a range of values for \textit{regul} between 10 and 500 and selected a value of 100, based on Cappellari\footnote{https://pypi.org/project/ppxf/} and consistent with \citet{Cappellari2023} and \citet{Woo2024}, to give a smoothed solution without over-blurring. We maintained this value of \textit{regul} throughout to facilitate comparison of results from all fittings.
Whilst pPXF fitted gas emission lines and stellar populations simultaneously, we only use results from the stellar population fits. We masked 
gaps between the RSS detector chips using the pPXF \textit{goodpixels} keyword. These fittings delivered Age$_L$ and [M/H]$_L$, calculated by pPXF as the total luminosity weighted mean of the ages and metallicities of the template SSP spectra contributing to the solution \citep{Cappellari2017}.

We estimated uncertainties in Age$_L$ and [M/H]$_L$ using Monte Carlo simulations. 100 test spectra were generated by adding random Gaussian noise, based on the S/N ratio, to each bin of the observed spectrum. Each test spectrum was fitted using pPXF and uncertainties were calculated as the standard deviation of the 100 outputted Age$_L$ and [M/H]$_L$ values. Note that this Monte Carlo approach accounts for random flux errors, but not for any unknown systematic errors in the fits. This is valid for our analysis, where we are making relative comparisons between galaxies and samples.

\begin{figure*}
    \centering
    \begin{subfigure}{0.49\textwidth}
        \centering
        \includegraphics[width=\columnwidth, alt={Alt text here}]{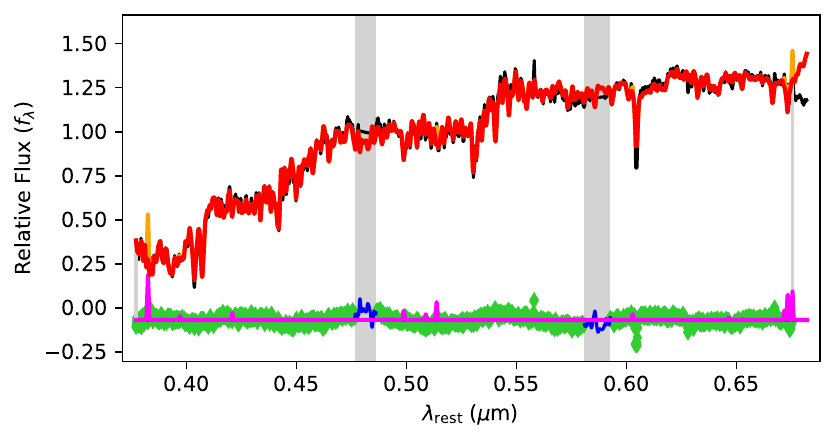}
    \end{subfigure}
    \hfill
    \begin{subfigure}{0.49\textwidth}
        \centering
        \includegraphics[width=0.96\columnwidth, alt={Alt text here}]{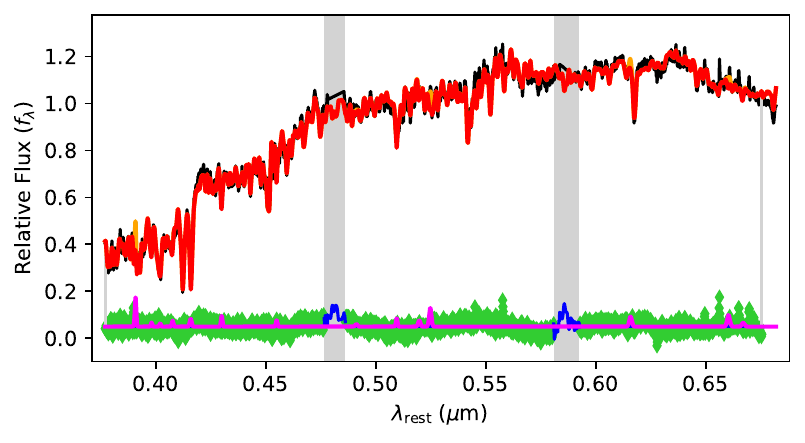}
    \end{subfigure}
    \vskip\baselineskip
    \begin{subfigure}{0.49\textwidth}
        \centering
        \includegraphics[width=\columnwidth, alt={Alt text here}]{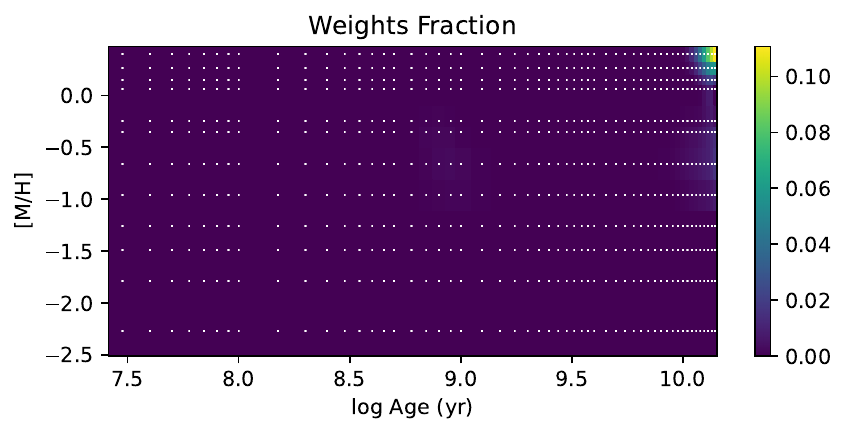}
    \end{subfigure}
    \hfill
    \begin{subfigure}{0.49\textwidth}
        \centering
        \includegraphics[width=\columnwidth, alt={This four panel figure shows examples of the output from pPXF. The upper two panels show x-y plots of the observed spectrum overlayed with the spectrum from combining SSPs in pPXF. Residuals show consistency between the observed and calculated spectra. The lower two panels show the weight of each SSP that contributed to the solution as a colour on a grid ot metallicity versus age. The lower left panel shows one old, high metallicity component and the lower right panel shows one old and one intermediate age stellar population component.}]{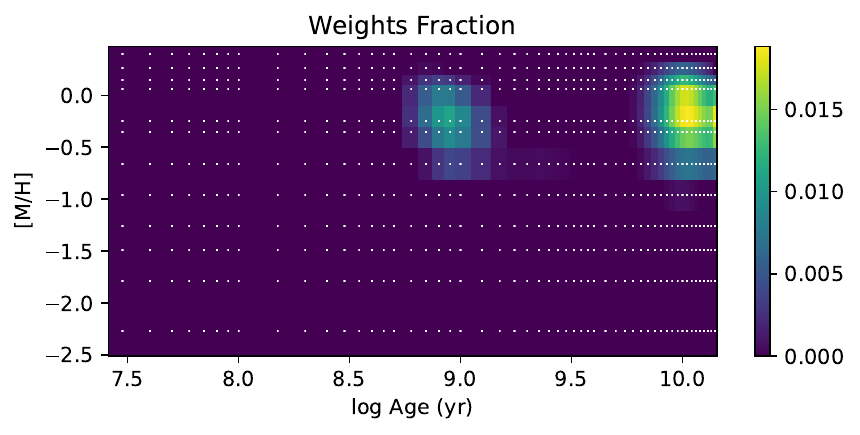}
    \end{subfigure}    
        \caption{Example outputs from luminosity weighted pPXF full spectrum fitting for GAMA422436 (left panels) and GAMA99687 (right panels). For each target the upper plot shows relative flux vs. wavelength with the observed spectrum in black, the best fit stellar spectrum generated by combining template SSPs in red, the gas emission fit in magenta, fitting residuals in green and masked areas in grey. The lower plots shows a grid of [M/H]$_L$ vs. log$_{10}$ Age$_L$, with white dots showing the position of each MILES template SSP. Post-regularization (luminosity weighted) template weight fraction at each grid point is represented by colour. See \citet{Cappellari2023} for a full explanation of pPXF plots. GAMA422436 contains a single old, metal rich stellar population at >10~Gyr. In contrast GAMA99687 contains two distinct stellar population groupings, at $\approx$0.9~Gyr and $\protect\goa$10~Gyr, both at sub-solar metallicity. A complete set of pPXF output plots are presented in Fig.~\ref{fig:ppxf_plots}.}
    \label{fig:fig_example_ppxf}
\end{figure*}

In addition to numerical averages for Age$_L$ and [M/H]$_L$ pPXF generates two plots from each fitting, with typical examples presented in Fig.~\ref{fig:fig_example_ppxf}, for central $R_e/8$ apertures. Upper panels show plots of the fit between the observed spectrum and the spectrum generated by pPXF by combining template SSPs. Lower panels show the weights fraction of SSPs used in the solution on an age vs. metallicity grid. The left side column shows a galaxy where there is a single old, metal-rich stellar population, the right side column is an example of a galaxy where there is an intermediate age and an old age component in the stellar population. In our sample of 15 dusty ETGs there are 15 with old population components of $\approx$10~Gyr or greater, 11 with intermediate components of 0.5~$\lessapprox$~Gyr~$\lessapprox$~2, i.e. GAMA79849, 85416, 99687, 136847, 227264, 272990, 298980, 560238, 569555, 570227 and 3576053, and 4 with young components of $\approx$0.1~Gyr or less, i.e. GAMA65075, 85416, 227264 and 298980. A complete set of pPXF age vs. metallicity grids for our 15 target ETGs are presented in Fig.~\ref{fig:ppxf_plots} of the Appendix. 

From the GAMA groups catalogue G3CGal version 10, \citep{Robotham2011}, available from the GAMA survey website, 11 of our dusty ETGs are in small groups containing between 2 to 19 galaxies, and 4 are isolated. We find examples of dusty ETGs with intermediate age populations both within groups (e.g. GAMA136847, 570227) and isolated (e.g. GAMA272990, 3576053). Therefore, there are no clear trends of dusty ETGs SFH with group membership from that catalogue.

To check if the intermediate age population components were artefacts from full spectrum fitting we used pPXF to fit synthetic test spectra which we generated by combining MILES SSPs. The SSPs used 
had either a metal-rich ([M/H]=+0.06) or metal-poor ([M/H]=-0.35) population. The SSPs were normalised, converted to a S/N ratio of 30 per wavelength bin by adding Gaussian noise, blurred to SALT instrument resolution of 5\r{A} and trimmed to our science spectra wavelengths. We generated test spectra  with only old population components of 13, 12 and 11~Gyr and test spectra with old population components of 13 and 12~Gyr plus a 2\% mass-weighted young component of 0.1~Gyr or intermediate age component of 1~Gyr. With these as input spectra, and using MILES SSPs for fitting, pPXF produced a solution containing only old population components in the first case, similar to the lower left panel of Fig.~\ref{fig:fig_example_ppxf}, and a solution with old and young or intermediate age population components in the second case. This recovery of the correct age components was true irrespective of the metallicity of the old and intermediate age components (see the top three rows in Fig.~\ref{fig:synthetic_spectra_plots}).
The metallicity was generally well recovered for an old model, and for the old component of composite age models, but not for the 1~Gyr component of composite models, where metallicity was overestimated. 
To assess the effect of early, extended star formation we generated another synthetic test spectrum containing equally mass-weighted components of 13, 8.5, 5.5, 3.5, 2.5, 1.5 and 1~Gyr, chosen to give approximately equal log$_{10}$(age)~increments across the age range, and fitted this with pPXF. The solution contained a spread of ages between $\approx$13 and 0.6~Gyr, but with an apparent dearth of stars around 3 to 5~Gyr. Again, pPXF returned a metallicity anomalously higher than the test spectrum for the youngest component. We present age vs. metallicity grids from these synthetic spectra fittings in Fig.~\ref{fig:synthetic_spectra_plots} of the Appendix. These tests help to show that pPXF is not producing solutions with spurious intermediate age components. This was our main goal for these tests, however, the simulations do reveal limitations in these pPXF fits to discrete models where the metallicities of younger components may not be accurately recovered. We note that intermediate age components in our fits to dusty ETG spectra are generally metal-poor rather than metal-rich.

Results from fitting the major axis spatially binned apertures using MILES and sMILES SSPs were used to plot major axis Age$_L$ and [M/H]$_L$ profiles for each target, with an example shown in the lower two panels of Fig.~\ref{fig:fig_4}. The full set of these profiles are available as Supplementary Material online.  
The plots show relatively close agreement in fits using MILES SSPs and sMILES SSPs, mostly within the estimated errors, as seen for example in Fig.~\ref{fig:fig_4}. The profiles generally show Age$_L$ is constant or decreasing with increasing radius, which aligns qualitatively with results within $R_e$ from previous investigations of ETGs, e.g. CALIFA (\citealt{Gonzalez2015}, their fig. 9), SAMI (\citealt{Ferreras2019}, their fig. 9) and MaNGA (\citealt{Liu2020} their fig. 13), while SAURON (\citealt{Kuntschner2010}, their fig. 12) shows age to be constant or increasing with radius within $R_e$. \citet{Katkov2015} produced major axis stellar population profiles of 9 ETGs using full spectrum fitting of longslit spectra from SALT and found 2 (NGC2917 and 3375) of the 9 ETGs to have age decreasing with increasing radius in the central regions. The profiles from our fittings generally show [M/H]$_L$ decreasing with increasing radius, which aligns qualitatively with results within $R_e$ from previous investigations of ETGs, e.g. SAURON, CALIFA (\citeauthor{Gonzalez2015}, their fig. 12) and MaNGA.

Age$_L$ and [M/H]$_L$ results for the central $R_e$/8 aperture of each target ETG, obtained using MILES and sMILES SSPs, are presented in Table~\ref{tab:table_4}. Results for GAMA298980 and 569555 appear anomalous, with unexpectedly low metallicities for their masses ([M/H]$_L$ of -0.73 and -1.09 dex respectively, from MILES SSP fits). These two targets show the highest H$\alpha$ equivalent widths, see Fig.~\ref{fig:Fig_1}, highest SFRs, see Fig.~\ref{fig:fig_2}, and they have strong emission lines in their spectra making it difficult to fit older stellar populations and absorption lines accurately. Therefore these two galaxies have been excluded from our subsequent plots. 

Central $R_e$/8 aperture Age$_L$ and [M/H]$_L$ results are plotted vs. ${\sigma_0}$ in Fig.~\ref{fig:fig_trends}. For comparison we include in Fig.~\ref{fig:fig_trends} (lower panels) trends from ATLAS$^{3D}$ which analysed spectra with a wavelength range from 4800 to 5380\r{A} \citep{Mcdermid2015} and trends from Lick indices SSP fitting of high S/N, stacked, passive galaxy spectra by \citet{Knowles2023}, based on SDSS stacked spectra from \citet{laBarbera2013}. For fits to our dusty ETGs we performed a 
Tukey biweight fit for each of our data sets using 
\textsc{statsmodels.robust.norms.TukeyBiweight} with cut-off parameter c=3.5. This provides a straight line fit to the trends that is more robust against individual outlier points than simple linear regression.

The upper left two panels of Fig.~\ref{fig:fig_trends} shows broad comparability between the Age$_L$ and [M/H]$_L$ results obtained from pPXF full spectrum fittings using either MILES or sMILES SSPs.
All Age$_L$ and [M/H]$_L$ results from pPXF fittings show a loose trend of increasing Age$_L$ and [M/H]$_L$ with increasing $\sigma_0$, qualitatively reproducing the positive correlations seen in ATLAS$^{3D}$ results \citep{Mcdermid2015}.  
The large scatter, particularly in Age$_L$, partly results from the intermediate age sub-comnponents required to fit most of these Dusty ETG spectra (see Appendix). Results from Lick indices SSP fitting are discussed in Section~\ref{subsec:analysis_alpha_lick}.

\subsection{Average age, metallicity and  \texorpdfstring{$\alpha$}{}-element abundance by full spectrum fitting}
\label{subsec:analysis_alpha_ppxf}  

We obtained values for [$\alpha$/Fe]$_L$ for the central $R_e$/8 aperture of each target ETG by repeating full spectrum fitting using pPXF as described in Section~\ref{subsec:analysis_population}. We used template SSPs from the three dimensional age-metallicity-$\alpha$-element abundance grid of sMILES SSPs \citep{Knowles2023}. We edited the pPXF \textsc{miles\_util.py} routine to add the $\alpha$-element abundance dimension thus enabling pPXF to assemble a three dimensional grid of SSPs (sMILES 3D) and we altered our pPXF setup routine to calculate and report luminosity weighted average $\alpha$-element abundance in addition to age and metallicity. From these fittings we obtained values for V, $\sigma_0$, Age$_L$, [M/H]$_L$ and [$\alpha$/Fe]$_L$ and their 1-sigma errors. 

\begin{figure*}
    \centering
    \includegraphics[width=\textwidth, alt={This figure consists of two rows each containing three plots. The upper row shows plots of results from full spectrum fitting and Lick index SSP fitting for age, metallicity and alpha-element abundance versus velocity dispersion and their trend lines. The trend lines show age and metallicity increasing with increasing velocity dispersion while alpha-element abundance appears to decrease with increasing velocity dispersion.}]{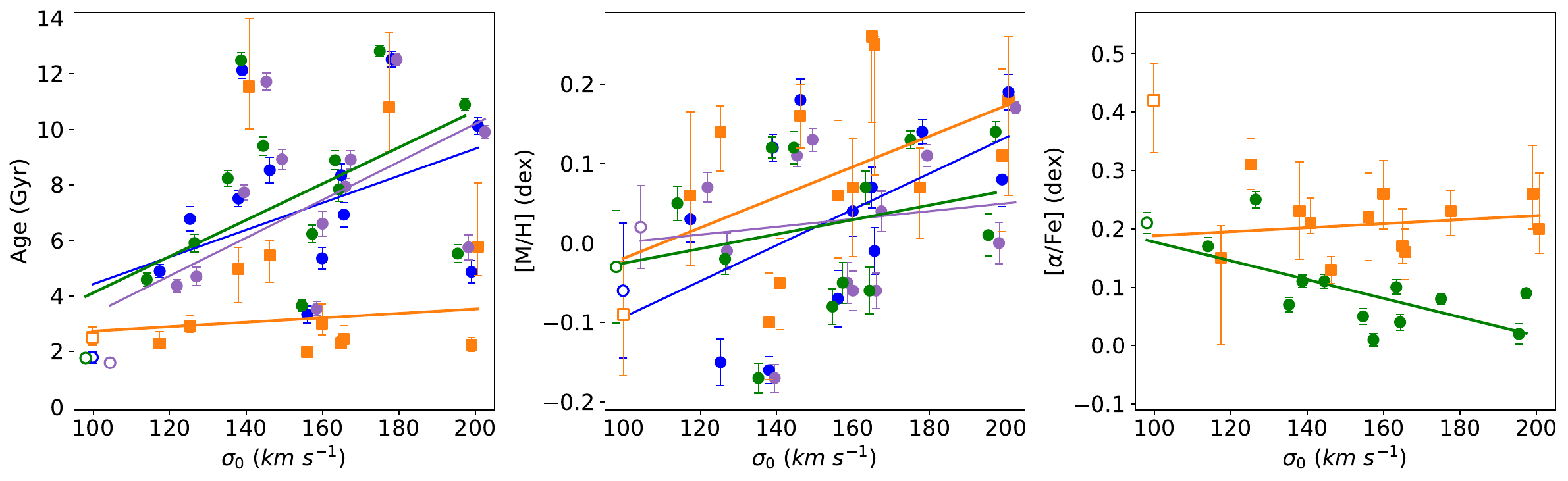}
    \includegraphics[width=\textwidth, alt={The lower row of three plots reproduces the trend lines for age, metallicity and alpha-element abundance versus velocity dispersion and adds their uncertainties, together with comparitor trend lines from ATLAS3D and La Barbera et al. (2013).}] {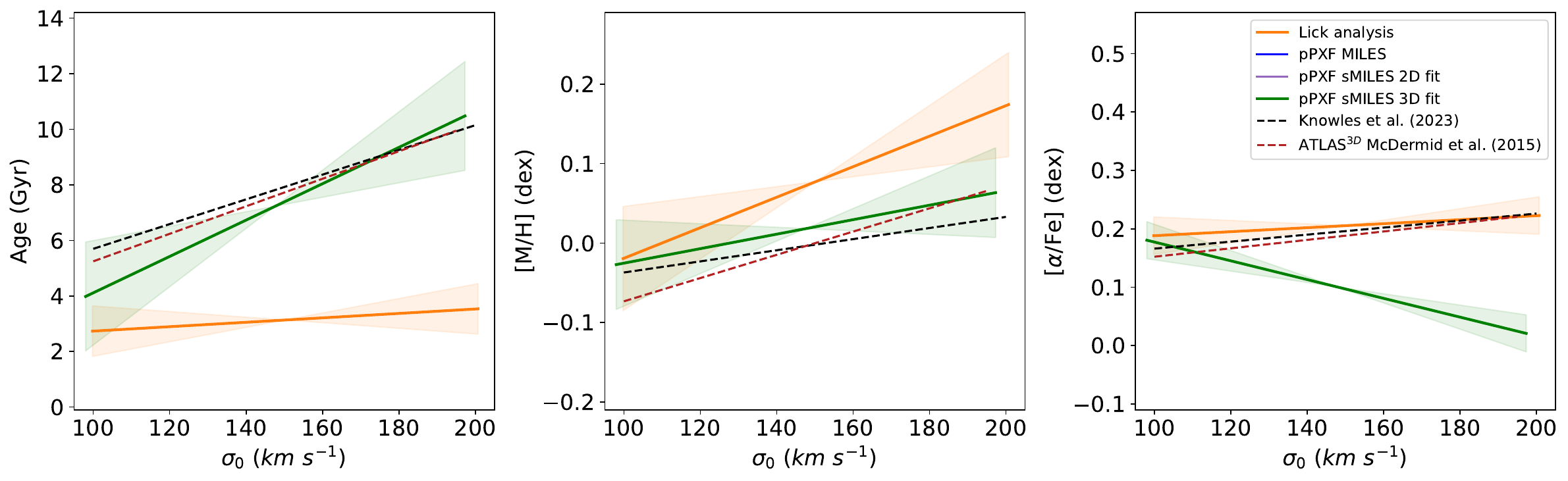}
    \caption{Upper panels show Age, [M/H] and [$\alpha$/Fe] from pPXF full spectrum fitting and Lick index SSP fitting of the central $R_e$/8 apertures versus ${\sigma_0}$. Results are shown as points with 1-sigma error bars, for 13 dusty ETGs, excluding the highest H$_\alpha$ emitters GAMA298980 and GAMA569555. Trend lines show Tukey biweight regression lines with a cut-off parameter of c=3.5 fitted to the points.  GAMA85416 is shown with open symbols. Results from pPXF full spectrum fitting are luminosity weighted. 
    Ages in the left panel show a loose trend, increasing with increasing ${\sigma_0}$ and there appears to be a systemic difference between ages from pPXF full spectrum fitting and those from Lick index SSP fitting. Metallicities in the centre panel show a loose trend, increasing with increasing ${\sigma_0}$. [$\alpha$/Fe] results from full spectrum fitting, in the right panel, show a loose trend, decreasing with increasing ${\sigma_0}$. [$\alpha$/Fe] from Lick index analysis appear to be systemically higher than from full spectrum fitting, with an accompanying change of slope. 
    The lower panels provide a comparison with best fit lines for ETGs in ATLAS$^{3D}$ (red dashed line) \citep{Mcdermid2015}, and for the \citet{laBarbera2013} stacked spectra analysed by \citet{Knowles2023} (black dashed line), from SSP fits to Lick indices.  
    Shaded areas in the lower panels show the 1-sigma confidence interval on the slope of each best fit line.} 
    \label{fig:fig_trends}
\end{figure*}

Values for $\sigma_0$, matched the results obtained in Section~\ref{subsec:analysis_kinematics} and \ref{subsec:analysis_population} within their 1-sigma uncertainties, see Fig.~\ref{fig:fig_3}. Results for Age$_L$, [M/H]$_L$ and [$\alpha$/Fe]$_L$ are presented in Table~\ref{tab:table_4} and plotted in Fig.~\ref{fig:fig_trends}. Age$_L$ and [M/H]$_L$ results generally match those from our previous full spectrum fittings (MILES and sMILES 2D). Our results show apparent trends of increasing Age$_L$ and [M/H]$_L$ with increasing $\sigma_0$, which agree qualitatively with the more well defined trends from the ATLAS$^{3D}$ project \citep{Mcdermid2015}. However, unlike ATLAS$^{3D}$, our pPXF full spectrum fitting results have a trend of decreasing [$\alpha$/Fe]$_L$ with increasing $\sigma_0$. \citet{Pernet2024} found that full spectrum fitting, e.g. using pPXF, with $\alpha$-element dependent SSPs, gives a decreasing trend of [$\alpha$/Fe] with increasing $\sigma$ (see their figure 1). This is in contrast to previous results that show an increasing trend of [$\alpha$/Fe] proxies with $\sigma$, for example in \citet{Johansson2012}, \citet{Conroy2014}, \citet{laBarbera2014} and \citet{Mcdermid2015}. 
\citeauthor{Pernet2024} conclude that the decreasing trend is an artefact of using full spectrum fitting with SSPs built from stellar population models which do not take account of variations in abundances of individual $\alpha$-elements. This is the likely cause of the decreasing trend in our [$\alpha$/Fe]$_L$ results obtained from pPXF full spectrum fitting over a large wavelength range with sMILES SSPs. 
Earlier studies, including those cited above, did not reveal this artefact because they were either relying mostly on Mg Lick indices to estimate [$\alpha$/Fe] (e.g. \citet{Johansson2012}, \citet{laBarbera2014}, \citet{Mcdermid2015}) or their full spectrum fits were constrained by their data to a narrow range, making the Mg features again their main source of information about [$\alpha$/Fe] (e.g. \citet{Mcdermid2015}). Varying individual elements \citet{Conroy2014} showed that [O/Fe] and [Mg/Fe] increase strongly with $\sigma$, but they did not show any fits varying [$\alpha$/Fe] as a whole.

\begin{landscape}
\begin{threeparttable}
    \centering
    \caption{Age$_L$, [M/H]$_L$ and [$\alpha$/Fe]$_L$ results from pPXF full spectrum fitting and Lick index SSP fitting, for the central $R_e$/8 apertures. Uncertainties are 1-sigma values calculated by Monte Carlo simulations.}
    \label{tab:table_4}
    \begin{tabular}{c @{\hspace{18pt}} c @{\hspace{6pt}} c @{\hspace{18pt}} c @{\hspace{6pt}} c @{\hspace{18pt}} c @{\hspace{6pt}} c @{\hspace{6pt}} c @{\hspace{18pt}} c @{\hspace{6pt}} c @{\hspace{6pt}} c}
    \toprule
    GAMA & \multicolumn{7}{c}{Full Spectrum Fitting} & \multicolumn{3}{c}{Lick index SSP fitting}\\
    ~ & \multicolumn{2}{c}{\centering MILES SSPs} & \multicolumn{2}{c}{\centering sMILES SSPs (2D grid)} & \multicolumn{3}{c}{\centering sMILES SSPs (3D grid)} & \multicolumn{3}{c}{~}\\
    ~ & Age$_L$  & [M/H]$_L$ & Age$_L$ & [M/H]$_L$ & Age$_L$ & [M/H]$_L$ & [$\alpha$/Fe]$_L$ & Age$_L$ & [M/H]$_L$ & [$\alpha$/Fe]$_L$ \\
     ~ & (Gyr) & (dex) & (Gyr) & (dex) & (Gyr) & (dex) & (dex)& (Gyr) & (dex) & (dex)\\ 
    \midrule
    65075   & 6.77 $\pm$ 0.44  & $-0.15$ $\pm$ 0.03 & 4.70 $\pm$ 0.33 & $-0.01$ $\pm$ 0.02 & 5.91 $\pm$ 0.31 & $-0.02$ $\pm$ 0.02 & 0.25 $\pm$ 0.01 & $2.94^{+0.36}_{-0.25}$ & $0.14^{+0.04}_{-0.05}$ & $0.32^{+0.05}_{-0.05}$ \\ \addlinespace[0.5em]
    79849   & 5.36 $\pm$ 0.39 & 0.04 $\pm$ 0.03 & 6.60 $\pm$ 0.45 & $-0.06$ $\pm$ 0.03 & 6.23 $\pm$ 0.32 & $-0.05$ $\pm$ 0.03 & 0.01 $\pm$ 0.01 & $2.87^{+0.68}_{-0.32}$ & $0.07^{+0.07}_{-0.09}$ & $0.25^{+0.08}_{-0.06}$ \\ \addlinespace[0.5em]
    85416   & 1.79 $\pm$ 0.20 & $-0.06$ $\pm$ 0.09 & 1.60 $\pm$ 0.17 & 0.02 $\pm$ 0.05 & 1.77 $\pm$ 0.18 & $-0.03$ $\pm$ 0.07 & 0.21 $\pm$ 0.02 & $2.50^{+0.23}_{-0.32}$ & $-0.10^{+0.08}_{-0.07}$ & $0.42^{+0.08}_{-0.11}$ \\ \addlinespace[0.5em]
    99687   & 7.51 $\pm$ 0.30 & $-0.16$ $\pm$ 0.02 & 7.73 $\pm$ 0.27 & $-0.17$ $\pm$ 0.02 & 8.24 $\pm$ 0.28 & $-0.17$ $\pm$ 0.02 & 0.07 $\pm$ 0.01 & $4.94^{+1.14}_{-1.06}$ & $-0.10^{+0.07}_{-0.07}$ & $0.23^{+0.10}_{-0.09}$ \\ \addlinespace[0.5em]
    136847  & 8.35 $\pm$ 0.40 & 0.07 $\pm$ 0.03 & 8.91 $\pm$ 0.33 & 0.04 $\pm$ 0.03 & 8.89 $\pm$ 0.34 & 0.07 $\pm$ 0.02 & 0.10 $\pm$ 0.01 & $2.29^{+0.17}_{-0.05}$ & $0.26^{+0.00}_{-0.11}$ & $0.17^{+0.04}_{-0.04}$ \\ \addlinespace[0.5em]
    227264  & 4.89 $\pm$ 0.24 & 0.03 $\pm$ 0.03 & 4.37 $\pm$ 0.22 & 0.07 $\pm$ 0.02 & 4.59 $\pm$ 0.23 & 0.05 $\pm$ 0.02 & 0.17 $\pm$ 0.02 & $2.32^{+0.36}_{-0.18}$ & $0.06^{+0.11}_{-0.10}$ & $0.15^{+0.05}_{-0.13}$ \\ \addlinespace[0.5em]
    227266  & 10.12 $\pm$ 0.29 & 0.19 $\pm$ 0.02 & 9.90 $\pm$ 0.23 & 0.17 $\pm$ 0.01 & 10.89 $\pm$ 0.21 & 0.14 $\pm$ 0.01 & 0.09 $\pm$ 0.01 & $6.09^{+1.98}_{-1.52}$ & $0.17^{+0.09}_{-0.12}$ & $0.21^{+0.11}_{-0.04}$ \\ \addlinespace[0.5em]
    272990  & 3.34 $\pm$ 0.31 & $-0.07$ $\pm$ 0.03 & 3.54 $\pm$ 0.25 & $-0.05$ $\pm$ 0.03 & 3.65 $\pm$ 0.21 & $-0.08$ $\pm$ 0.02 & 0.05 $\pm$ 0.01 & $1.98^{+0.11}_{-0.22}$ & $0.06^{+0.12}_{-0.06}$ & $0.22^{+0.07}_{-0.07}$ \\ \addlinespace[0.5em]
    298980  & 1.52 $\pm$ 0.30 & $-0.73$ $\pm$ 0.11 & 1.13 $\pm$ 0.16 & $-0.43$ $\pm$ 0.08 & 1.22 $\pm$ 0.11 & $-0.43$ $\pm$ 0.05 & 0.18 $\pm$ 0.02 & $2.73^{+1.62}_{-0.93}$ & $-0.49^{+0.29}_{-0.33}$ & $0.60^{+0.00}_{-0.00}$ \\ \addlinespace[0.5em]
    422436  & 12.12 $\pm$ 0.28 & 0.12  $\pm$ 0.02 & 11.72 $\pm$ 0.30 & 0.11 $\pm$ 0.01 & 12.48 $\pm$ 0.29 & 0.12 $\pm$ 0.01 & 0.11 $\pm$ 0.01 & $11.50^{+2.50}_{-1.50}$ & $-0.04^{+0.06}_{-0.06}$ & $0.22^{+0.03}_{-0.03}$ \\ \addlinespace[0.5em]
    546040  & 12.52 $\pm$ 0.28 & 0.14  $\pm$ 0.01 & 12.51 $\pm$ 0.20 & 0.11 $\pm$ 0.01 & 12.81 $\pm$ 0.20 & 0.13 $\pm$ 0.01 & 0.08 $\pm$ 0.01 & $10.50^{+3.00}_{-1.50}$ & $0.08^{+0.06}_{-0.05}$ & $0.22^{+0.05}_{-0.04}$ \\ \addlinespace[0.5em]
    560238  & 8.53 $\pm$ 0.46 & 0.18  $\pm$ 0.02 & 8.92 $\pm$ 0.37 & 0.13 $\pm$ 0.01 & 9.40 $\pm$ 0.34 & 0.12 $\pm$ 0.02 & 0.11 $\pm$ 0.01 & $5.49^{+0.37}_{-0.99}$ & $0.17^{+0.04}_{-0.04}$ & $0.13^{+0.02}_{-0.02}$ \\ \addlinespace[0.5em]
    569555  & 2.78 $\pm$ 0.28 & $-1.09$ $\pm$ 0.07 & 2.01 $\pm$ 0.22 & $-0.85$ $\pm$ 0.06 & 2.02 $\pm$ 0.26 & $-0.82$ $\pm$ 0.06 & $-0.04$ $\pm$ 0.03 & $1.91^{+0.72}_{-0.41}$ & $-0.66^{+0.20}_{-0.29}$ & $0.44^{+0.16}_{-0.37}$ \\ \addlinespace[0.5em]
    570227  & 6.93 $\pm$ 0.44 & $-0.01$ $\pm$ 0.04 & 7.94 $\pm$ 0.30 & $-0.06$ $\pm$ 0.02 & 7.84 $\pm$ 0.43 & $-0.06$ $\pm$ 0.03 & 0.04 $\pm$ 0.01 & $2.47^{+0.46}_{-0.14}$ & $0.23^{+0.03}_{-0.15}$ & $0.16^{+0.04}_{-0.04}$ \\ \addlinespace[0.5em]
    3576053 & 4.87 $\pm$ 0.40  & 0.08  $\pm$ 0.03 & 5.75 $\pm$ 0.44 & 0.00 $\pm$ 0.03 & 5.53 $\pm$ 0.32 & 0.01 $\pm$ 0.03 & 0.02 $\pm$ 0.02 & $2.25^{+0.25}_{-0.17}$ & $0.09^{+0.10}_{-0.09}$ & $0.28^{+0.06}_{-0.08}$ \\ \bottomrule
    \end{tabular}
\end{threeparttable}
\end{landscape}

\subsection{Average age, metallicity and \texorpdfstring{$\alpha$}{}-element abundance from Lick index SSP fitting}
\label{subsec:analysis_alpha_lick}  

\begin{figure}
    \centering
    \includegraphics[width=\columnwidth, alt={The figure shows an example of diagnostic plots produced by the Lick index SSP fitting program. Upper panels plot stellar population age, metallicity and alpha-element abundance results from Monte Carlo analysis of perturbed input spectra.}]{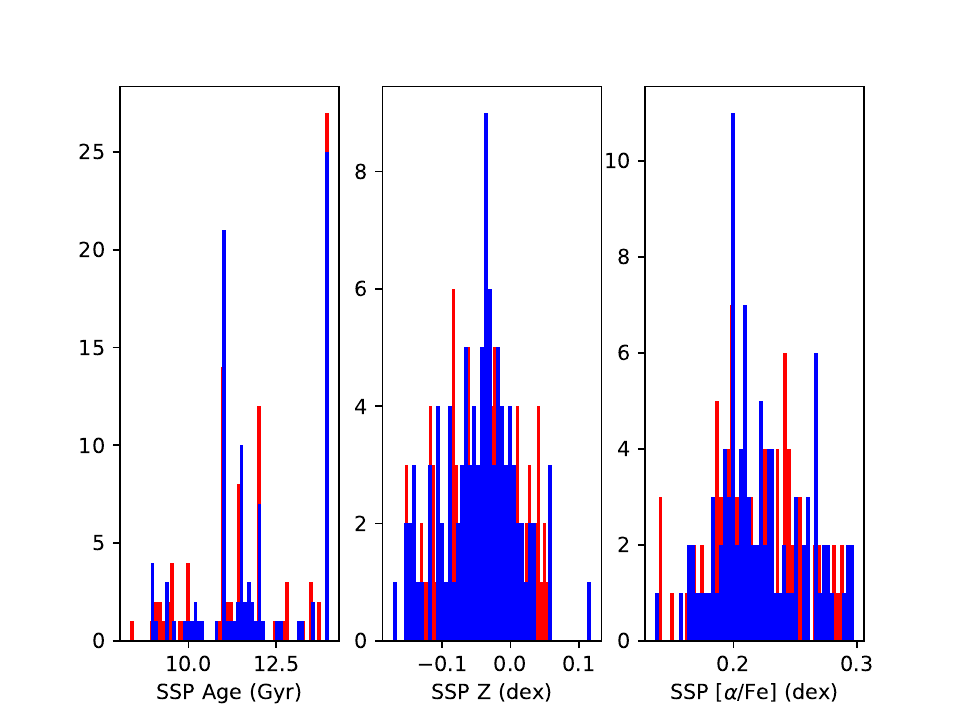}
    \includegraphics[width=\columnwidth, alt={Lower panels plot contours of reduced chi-squared for stellar population age vs. metallicity and metallicity vs. alpha element abundance.}]{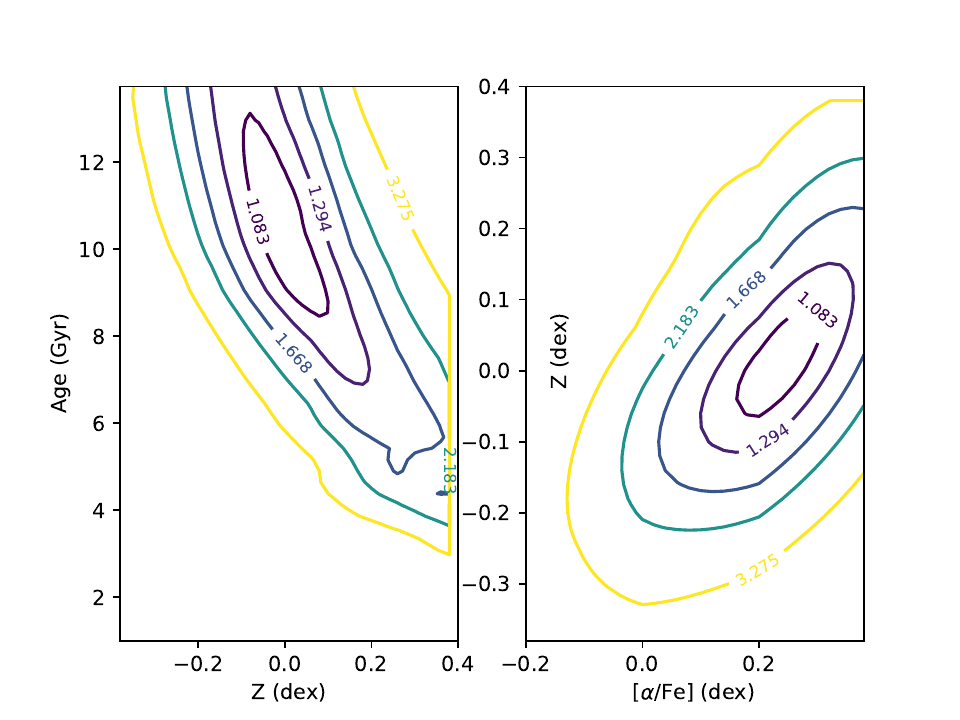}
    \caption{Example diagnostic plots from Lick index SSP fitting of GAMA422436. The upper panel shows fitting results from each of two Monte Carlo simulations of 100 cycles using a perturbed input spectrum. Blue data were generated from fits starting at 1~Gyr, red data from fits starting at an 7~Gyr. The lower panel shows Age vs. metallicity and metallicity vs. [$\alpha$/Fe] contour plots of normalised reduced $\chi^2$ across these parameter spaces. Degeneracies between parameters can be seen for Age and metallicity. In these plots Z=[M/H] in fits to sMILES SSPs (as defined in Section~\ref{sec:analysis}).
    These 2-parameter plots are shown at the best fit for the third parameter in each case (see Lick index SSP fitting results shown in Table~\ref{tab:table_4}).}
    \label{fig:fig_perturbations}
\end{figure}

We fitted values for age, [M/H] and [$\alpha$/Fe] from the central $R_e$/8 apertures by using Lick indices \citep{Worthey1994} to select the best fitting SSP. We adapted the process and Python code developed by \citet{Knowles2023}, described in their section~4.2. We used the following 15 Lick indices - H$\delta_A$, H$\delta_F$, G4300, H$\gamma_A$, H$\gamma_F$, Fe4383, Ca4455, Fe4531, H$\beta$, Mg$b$, Fe5270, Fe5335, Fe5406, Fe5709 and Fe5782, as used in \citet{Knowles2023}. These indcices mostly avoided 
SALT chip-gap wavelengths in the rest-frame corrected spectra of our Dusty ETGs, due to our careful choice of grating setup.
Sensitivity to [$\alpha$/Fe] is dominated by [Mg/Fe] in these Lick indices fits.
The code continuum normalises the target spectrum and SSPs, measures Lick line indices and performs multiple searches from different starting points using a Powell minimisation algorithm to determine the single best fitting SSP, see \citeauthor{Knowles2023} for a full explanation. An initial fit was used to identify the single sMILES SSP which best fits the Lick indices from the observed spectrum. This provided a base line for removal of emission lines. SSP fitting was then performed on the emission line removed observed spectrum using sMILES SSPs, to give an Age-[M/H]-[$\alpha$/Fe] solution. 1-sigma errors were estimated by perturbing the emission line removed observed spectrum with randomised flux errors and performing a Monte Carlo simulation, see the example diagnostic plots in Fig.~\ref{fig:fig_perturbations}. The upper panel shows number of fitting results on the y-axis for each of Age, [M/H] and [$\alpha$/Fe] from each of two Monte Carlo simulations of 100 cycles using the perturbed input spectrum, see \citet{Knowles2023}. The lower panel shows contour plots of reduced $\chi^2$, normalised by the best fit, for Age vs. [M/H] and [M/H] vs. [$\alpha$/Fe], from the Lick index SSP fitting. Each contour line is generated from the reduced $\chi^2$ achieved from fitting the individual SSPs represented on the x and y axes. The zone with lowest reduced $\chi^2$ represents the best fit Age-[M/H]-[$\alpha$/Fe] solution.

Lick index SSP fitting results are presented in Table~\ref{tab:table_4} and plotted in Fig.~\ref{fig:fig_trends} for comparison with results from pPXF full spectrum fitting. The Age plot shows a loose positive trend with increasing $\sigma_0$, with wide scatter of the results around the best fit lines. The metallicity plot shows a positive trend with increasing $\sigma_0$, again with wide scatter of the results around the best fit lines.  
However, our Lick index SSP fitting delivers systemically younger ages and higher metallicities when compared with our results from full spectrum fitting and the \citet{Mcdermid2015} results for ETGs in ATLAS$^{3D}$. 
Lick indices SSP fits to composite populations are known to be biased to the youngest components, in fitted Age, and to the older components, in fitted metallicity \citep{Serra2007}.
Our Lick index SSP fits may be dominated by young stellar population components, 
seen in the age vs. metallicity grids from pPXF composite fits in Fig.~\ref{fig:ppxf_plots}. 
This effect would not be seen in Lick index SSP fits of passive ETGs, such as in the \citet{laBarbera2013} data set. In the right panel of Fig.~\ref{fig:fig_trends} the best fit line to the [$\alpha$/Fe] abundance results from Lick index SSP fitting shows a shallow  positive trend with increasing $\sigma_0$, again with wide scatter.
The shallow slope of the best fit line indicates only a loose trend with $\sigma_0$. A positive trend  was also seen in the ATLAS$^{3D}$ results and \citet{Katkov2015} also reported a positive trend between [$\alpha$/Fe] and $\sigma_0$ from Lick index analysis and full spectrum fitting with \textsc{NBursts} for 21 ETGs, 9 of which were observed using SALT.

\citet{Knowles2023} applied the Lick index SSP fitting process to spectra from 24781 nearby passive ETGs (0.05 < z < 0.095) with 
$\sigma_0$ between 100 and 300 km~s$^{-1}$, from \citet{laBarbera2013, laBarbera2014, laBarbera2015}, stacked into bins of $\sigma_0$ to provide 18 stacked spectra with S/N ratios greater than several hundred. From these stacked SDSS spectra \citeauthor{Knowles2023} extracted best fitting SSP results for Age, [M/H] and [$\alpha$/Fe]. As a control sample we also plot the \citeauthor{Knowles2023} trend lines in Fig.~\ref{fig:fig_trends}. Our dusty ETGs are compared with \citeauthor{Knowles2023} in more detail in the Supplementary Material.
These best fit lines show close agreement between Age, [M/H] and [$\alpha$/Fe] results from \citeauthor{Knowles2023} with results from ATLAS$^{3D}$ \citep{Mcdermid2015}. Without the outlier GAMA85416, [$\alpha$/Fe] values and trend from our results show close agreement with the \citeauthor{Knowles2023} and \citeauthor{Mcdermid2015} trendlines. Like GAMA298980 and 569555, GAMA85416 has relatively high H$\alpha$ equivalent width, high SFR and strong emission lines in its spectra, making it difficult to fit the older stellar populations and absorption lines accurately.

Differences in age and metallicity results from our spectra vs. \citeauthor{Knowles2023} results are unlikely to be due to the Lick index SSP fitting method, because the same method was used for both analyses, but may be due to 
differences in average properties between our dusty ETGs and the passive ETGs from \citet{Knowles2023}. 

\section{Discussion}
\label{sec:discussion}  

In this section we discuss the results for our sample of dusty ETGs, where possible comparing them to reported results for non-dusty ETGs to put them into a wider context.

\subsection{Age, metallicity and \texorpdfstring{$\alpha$}{}-element abundance ratios}
\label{subsec:discussion_population}  

From our pPXF fits for Age$_L$ and [M/H]$_L$, 12 of the dusty ETGs in our sample of 15 showed evidence of younger stellar population components. After excluding results from GAMA298980 and 569555, our results from full spectrum fitting show both Age$_L$ and [M/H]$_L$ increasing with increasing $\sigma_0$. These trends are well known for ETGs in general, e.g. see \citet{Gallazzi2006}, \citet{Conroy2014}, \citet{Mcdermid2015}, and \citet{Liu2020}. However, when we compare our age and metallicity results from Lick index SSP fitting with results derived using the same method for high S/N ratio stacked spectra of passive galaxies \citep{Knowles2023} and with our results from full spectrum fitting (see Fig.~\ref{fig:fig_trends}
), we find our Lick index SSP fitting gives younger age populations and slightly higher metallicity, for our dusty ETGs sample compared with passive ETGs. We suggest this discrepancy is due to the sensitivity of Lick index SSP fitting to the presence of younger stellar population components aged $\lessapprox$2.5~Gyr \citep{Serra2007}. Tests using different Balmer lines in the SSP fit support this suggestion (details are given in the Supplementary Material).

Excluding GAMA85416, which has a high [$\alpha$/Fe] enhancement for its $\sigma_0$, our [$\alpha$/Fe] results from Lick index SSP fitting reproduce those of \citet{Mcdermid2015} and \citet{Knowles2023}, with a slightly positive trend with increasing ${\sigma_0}$, indicating that we do not detect any difference in star formation durations for dusty ETGs compared to other ETGs of similar ${\sigma_0}$. The negative trend with increasing ${\sigma_0}$ in our [$\alpha$/Fe]$_L$ results from full spectrum fitting can be explained by a bias introduced by full spectrum fitting using pPXF with $\alpha$-enhanced SSPs \citep{Pernet2024}.

\subsection{Check using SDSS stacked spectra}
\label{subsec:SDSSstacks}

As a check of our findings, we used the SDSS spectroscopic database (data release 18) to search for GAMA galaxies classified as ETGs (E or S0) in our Parent sample. We found 710 matches, with 360 having stellar velocity dispersion in the range 100 to 200 km s$^{-1}$ (similar to our SALT sample). Amongst these we selected the dustiest and non-dustiest using the following criteria: \\

Dustiest subset:

\begin{description}
    \item Detection $>5\sigma$ at 250$\mu$m from Herschel-ATLAS.
    \item Dust mass of $>5\times 10^{6}$ $M_{\sun}$ from GAMA \textsc{MagPhys} SED fits.
    \item These critera resulted in 29 galaxies with good quality SDSS spectra (S/N ratio $>10$).
\end{description}

Non-dustiest subset:

\begin{description}
    \item Non-detection at 250$\mu$m from Herschel-ATLAS.
    \item Dust mass of $<3\times 10^{-5}$ $M_{\sun}$ from GAMA \textsc{MagPhys} SED fits.
    \item These criteria resulted in 51 galaxies with good quality SDSS spectra (S/N ratio $>10$).
\end{description}

\begin{figure*}
    \centering
    \includegraphics[width=0.498\textwidth, alt={Contour plots showing stellar population parameter fits to stacked SDSS}]{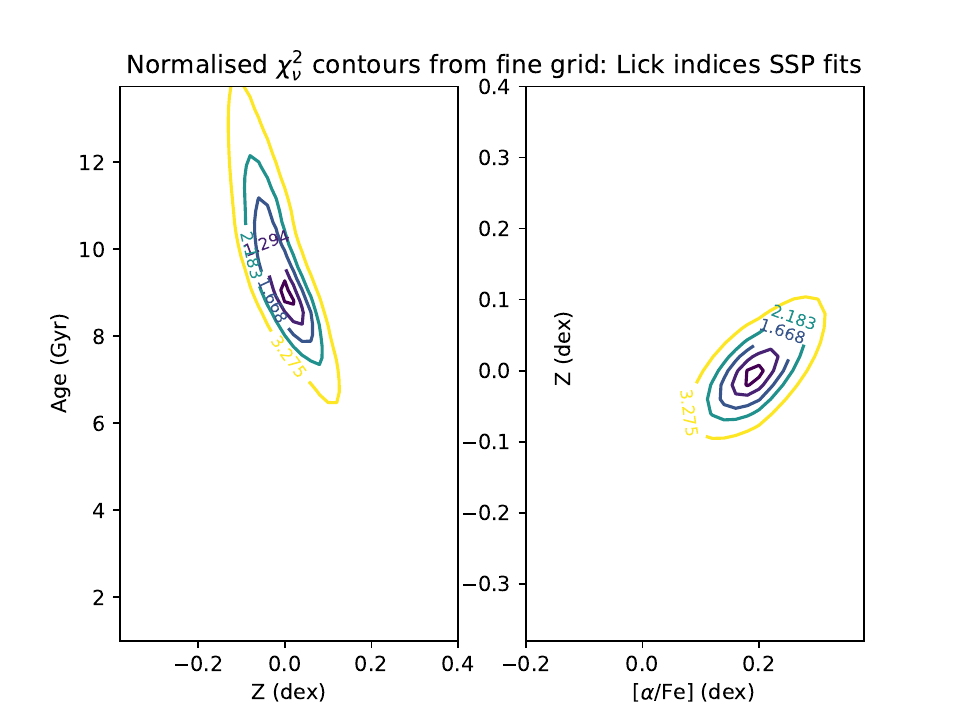}
    \includegraphics[width=0.498\textwidth, alt={Contour plots showing stellar population parameter fits to stacked SDSS}]{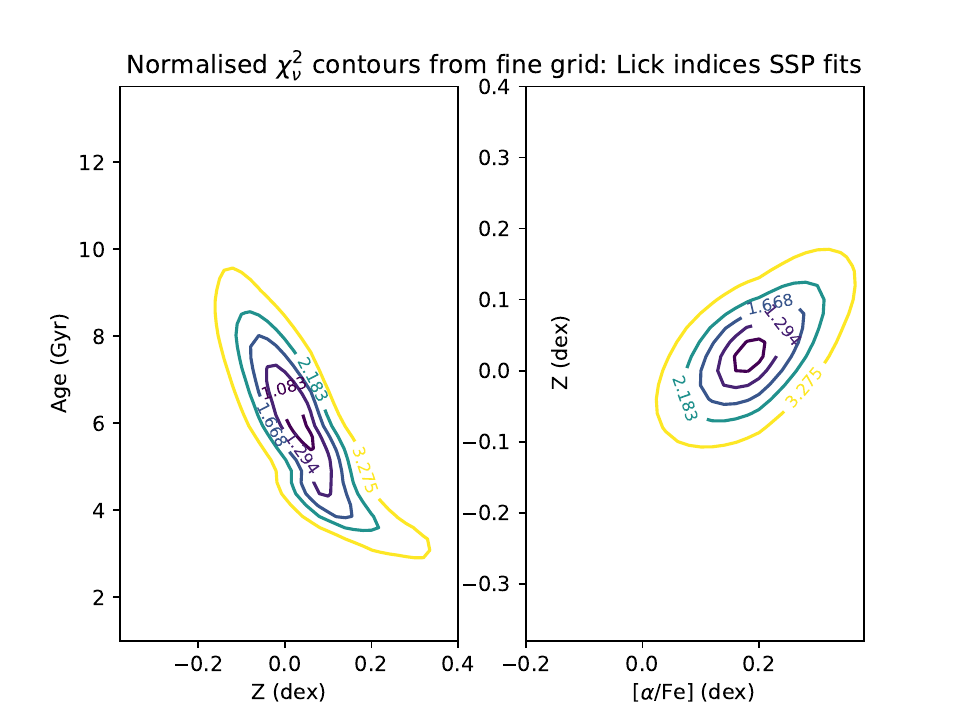}
    \caption{Lick indices SSP fits to SDSS stacks for non-dusty ETG (2 left panels) and dusty ETGs (2 right panels). See description of contours in Fig.~\ref{fig:fig_perturbations}. Inner contours show 68\% confidence levels for 3 fitted parameters, whilst other contours illustrate the shape of chi-squared space further out from the best fits.}
    \label{fig:fig_SDSS_contours}
\end{figure*}

The above two spectral datasets were stacked to maximize signal-to-noise in the SDSS optical spectra, leading to a dusty and non-dusty stacked spectrum respectively. The stacking software used  corrects for Galactic extinction, rebins to a linear wavelength scale in rest-frame air wavelengths, and outputs a median average of the input spectra (F. La Barbera - private communication; \citealt{laBarbera2013}).
These stacked spectra have S/N ratio per 0.5 Angstrom bin of $\approx 57$ and $\approx 131$ for the dusty and non-dusty stacked spectra respectively, calculated in the 4500 to 5500 Angstrom wavelength range. We fitted these two stacked spectra with our SSP fitting software (see Section~\ref{subsec:analysis_alpha_lick}) to determine how their luminosity weighted stellar population properties compared. After correcting the spectra for residual line emission the dusty ETG stack was found to have a luminosity-weighted average age of 6.12$\pm$0.59 Gyr and the non-dusty ETG stack was found to have age of 8.96$\pm$0.21 Gyr. Thus, the dusty ETG SDSS stack is 2.8$\pm$0.6 Gyr younger than the non-dusty ETG SDSS stack, supporting our finding from the SALT spectra that dusty ETGs generally have a population of younger stars than non-dusty ETGs. Fig.~\ref{fig:fig_SDSS_contours} shows the difference between SSP fits to these two SDSS stacks. The left panels show old age, non-dusty ETGs; the right panels show younger age, dusty ETGs. No difference is found in mean metallicity or [$\alpha$/Fe] ratio.

\begin{figure*}
    \centering
    \includegraphics[width=0.49\textwidth, alt={This figure shows results from pPXF fitting to stacked SDSS spectra. The spectral fit plane is shown.}]{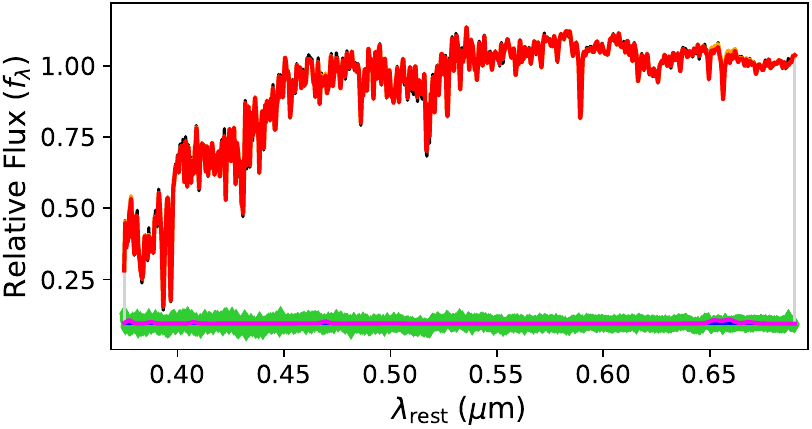}
    \includegraphics[width=0.49\textwidth, alt={This figure shows results from pPXF fitting to stacked SDSS spectra. The spectral fit plane is shown.}]{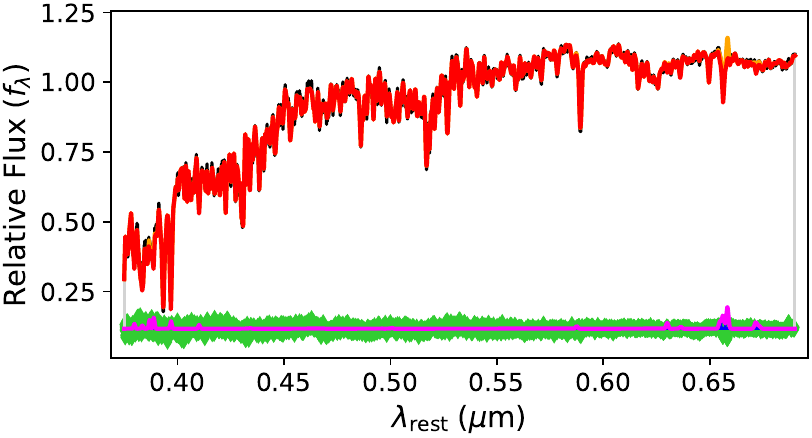}
    \includegraphics[width=0.49\textwidth, alt={This figure shows results from pPXF fitting to stacked SDSS spectra. The age versus metallicity plane is shown.}]{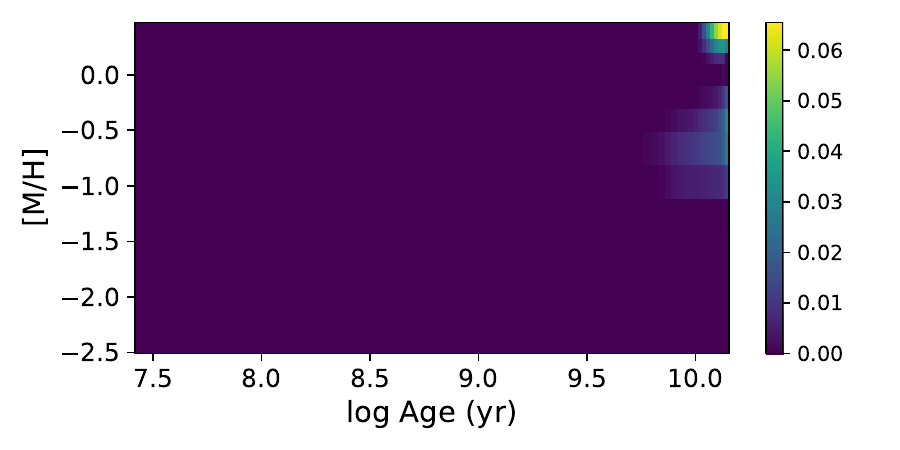}
    \includegraphics[width=0.49\textwidth, alt={This figure shows results from pPXF fitting to stacked SDSS spectra. The age versus metallicity plane is shown.}]{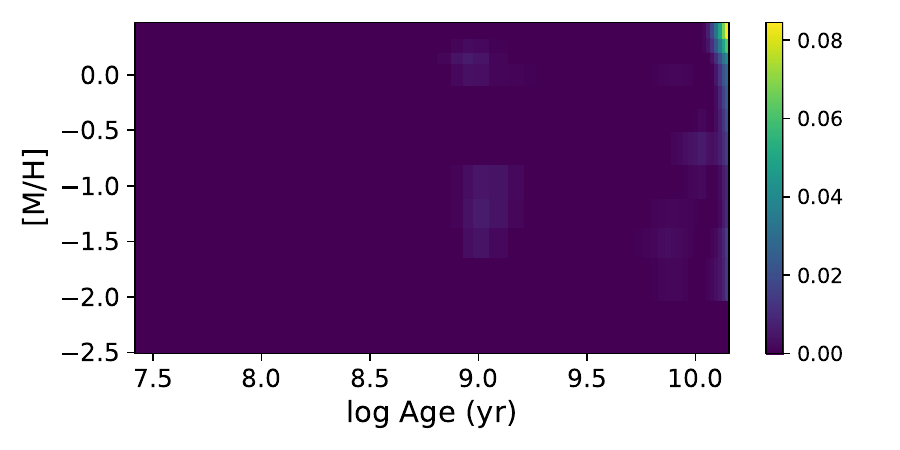}   
    \includegraphics[width=0.49\textwidth, alt={Alternative grid scaling used.}]{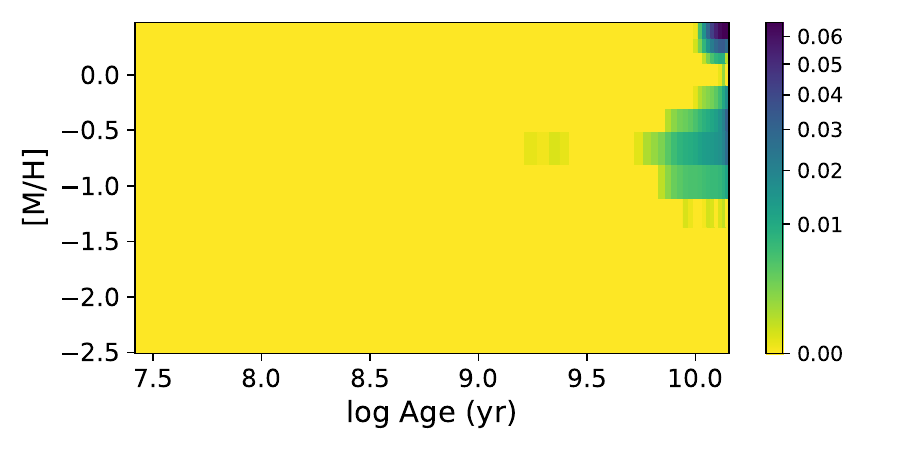}
    \includegraphics[width=0.49\textwidth, alt={Alternative grid scaling used.}]{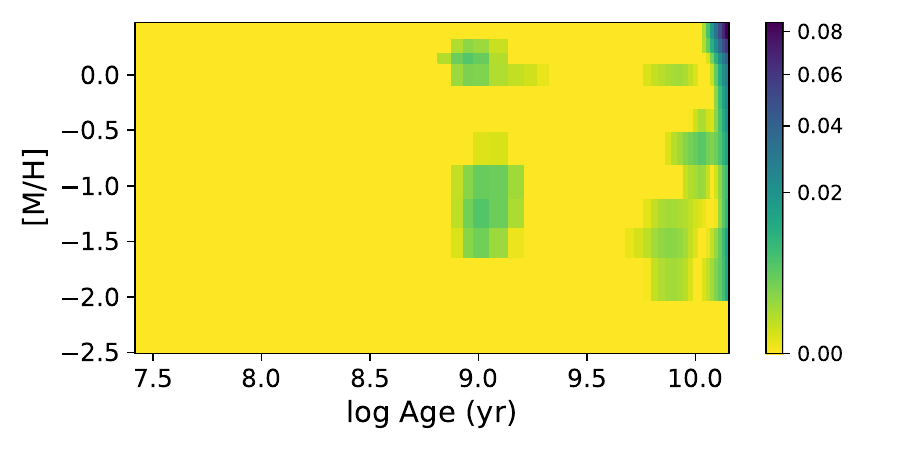}
    \caption{Top: Luminosity-weighted pPXF fits to SDSS stacked spectra for non-dusty ETGs (left panels) and dusty ETGs (right panels). Middle: Grid showing SSPs fitting these spectra (linearly scaled with default pPXF colour palette). See description in Fig.~\ref{fig:fig_example_ppxf}. Bottom: Alternative scaling (square-root power law, with reversed colour coding) to highlight where the fitted SSPs are for these SDSS stacked spectra. As well as old stars in both stacks, an intermediate age population shows up in the dusty ETGs stack but is missing in the non-dusty ETGs.}
    \label{fig:fig_SDSS_ppxf}
\end{figure*}

To investigate their star formation histories we used pPXF to map the age and metallicity distributions for these two SDSS stacks (see Section~\ref{subsec:analysis_population}). The results are shown in Fig.~\ref{fig:fig_SDSS_ppxf} where we again see evidence of a low-level, intermediate age contribution to the stellar population in the dusty ETGs  
(right panels). The non-dusty ETG stack shows no such intermediate age population (left panels). For these SDSS stacked spectra we also show grids scaled to highlight the contributions in these grid plots (lowest panels in Fig.~\ref{fig:fig_SDSS_ppxf}).

Therefore, we find evidence of intermediate age ($\approx$1 Gyr) populations in dusty ETGs from two independent sources of optical spectra (SALT and SDSS).

\subsection{Origins of Dust}
\label{subsec:dust}  
In this paper we set out to investigate the origins of dust in our sample of dusty ETGs by looking for differences in their stellar kinematics and stellar populations when compared to non-dusty ETGs. 
From our stellar population full spectrum fits we find evidence of intermediate-age populations, as well as older populations, in most of these dusty ETGs. Intermediate age populations ($\approx$0.5 to $\approx$2~Gyr) could be associated with the acquisition of ISM, including dust, into these galaxies. We use the pPXF age vs. metallicity grids, see Fig.~\ref{fig:ppxf_plots}, to test the following simple evolutionary scenarios.
A caveat to note is that 
metallicities are not well constrained in younger components of composite stellar populations. See also \citet{Cappellari2023}, their figure 15 and discussion.

\begin{enumerate}
    \item An early burst of star formation, then quiescence providing for self-enrichment of the ISM with metals during early star formation and enrichment of the ISM with dust expelled by evolved stars. This scenario may apply to GAMA227266, 422436 and 546040 where their pPXF plots show single, old, high-metallicity stellar populations. The origin of the dust is unclear in these three galaxies, with no evidence of any young stars. 
    \item An early burst of star formation plus an intermediate age burst of star formation, then quiescence. This scenario is as (i) above with the addition of a second burst of star formation $\approx$1~Gyr ago, triggered by secular evolution and using ISM enriched by evolution of the older stars. In this scenario the more recently formed stars would have higher metallicity than the old population. While many of the age vs. metallicity grids in Fig.~\ref{fig:ppxf_plots} show intermediate age stellar population components at 0.5~$\lessapprox$~Gyr~$\lessapprox$~2, none of these have higher metallicity than the older stellar population components.
    \item As (ii) above but with the second burst of star formation triggered by externally accreted gas e.g. from a minor merger, with metallicity of the gas determined by the SFH of the merging (secondary) galaxy. In this scenario more recently formed stars could have higher, lower or identical metallicity than the old population. 
    Notwithstanding the uncertainties in metallicities noted above, 
    from our age vs. metallicity grids in Fig.~\ref{fig:ppxf_plots} we attempted to estimate mean metallicity of the intermediate age stellar population component for these galaxies. From this, using the stellar mass vs. metallicity scaling relation from \citet{Trussler2021}, their fig. 2, we estimated the stellar mass of the postulated secondary galaxy. Using stellar masses of our dusty ETGs from Table~\ref{tab:table_1}, we estimated the stellar mass ratio between the primary and secondary galaxies, see Table~\ref{tab:table_6}. All ratios are >4:1 which is widely accepted as the mass ratio boundary between major and minor mergers, see for example \citet{Man2016}. The postulated secondary galaxy could have brought in dust and gas, which would have to survive since that intermediate age about 0.5 to 2~Gyr ago. This timescale is consistent with the timescale for ISM removal in ETGs of $\approx$2.3~Gyr, measured by \citet{Michalowski2024}.
    \item Early continuous star formation followed by quenching and secular evolution e.g. star formation in a late-type galaxy followed by quenching and morphological shift to ETG. This scenario is similar to scenario (i) but with star formation and
    ISM enrichment 
    happening over a more extended timescale.  
    Our tests with synthetic spectra in Section~\ref{subsec:analysis_population} indicate that continuous star formation will show in the top right corner of the age vs. metallicity grid as older and younger foci connected by a less dense zone, see lower right panel in Fig.~\ref{fig:synthetic_spectra_plots}. While this pattern appears in some of our observed age vs. metallicity grids in Fig.~\ref{fig:ppxf_plots} all these observed grids (e.g. GAMA79849, 272990, 570227), have an additional intermediate age component, making this an unlikely scenario for evolution of any of the ETGs in our sample.
    \item Early continuous star formation as (iv) above but with the second burst of star formation triggered by externally accreted gas. The two connected foci discussed in (iv) plus a separate intermediate population component is seen strongly in the age vs. metallicity grid for GAMA272990 and less strongly in the grids for GAMA79849 and 570227. Therefore this scenario may represent what happened in these three dusty ETG galaxies.
\end{enumerate}

\begin{table}
    \centering
    \caption{Estimation of the mass ratio of Primary to Secondary galaxies for the minor merger scenario discussed in Section~\ref{subsec:dust} item (iii).}
    \label{tab:table_6}
    \begin{tabular}{c c c c}
    \toprule
    GAMA & Metallicity of & Mass of & Mass Ratio \\ 
    ~ & Secondary & Secondary & Primary:Secondary \\
    ~ & [M/H] & $\times10^{10}$ ($M_{\sun}$) & ~ \\ 
    \midrule
    79849 & -0.25 & 1.0 & 4.5:1 \\
    99687 & -0.2 & 0.53 & 10:1 \\
    136847 & -0.2 & 0.53 & 8:1 \\
    227264 & -0.2 & 0.53 & 5:1 \\
    272990 & -0.1 & 0.35 & 9:1 \\
    570227 & -0.2 & 0.53 & 8:1 \\
    3576053 & -0.2 & 0.53 & 41:1 \\
    \bottomrule
    \end{tabular}
\end{table}

SFHs in the majority of dusty ETGs in our sample fit with formation scenarios which include ISM injection e.g. from an ISM-rich minor merger, which would support the theory that at least part of the dust is externally sourced. In scenarios (i), (ii) and (iii) we would expect the population from an old star formation burst to be [$\alpha$/Fe] enhanced, while in scenarios (iv) and (v) extended early star formation should give a population less [$\alpha$/Fe] enhanced. Three dimensional age vs. metallicity vs. [$\alpha$/Fe] grids from pPXF full spectrum fitting would allow determination of Age$_L$, [M/H]$_L$ and [$\alpha$/Fe]$_L$ for each stellar population component however, the bias in [$\alpha$/Fe]$_L$ results from pPXF full spectrum fitting with $\alpha$-enhanced SSPs shown by \citet{Pernet2024}, and shown in our Fig.~\ref{fig:fig_trends}, prevent use of our data in this way. We note that our [$\alpha$/Fe] results from Lick index SSP fitting match those of \citet{laBarbera2013} and \citet{Mcdermid2015}, see Fig.~\ref{fig:fig_trends}~and  Supplementary Material,  
and all of the ETGs in our sample show an overall [$\alpha$/Fe]$_L$ >0, see Table~\ref{tab:table_4} and Fig.~\ref{fig:fig_trends}. 

We reviewed published results to see if we could find similar analyses of the SFH of ETGs that had not been selected for their dusty nature. \citet{Mcdermid2015} studied the SFH in a large sample of ETGs from the ATLAS$^{3D}$ survey. They show differential mass fractions versus time for galaxies grouped into dynamical mass bins (\citeauthor{Mcdermid2015}, their fig. 14). Their plot gives a broad indication that the SFH timescale of lower mass galaxies extend over longer times, however it does not indicate detailed histories of individual galaxies. The galaxies in our current sample correspond to intermediate mass bins in their data, which do not show star formation younger than 3~Gyrs.

More recently, \citet{Johnston2022} analysed a sample of 78 individual lenticular galaxies from their BUDDI-MANGA project. They show that on average, for lenticular galaxies with stellar masses >10$^{10}$ $M_{\sun}$, stellar mass formed before 3~Gyr ago, with most of the mass forming at much earlier times (\citeauthor{Johnston2022}, their fig.  10). It is a similar picture for their sample of Sa bulges (\citet{Jegatheesan2024}, their fig. 6, top middle panel). The same group have also studied three individual elliptical galaxies, from MUSE data and plotted their results in a similar way to our Fig.~\ref{fig:ppxf_plots}. Their three ellipticals have older star formation that shows no signs of intermediate age populations as young as $\approx$1~Gyr (see \citet{Jegatheesan2025}, their fig. 9).

Therefore, our sample of dusty ETGs reveals more intermediate age contributions to the stellar mass than for other samples of ETGs that are not selected by their levels of dust content. However, from a sample of ETGs selected for their HI detections, \citet{Boardman2017} show maps which indicate the presence of younger populations (their figs. 27, 28 and 29). 

\subsection{GAMA272990}
\label{subsec:272990}  
GAMA272990 is particularly interesting because previous work has shown it to have a clear E morphology, with evidence for high molecular gas and dust content and here we show that it has younger components in the stellar population. Using submillimeter observations of CO from the Atacama Large Millimetre/submillimeter Array (ALMA) \citet{Sansom2019} showed the presence of a massive, extended rotating molecular gas disc of $\approx$10 arcsec diameter. Further analysis by \citet{Glass2022} confirmed that the molecular gas disc is star forming and has asymmetry, suggesting a minor merger in relatively recent times.

Our stellar kinematic results show that GAMA272990 sits below the line of isotropic oblate spheroids that are flattened only by rotation, as expected for elliptical galaxies \citep{Cappellari2016}. Our major axis profiles show stellar $\sigma$, Age$_L$ and [M/H]$_L$ all reducing with increasing radius, with no discontinuities within $R_e$ (see Fig.~\ref{fig:fig_4}). 
Full spectrum fitting revealed the presence of three stellar populations within the central $R_e$/8  
aperture, one old population at $\goa$10~Gyr and two intermediate age components, one at $\approx$3~Gyr and one aged between 0.5 and 1.5~Gyr, see Fig.~\ref{fig:ppxf_plots}. Bridging between the $\goa$10~Gyr and $\approx$3~Gyr populations suggests some level of early ongoing star formation as discussed in Section~\ref{subsec:dust} scenario (v). Lick index SSP fitting indicates that [$\alpha$/Fe] is enhanced within the central $R_e$/8 radius aperture. Our results support the conclusion of \citet{Glass2022} that GAMA272990 has undergone a gas-rich merger resulting in the molecular gas disc structure, elliptical morphology and rejuvenated star formation. We therefore
suggest a link between that merger, which contributed to the molecular gas, and the youngest age stellar population component, see the age vs. metallicity grid for GAMA272990 in  Fig.~\ref{fig:ppxf_plots}.

\section{Conclusions}
\label{sec:conclusions}  

We obtained major-axis, longslit optical spectra of 15 particularly dusty ETGs using SALT. Their classification as ETGs was based on morphology determined using carefully analysed survey data, including SDSS images \citep{Moffett2016} plus information from KiDS images \citep{Glass2024}. For each target we extracted S/N based profile apertures along the major axis, mostly within $\approx$$R_e$, and an aperture covering the central $R_e/8$ radius. From the major axis apertures we extracted kinematic profiles, i.e. V and ${\sigma}$ and stellar population profiles, i.e. Age$_L$, [M/H]$_L$, using full spectrum fitting. From the central $R_e/8$ radius aperture we extracted these parameters plus [$\alpha$/Fe]$_L$, using both full spectrum composite fits and Lick index SSP fits. Our main findings are: 

\begin{enumerate}
    \item Results from full spectrum fitting show the dusty ETGs in our sample have comparable luminosity weighted population ages compared to passive galaxies at the same ${\sigma_0}$. However, ages from Lick index SSP fitting are systematically younger than from full spectrum fitting. pPXF full spectrum fitting provides evidence of younger stellar population components in 12 of the 15 targets, see Fig. \ref{fig:ppxf_plots}, i.e. all except GAMA227266, 422436 and 546040, but with only weak evidence of a younger stellar population component in GAMA560238.
    \item Full spectrum fitting and Lick index SSP fitting of the central $R_e/8$ aperture of our dusty ETGs both show stellar population average age and metallicity increasing with ${\sigma_0}$, but with large scatter, see Fig.~\ref{fig:fig_trends}. Average age estimates are systematically younger from SSP fitting because that is more sensitive to younger populations.
    \item We tested stacked SDSS spectral data for dusty and non-dusty ETGs from our GAMA Parent Sample, to provide a test of our results using different data and to allow us to select a control case. These SDSS spectra show evidence of an intermediate age contribution to the stellar population in the dusty ETGs stack, which is not present in the non-dusty ETGs stack.
   \item Several history scenarios were considered against our findings of intermediate age populations in most dusty ETGs in our sample. A past influx of gas from an interaction or merger could have produced intermediate age stars, but more work is needed to understand why dust would still be present.
   \item Three of our 15 dusty ETGs observed with SALT optical spectroscopy show no evidence for anything other than an old stellar population. The high level of dust found in these three galaxies is unexpected, given the old age of the population. Their gas and dust masses will be explored further in {\textcolor{blue}{Glass et al. in preparation.}}
    \item Using pPXF age vs. metallicity grids (see Fig.~\ref{fig:ppxf_plots}) we find that the contributions from intermediate age stellar population components have lower metallicities than the host galaxies (but note that metallicities are uncertain in the intermediate age component). These lower metallicities and intermediate age components, along with our tests of simple star formation scenarios, suggest that at least some of the dust in our dusty ETG sample is externally sourced, e.g. via minor mergers. 
\end{enumerate}

Future spectroscopic studies of these ETGs and the parent GAMA sample, at high S/N (e.g. 4MOST-WAVES survey\footnote{https://wavesurvey.org/} or DESI survey\footnote{https://www.desi.lbl.gov/}) and an improved understanding of the use of full spectrum fitting {using SSPs with individual $\alpha$-element enhancements, are needed to clarify these findings. 

\section*{Acknowledgements}
GAMA is a joint European-Australasian project based around a spectroscopic campaign using the Anglo-Australian Telescope. The GAMA input catalogue is based on data taken from the Sloan Digital Sky Survey and the UKIRT Infrared Deep Sky Survey. Complementary imaging of the GAMA regions is being obtained by a number of independent survey programmes including GALEX MIS, VST KiDS, VISTA VIKING, WISE, Herschel-ATLAS, GMRT and ASKAP providing UV to radio coverage. GAMA is funded by the STFC (UK), the ARC (Australia), the AAO, and the participating institutions listed on the GAMA survey web portal at www.gama-survey.org/.

The spectra analysed in this paper were obtained with the Southern African Large Telescope (SALT), under programmes 2019-1-SCI-012, 2019-2-SCI-003, 2020-1-SCI-006 and 2020-2-SCI-005 (P.I. A E Sansom). Spectra for GAMA65075 were obtained from the SALT Data Archive at https://ssda.saao.ac.za.

Full spectrum fitting for analysis of stellar kinematics and populations made use of the pPXF package \citep{Cappellari2017} using SSP libraries available from the MILES website at miles.iac.es. 

We acknowledge the assistance of J. Bannister with configuration of pPXF for [$\alpha$/Fe] analysis, and A. T. Knowles, I. Ferreras and C. C. Popescu for their helpful comments. We thank F. La Barbera for the use of his SDSS spectral stacking software. Thanks go to an anonymous referee for their comments, which helped to improve this paper.


\section*{Data Availability}
\label{sec:data}
Some parameters used within this work are available through GAMA public data releases. Raw data used to generate the longslit spectra are available from the SALT Data Archive. Specific data products generated for this paper will be shared on reasonable request to the authors.


\bibliographystyle{mnras}
\bibliography{references_r3} 

\begin{thebibliography}{}
\makeatletter
\relax
\def\mn@urlcharsother{\let\do\@makeother \do\$\do\&\do\#\do\^\do\_\do\%\do\~}
\def\mn@doi{\begingroup\mn@urlcharsother \@ifnextchar [ {\mn@doi@}
  {\mn@doi@[]}}
\def\mn@doi@[#1]#2{\def\@tempa{#1}\ifx\@tempa\@empty \href
  {http://dx.doi.org/#2} {doi:#2}\else \href {http://dx.doi.org/#2} {#1}\fi
  \endgroup}
\def\mn@eprint#1#2{\mn@eprint@#1:#2::\@nil}
\def\mn@eprint@arXiv#1{\href {http://arxiv.org/abs/#1} {{\tt arXiv:#1}}}
\def\mn@eprint@dblp#1{\href {http://dblp.uni-trier.de/rec/bibtex/#1.xml}
  {dblp:#1}}
\def\mn@eprint@#1:#2:#3:#4\@nil{\def\@tempa {#1}\def\@tempb {#2}\def\@tempc
  {#3}\ifx \@tempc \@empty \let \@tempc \@tempb \let \@tempb \@tempa \fi \ifx
  \@tempb \@empty \def\@tempb {arXiv}\fi \@ifundefined
  {mn@eprint@\@tempb}{\@tempb:\@tempc}{\expandafter \expandafter \csname
  mn@eprint@\@tempb\endcsname \expandafter{\@tempc}}}

\bibitem[\protect\citeauthoryear{{Agius} et~al.,}{{Agius}
  et~al.}{2013}]{Agius2013}
{Agius} N.~K.,  et~al., 2013, \mn@doi [\mnras] {10.1093/mnras/stt310}, \href
  {https://ui.adsabs.harvard.edu/abs/2013MNRAS.431.1929A} {431, 1929}

\bibitem[\protect\citeauthoryear{{Agius} et~al.,}{{Agius}
  et~al.}{2015}]{Agius2015}
{Agius} N.~K.,  et~al., 2015, \mn@doi [\mnras] {10.1093/mnras/stv1191}, \href
  {https://ui.adsabs.harvard.edu/abs/2015MNRAS.451.3815A} {451, 3815}

\bibitem[\protect\citeauthoryear{{Bassett} et~al.,}{{Bassett}
  et~al.}{2017}]{Bassett2017}
{Bassett} R.,  et~al., 2017, \mn@doi [\mnras] {10.1093/mnras/stx1000}, \href
  {https://ui.adsabs.harvard.edu/abs/2017MNRAS.470.1991B} {470, 1991}

\bibitem[\protect\citeauthoryear{{Boardman} et~al.,}{{Boardman}
  et~al.}{2017}]{Boardman2017}
{Boardman} N.~F.,  et~al., 2017, \mn@doi [\mnras] {10.1093/mnras/stx1835},
  \href {https://ui.adsabs.harvard.edu/abs/2017MNRAS.471.4005B} {471, 4005}

\bibitem[\protect\citeauthoryear{{Bournaud}, {Jog}  \& {Combes}}{{Bournaud}
  et~al.}{2005}]{Bournaud2005}
{Bournaud} F.,  {Jog} C.~J.,   {Combes} F.,  2005, \mn@doi [\aap]
  {10.1051/0004-6361:20042036}, \href
  {https://ui.adsabs.harvard.edu/abs/2005A&A...437...69B} {437, 69}

\bibitem[\protect\citeauthoryear{{Bournaud}, {Jog}  \& {Combes}}{{Bournaud}
  et~al.}{2007}]{Bournaud2007}
{Bournaud} F.,  {Jog} C.~J.,   {Combes} F.,  2007, \mn@doi [\aap]
  {10.1051/0004-6361:20078010}, \href
  {https://ui.adsabs.harvard.edu/abs/2007A&A...476.1179B} {476, 1179}

\bibitem[\protect\citeauthoryear{{Buckley}, {Swart}  \& {Meiring}}{{Buckley}
  et~al.}{2006}]{Buckley2006}
{Buckley} D. A.~H.,  {Swart} G.~P.,   {Meiring} J.~G.,  2006, in {Stepp} L.~M.,
   ed.,  Society of Photo-Optical Instrumentation Engineers (SPIE) Conference
  Series Vol. 6267, Society of Photo-Optical Instrumentation Engineers (SPIE)
  Conference Series. p. 62670Z, \mn@doi{10.1117/12.673750}

\bibitem[\protect\citeauthoryear{{Buckley} et~al.,}{{Buckley}
  et~al.}{2008}]{Buckley2008}
{Buckley} D.~A.~H.,  et~al., 2008, in {McLean} I.~S.,  {Casali} M.~M.,  eds,
  Society of Photo-Optical Instrumentation Engineers (SPIE) Conference Series
  Vol. 7014, Ground-based and Airborne Instrumentation for Astronomy II. p.
  701407, \mn@doi{10.1117/12.789438}

\bibitem[\protect\citeauthoryear{{Bundy} et~al.,}{{Bundy}
  et~al.}{2015}]{Bundy2015}
{Bundy} K.,  et~al., 2015, \mn@doi [\apj] {10.1088/0004-637X/798/1/7}, \href
  {https://ui.adsabs.harvard.edu/abs/2015ApJ...798....7B} {798, 7}

\bibitem[\protect\citeauthoryear{{Burgh}, {Nordsieck}, {Kobulnicky},
  {Williams}, {O'Donoghue}, {Smith}  \& {Percival}}{{Burgh}
  et~al.}{2003}]{Burgh2003}
{Burgh} E.~B.,  {Nordsieck} K.~H.,  {Kobulnicky} H.~A.,  {Williams} T.~B.,
  {O'Donoghue} D.,  {Smith} M.~P.,   {Percival} J.~W.,  2003, in {Iye} M.,
  {Moorwood} A. F.~M.,  eds,  Society of Photo-Optical Instrumentation
  Engineers (SPIE) Conference Series Vol. 4841, Instrument Design and
  Performance for Optical/Infrared Ground-based Telescopes. pp 1463--1471,
  \mn@doi{10.1117/12.460312}

\bibitem[\protect\citeauthoryear{Burkert \& Naab}{Burkert \&
  Naab}{2003}]{Burkert2003}
Burkert A.,  Naab T.,  2003, Major Mergers and the Origin of Elliptical
  Galaxies.
Springer Berlin Heidelberg, Berlin, Heidelberg, pp 327--339,
  \mn@doi{10.1007/978-3-540-45040-5_27}, \url
  {https://doi.org/10.1007/978-3-540-45040-5_27}

\bibitem[\protect\citeauthoryear{{Cappellari}}{{Cappellari}}{2016}]{Cappellari2016}
{Cappellari} M.,  2016, \mn@doi [\araa] {10.1146/annurev-astro-082214-122432},
  \href {https://ui.adsabs.harvard.edu/abs/2016ARA&A..54..597C} {54, 597}

\bibitem[\protect\citeauthoryear{{Cappellari}}{{Cappellari}}{2017}]{Cappellari2017}
{Cappellari} M.,  2017, \mn@doi [\mnras] {10.1093/mnras/stw3020}, \href
  {https://ui.adsabs.harvard.edu/abs/2017MNRAS.466..798C} {466, 798}

\bibitem[\protect\citeauthoryear{{Cappellari}}{{Cappellari}}{2023}]{Cappellari2023}
{Cappellari} M.,  2023, \mn@doi [\mnras] {10.1093/mnras/stad2597}, \href
  {https://ui.adsabs.harvard.edu/abs/2023MNRAS.526.3273C} {526, 3273}

\bibitem[\protect\citeauthoryear{{Cappellari} \& {Emsellem}}{{Cappellari} \&
  {Emsellem}}{2004}]{Cappellari2004}
{Cappellari} M.,  {Emsellem} E.,  2004, \mn@doi [\pasp] {10.1086/381875}, \href
  {https://ui.adsabs.harvard.edu/abs/2004PASP..116..138C} {116, 138}

\bibitem[\protect\citeauthoryear{{Cappellari} et~al.,}{{Cappellari}
  et~al.}{2011}]{Cappellari2011}
{Cappellari} M.,  et~al., 2011, \mn@doi [\mnras]
  {10.1111/j.1365-2966.2010.18174.x}, \href
  {https://ui.adsabs.harvard.edu/abs/2011MNRAS.413..813C} {413, 813}

\bibitem[\protect\citeauthoryear{{Catala} et~al.,}{{Catala}
  et~al.}{2013}]{Catala2013}
{Catala} L.,  et~al., 2013, \mn@doi [\mnras] {10.1093/mnras/stt1602}, \href
  {https://ui.adsabs.harvard.edu/abs/2013MNRAS.436..590C} {436, 590}

\bibitem[\protect\citeauthoryear{{Cid Fernandes}, {Stasi{\'n}ska},
  {Schlickmann}, {Mateus}, {Vale Asari}, {Schoenell}  \& {Sodr{\'e}}}{{Cid
  Fernandes} et~al.}{2010}]{Cid-Fernandes2010}
{Cid Fernandes} R.,  {Stasi{\'n}ska} G.,  {Schlickmann} M.~S.,  {Mateus} A.,
  {Vale Asari} N.,  {Schoenell} W.,   {Sodr{\'e}} L.,  2010, \mn@doi [\mnras]
  {10.1111/j.1365-2966.2009.16185.x}, \href
  {https://ui.adsabs.harvard.edu/abs/2010MNRAS.403.1036C} {403, 1036}

\bibitem[\protect\citeauthoryear{{Cid Fernandes}, {Stasi{\'n}ska}, {Mateus}  \&
  {Vale Asari}}{{Cid Fernandes} et~al.}{2011}]{Cid-Fernandes2011}
{Cid Fernandes} R.,  {Stasi{\'n}ska} G.,  {Mateus} A.,   {Vale Asari} N.,
  2011, \mn@doi [\mnras] {10.1111/j.1365-2966.2011.18244.x}, \href
  {https://ui.adsabs.harvard.edu/abs/2011MNRAS.413.1687C} {413, 1687}

\bibitem[\protect\citeauthoryear{{Clemens} et~al.,}{{Clemens}
  et~al.}{2010}]{Clemens2010}
{Clemens} M.~S.,  et~al., 2010, \mn@doi [\aap] {10.1051/0004-6361/201014533},
  \href {https://ui.adsabs.harvard.edu/abs/2010A&A...518L..50C} {518, L50}

\bibitem[\protect\citeauthoryear{{Conroy}}{{Conroy}}{2013}]{Conroy2013}
{Conroy} C.,  2013, \mn@doi [\araa] {10.1146/annurev-astro-082812-141017},
  \href {https://ui.adsabs.harvard.edu/abs/2013ARA&A..51..393C} {51, 393}

\bibitem[\protect\citeauthoryear{{Conroy}, {Graves}  \& {van Dokkum}}{{Conroy}
  et~al.}{2014}]{Conroy2014}
{Conroy} C.,  {Graves} G.~J.,   {van Dokkum} P.~G.,  2014, \mn@doi [\apj]
  {10.1088/0004-637X/780/1/33}, \href
  {https://ui.adsabs.harvard.edu/abs/2014ApJ...780...33C} {780, 33}

\bibitem[\protect\citeauthoryear{{Conselice}}{{Conselice}}{2014}]{Conselice2014}
{Conselice} C.~J.,  2014, \mn@doi [\araa]
  {10.1146/annurev-astro-081913-040037}, \href
  {https://ui.adsabs.harvard.edu/abs/2014ARA&A..52..291C} {52, 291}

\bibitem[\protect\citeauthoryear{{Cortese} et~al.,}{{Cortese}
  et~al.}{2012}]{Cortese2012}
{Cortese} L.,  et~al., 2012, \mn@doi [\aap] {10.1051/0004-6361/201118499},
  \href {https://ui.adsabs.harvard.edu/abs/2012A&A...540A..52C} {540, A52}

\bibitem[\protect\citeauthoryear{{Crawford} et~al.,}{{Crawford}
  et~al.}{2010}]{Crawford2010}
{Crawford} S.~M.,  et~al., 2010, in {Silva} D.~R.,  {Peck} A.~B.,   {Soifer}
  B.~T.,  eds,  Society of Photo-Optical Instrumentation Engineers (SPIE)
  Conference Series Vol. 7737, Observatory Operations: Strategies, Processes,
  and Systems III. p. 773725, \mn@doi{10.1117/12.857000}

\bibitem[\protect\citeauthoryear{{Crocker}, {Bureau}, {Young}  \&
  {Combes}}{{Crocker} et~al.}{2011}]{Crocker2011}
{Crocker} A.~F.,  {Bureau} M.,  {Young} L.~M.,   {Combes} F.,  2011, \mn@doi
  [\mnras] {10.1111/j.1365-2966.2010.17537.x}, \href
  {https://ui.adsabs.harvard.edu/abs/2011MNRAS.410.1197C} {410, 1197}

\bibitem[\protect\citeauthoryear{{Davies} et~al.,}{{Davies}
  et~al.}{2019}]{Davies2019}
{Davies} J.~I.,  et~al., 2019, \mn@doi [\aap] {10.1051/0004-6361/201935547},
  \href {https://ui.adsabs.harvard.edu/abs/2019A&A...626A..63D} {626, A63}

\bibitem[\protect\citeauthoryear{{Davis} et~al.,}{{Davis}
  et~al.}{2011}]{Davis2011}
{Davis} T.~A.,  et~al., 2011, \mn@doi [\mnras]
  {10.1111/j.1365-2966.2011.19355.x}, \href
  {https://ui.adsabs.harvard.edu/abs/2011MNRAS.417..882D} {417, 882}

\bibitem[\protect\citeauthoryear{{Davis} et~al.,}{{Davis}
  et~al.}{2015}]{Davis2015}
{Davis} T.~A.,  et~al., 2015, \mnras, 449, 3503

\bibitem[\protect\citeauthoryear{{Draine} \& {Salpeter}}{{Draine} \&
  {Salpeter}}{1979}]{Draine1979}
{Draine} B.~T.,  {Salpeter} E.~E.,  1979, \mn@doi [\apj] {10.1086/157206},
  \href {https://ui.adsabs.harvard.edu/abs/1979ApJ...231..438D} {231, 438}

\bibitem[\protect\citeauthoryear{{Driver} et~al.,}{{Driver}
  et~al.}{2009}]{Driver2009}
{Driver} S.~P.,  et~al., 2009, \mn@doi [Astronomy and Geophysics]
  {10.1111/j.1468-4004.2009.50512.x}, \href
  {https://ui.adsabs.harvard.edu/abs/2009A&G....50e..12D} {50, 5.12}

\bibitem[\protect\citeauthoryear{{Driver} et~al.,}{{Driver}
  et~al.}{2022}]{Driver2022}
{Driver} S.~P.,  et~al., 2022, \mn@doi [\mnras] {10.1093/mnras/stac472}, \href
  {https://ui.adsabs.harvard.edu/abs/2022MNRAS.513..439D} {513, 439}

\bibitem[\protect\citeauthoryear{{Eales} et~al.,}{{Eales}
  et~al.}{2010}]{Eales2010}
{Eales} S.,  et~al., 2010, \mn@doi [\pasp] {10.1086/653086}, \href
  {https://ui.adsabs.harvard.edu/abs/2010PASP..122..499E} {122, 499}

\bibitem[\protect\citeauthoryear{{Eales} et~al.,}{{Eales}
  et~al.}{2015}]{Eales2015}
{Eales} S.,  et~al., 2015, \mn@doi [\mnras] {10.1093/mnras/stv1300}, \href
  {https://ui.adsabs.harvard.edu/abs/2015MNRAS.452.3489E} {452, 3489}

\bibitem[\protect\citeauthoryear{{Falc{\'o}n-Barroso},
  {S{\'a}nchez-Bl{\'a}zquez}, {Vazdekis}, {Ricciardelli}, {Cardiel}, {Cenarro},
  {Gorgas}  \& {Peletier}}{{Falc{\'o}n-Barroso} et~al.}{2011}]{Falcon2011}
{Falc{\'o}n-Barroso} J.,  {S{\'a}nchez-Bl{\'a}zquez} P.,  {Vazdekis} A.,
  {Ricciardelli} E.,  {Cardiel} N.,  {Cenarro} A.~J.,  {Gorgas} J.,
  {Peletier} R.~F.,  2011, \mn@doi [\aap] {10.1051/0004-6361/201116842}, \href
  {https://ui.adsabs.harvard.edu/abs/2011A&A...532A..95F} {532, A95}

\bibitem[\protect\citeauthoryear{{Ferreras} et~al.,}{{Ferreras}
  et~al.}{2019}]{Ferreras2019}
{Ferreras} I.,  et~al., 2019, \mn@doi [\mnras] {10.1093/mnras/stz2095}, \href
  {https://ui.adsabs.harvard.edu/abs/2019MNRAS.489..608F} {489, 608}

\bibitem[\protect\citeauthoryear{{French} et~al.,}{{French}
  et~al.}{2023}]{French2023}
{French} K.~D.,  et~al., 2023, \mn@doi [\apj] {10.3847/1538-4357/aca46e}, \href
  {https://ui.adsabs.harvard.edu/abs/2023ApJ...942...25F} {942, 25}

\bibitem[\protect\citeauthoryear{{Gallazzi}, {Charlot}, {Brinchmann}  \&
  {White}}{{Gallazzi} et~al.}{2006}]{Gallazzi2006}
{Gallazzi} A.,  {Charlot} S.,  {Brinchmann} J.,   {White} S. D.~M.,  2006,
  \mn@doi [\mnras] {10.1111/j.1365-2966.2006.10548.x}, \href
  {https://ui.adsabs.harvard.edu/abs/2006MNRAS.370.1106G} {370, 1106}

\bibitem[\protect\citeauthoryear{{Ge}, {Yan}, {Cappellari}, {Mao}, {Li}  \&
  {Lu}}{{Ge} et~al.}{2018}]{Ge2018}
{Ge} J.,  {Yan} R.,  {Cappellari} M.,  {Mao} S.,  {Li} H.,   {Lu} Y.,  2018,
  \mn@doi [\mnras] {10.1093/mnras/sty1245}, \href
  {https://ui.adsabs.harvard.edu/abs/2018MNRAS.478.2633G} {478, 2633}

\bibitem[\protect\citeauthoryear{{Glass}}{{Glass}}{2024}]{Glass2024}
{Glass} D. H.~W.,  2024, Ph{D} thesis, University of Central Lancashire,
  Preston, UK, http://doi.org/10.17030/uclan.thesis.00042515

\bibitem[\protect\citeauthoryear{{Glass}, {Sansom}, {Davis}  \&
  {Popescu}}{{Glass} et~al.}{2022}]{Glass2022}
{Glass} D. H.~W.,  {Sansom} A.~E.,  {Davis} T.~A.,   {Popescu} C.~C.,  2022,
  \mn@doi [\mnras] {10.1093/mnras/stac3001}, \href
  {https://ui.adsabs.harvard.edu/abs/2022MNRAS.517.5524G} {517, 5524}

\bibitem[\protect\citeauthoryear{{Gonz{\'a}lez Delgado} et~al.,}{{Gonz{\'a}lez
  Delgado} et~al.}{2015}]{Gonzalez2015}
{Gonz{\'a}lez Delgado} R.~M.,  et~al., 2015, \mn@doi [\aap]
  {10.1051/0004-6361/201525938}, \href
  {https://ui.adsabs.harvard.edu/abs/2015A&A...581A.103G} {581, A103}

\bibitem[\protect\citeauthoryear{{Goudfrooij} \& {de Jong}}{{Goudfrooij} \& {de
  Jong}}{1995}]{Goudfrooij1995}
{Goudfrooij} P.,  {de Jong} T.,  1995, \mn@doi [\aap]
  {10.48550/arXiv.astro-ph/9504011}, \href
  {https://ui.adsabs.harvard.edu/abs/1995A&A...298..784G} {298, 784}

\bibitem[\protect\citeauthoryear{{Griffith}, {Martini}  \& {Conroy}}{{Griffith}
  et~al.}{2019}]{Griffith2019}
{Griffith} E.,  {Martini} P.,   {Conroy} C.,  2019, \mn@doi [\mnras]
  {10.1093/mnras/sty3405}, \href
  {https://ui.adsabs.harvard.edu/abs/2019MNRAS.484..562G} {484, 562}

\bibitem[\protect\citeauthoryear{{Holwerda} et~al.,}{{Holwerda}
  et~al.}{2024}]{Holwerda2024}
{Holwerda} B.~W.,  et~al., 2024, \mn@doi [\pasa] {10.1017/pasa.2024.109}, \href
  {https://ui.adsabs.harvard.edu/abs/2024PASA...41..115H} {41, e115}

\bibitem[\protect\citeauthoryear{{Jegatheesan}, {Johnston}, {H{\"a}u{\ss}ler}
  \& {Nedkova}}{{Jegatheesan} et~al.}{2024}]{Jegatheesan2024}
{Jegatheesan} K.,  {Johnston} E.~J.,  {H{\"a}u{\ss}ler} B.,   {Nedkova} K.~V.,
  2024, \mn@doi [\aap] {10.1051/0004-6361/202347372}, \href
  {https://ui.adsabs.harvard.edu/abs/2024A&A...684A..32J} {684, A32}

\bibitem[\protect\citeauthoryear{{Jegatheesan}, {Johnston}, {H{\"a}u{\ss}ler},
  {Lassen}, {Riffel}  \& {Chies-Santos}}{{Jegatheesan}
  et~al.}{2025}]{Jegatheesan2025}
{Jegatheesan} K.,  {Johnston} E.~J.,  {H{\"a}u{\ss}ler} B.,  {Lassen} A.~E.,
  {Riffel} R.,   {Chies-Santos} A.~L.,  2025, \mn@doi [\aap]
  {10.1051/0004-6361/202452137}, \href
  {https://ui.adsabs.harvard.edu/abs/2025A&A...694A.224J} {694, A224}

\bibitem[\protect\citeauthoryear{{Johansson}, {Thomas}  \&
  {Maraston}}{{Johansson} et~al.}{2012}]{Johansson2012}
{Johansson} J.,  {Thomas} D.,   {Maraston} C.,  2012, \mn@doi [\mnras]
  {10.1111/j.1365-2966.2011.20316.x}, \href
  {https://ui.adsabs.harvard.edu/abs/2012MNRAS.421.1908J} {421, 1908}

\bibitem[\protect\citeauthoryear{Johnston et~al.,}{Johnston
  et~al.}{2022}]{Johnston2022}
Johnston E.~J.,  et~al., 2022, \mn@doi [Monthly Notices of the Royal
  Astronomical Society] {10.1093/mnras/stac1447}, 514, 6141

\bibitem[\protect\citeauthoryear{{Katkov}, {Kniazev}  \& {Sil'chenko}}{{Katkov}
  et~al.}{2015}]{Katkov2015}
{Katkov} I.~Y.,  {Kniazev} A.~Y.,   {Sil'chenko} O.~K.,  2015, \mn@doi [\aj]
  {10.1088/0004-6256/150/1/24}, \href
  {https://ui.adsabs.harvard.edu/abs/2015AJ....150...24K} {150, 24}

\bibitem[\protect\citeauthoryear{{Katkov}, {Kniazev}, {Kasparova}  \&
  {Sil'chenko}}{{Katkov} et~al.}{2019}]{Katkov2019}
{Katkov} I.~Y.,  {Kniazev} A.~Y.,  {Kasparova} A.~V.,   {Sil'chenko} O.~K.,
  2019, \mn@doi [\mnras] {10.1093/mnras/sty3268}, \href
  {https://ui.adsabs.harvard.edu/abs/2019MNRAS.483.2413K} {483, 2413}

\bibitem[\protect\citeauthoryear{{Kaviraj} et~al.,}{{Kaviraj}
  et~al.}{2012}]{Kaviraj2012}
{Kaviraj} S.,  et~al., 2012, \mn@doi [\mnras]
  {10.1111/j.1365-2966.2012.20957.x}, \href
  {https://ui.adsabs.harvard.edu/abs/2012MNRAS.423...49K} {423, 49}

\bibitem[\protect\citeauthoryear{{Kelvin} et~al.,}{{Kelvin}
  et~al.}{2014}]{Kelvin2014}
{Kelvin} L.~S.,  et~al., 2014, \mn@doi [\mnras] {10.1093/mnras/stu1507}, \href
  {https://ui.adsabs.harvard.edu/abs/2014MNRAS.444.1647K} {444, 1647}

\bibitem[\protect\citeauthoryear{{Knowles}, {Sansom}, {Allende Prieto}  \&
  {Vazdekis}}{{Knowles} et~al.}{2021}]{Knowles2021}
{Knowles} A.~T.,  {Sansom} A.~E.,  {Allende Prieto} C.,   {Vazdekis} A.,  2021,
  \mn@doi [\mnras] {10.1093/mnras/stab1001}, \href
  {https://ui.adsabs.harvard.edu/abs/2021MNRAS.504.2286K} {504, 2286}

\bibitem[\protect\citeauthoryear{{Knowles}, {Sansom}, {Vazdekis}  \& {Allende
  Prieto}}{{Knowles} et~al.}{2023}]{Knowles2023}
{Knowles} A.~T.,  {Sansom} A.~E.,  {Vazdekis} A.,   {Allende Prieto} C.,  2023,
  \mn@doi [\mnras] {10.1093/mnras/stad1647}, \href
  {https://ui.adsabs.harvard.edu/abs/2023MNRAS.523.3450K} {523, 3450}

\bibitem[\protect\citeauthoryear{{Kokusho}, {Kaneda}, {Bureau}, {Suzuki},
  {Murata}, {Kondo}  \& {Yamagishi}}{{Kokusho} et~al.}{2017}]{Kokusho2017}
{Kokusho} T.,  {Kaneda} H.,  {Bureau} M.,  {Suzuki} T.,  {Murata} K.,  {Kondo}
  A.,   {Yamagishi} M.,  2017, \mn@doi [\aap] {10.1051/0004-6361/201630158},
  \href {https://ui.adsabs.harvard.edu/abs/2017A&A...605A..74K} {605, A74}

\bibitem[\protect\citeauthoryear{{Kokusho} et~al.,}{{Kokusho}
  et~al.}{2019}]{Kokusho2019}
{Kokusho} T.,  et~al., 2019, \mn@doi [\aap] {10.1051/0004-6361/201833911},
  \href {https://ui.adsabs.harvard.edu/abs/2019A&A...622A..87K} {622, A87}

\bibitem[\protect\citeauthoryear{{Kuntschner} et~al.,}{{Kuntschner}
  et~al.}{2010}]{Kuntschner2010}
{Kuntschner} H.,  et~al., 2010, \mn@doi [\mnras]
  {10.1111/j.1365-2966.2010.17161.x}, \href
  {https://ui.adsabs.harvard.edu/abs/2010MNRAS.408...97K} {408, 97}

\bibitem[\protect\citeauthoryear{{La Barbera}, {Ferreras}, {Vazdekis}, {de la
  Rosa}, {de Carvalho}, {Trevisan}, {Falc{\'o}n-Barroso}  \&
  {Ricciardelli}}{{La Barbera} et~al.}{2013}]{laBarbera2013}
{La Barbera} F.,  {Ferreras} I.,  {Vazdekis} A.,  {de la Rosa} I.~G.,  {de
  Carvalho} R.~R.,  {Trevisan} M.,  {Falc{\'o}n-Barroso} J.,   {Ricciardelli}
  E.,  2013, \mn@doi [\mnras] {10.1093/mnras/stt943}, \href
  {https://ui.adsabs.harvard.edu/abs/2013MNRAS.433.3017L} {433, 3017}

\bibitem[\protect\citeauthoryear{{La Barbera}, {Pasquali}, {Ferreras},
  {Gallazzi}, {de Carvalho}  \& {de la Rosa}}{{La Barbera}
  et~al.}{2014}]{laBarbera2014}
{La Barbera} F.,  {Pasquali} A.,  {Ferreras} I.,  {Gallazzi} A.,  {de Carvalho}
  R.~R.,   {de la Rosa} I.~G.,  2014, \mn@doi [\mnras] {10.1093/mnras/stu1626},
  \href {https://ui.adsabs.harvard.edu/abs/2014MNRAS.445.1977L} {445, 1977}

\bibitem[\protect\citeauthoryear{{La Barbera}, {Ferreras}  \& {Vazdekis}}{{La
  Barbera} et~al.}{2015}]{laBarbera2015}
{La Barbera} F.,  {Ferreras} I.,   {Vazdekis} A.,  2015, \mn@doi [\mnras]
  {10.1093/mnrasl/slv029}, \href
  {https://ui.adsabs.harvard.edu/abs/2015MNRAS.449L.137L} {449, L137}

\bibitem[\protect\citeauthoryear{{Lee}, {Hwang}, {Hwang}, {Lee}  \&
  {Kim}}{{Lee} et~al.}{2023}]{Lee2023}
{Lee} Y.~H.,  {Hwang} H.~S.,  {Hwang} N.,  {Lee} J.~C.,   {Kim} K.-B.,  2023,
  \mn@doi [\apj] {10.3847/1538-4357/ace1ea}, \href
  {https://ui.adsabs.harvard.edu/abs/2023ApJ...953...88L} {953, 88}

\bibitem[\protect\citeauthoryear{{Le{\'s}niewska}, {Micha{\l}owski}, {Gall},
  {Hjorth}, {Nadolny}, {Ryzhov}  \& {Solar}}{{Le{\'s}niewska}
  et~al.}{2023}]{Lesniewska2023}
{Le{\'s}niewska} A.,  {Micha{\l}owski} M.~J.,  {Gall} C.,  {Hjorth} J.,
  {Nadolny} J.,  {Ryzhov} O.,   {Solar} M.,  2023, \mn@doi [\apj]
  {10.3847/1538-4357/acdcfc}, \href
  {https://ui.adsabs.harvard.edu/abs/2023ApJ...953...27L} {953, 27}

\bibitem[\protect\citeauthoryear{{Liske} et~al.,}{{Liske}
  et~al.}{2015}]{Liske2015}
{Liske} J.,  et~al., 2015, \mn@doi [\mnras] {10.1093/mnras/stv1436}, \href
  {https://ui.adsabs.harvard.edu/abs/2015MNRAS.452.2087L} {452, 2087}

\bibitem[\protect\citeauthoryear{{Liu}}{{Liu}}{2020}]{Liu2020}
{Liu} Y.,  2020, \mn@doi [\mnras] {10.1093/mnras/staa2012}, \href
  {https://ui.adsabs.harvard.edu/abs/2020MNRAS.497.3011L} {497, 3011}

\bibitem[\protect\citeauthoryear{{Man}, {Zirm}  \& {Toft}}{{Man}
  et~al.}{2016}]{Man2016}
{Man} A.~W.,  {Zirm} A.~W.,   {Toft} S.,  2016, \apj, 830, 89

\bibitem[\protect\citeauthoryear{{Martin}, {Kaviraj}, {Devriendt}, {Dubois}  \&
  {Pichon}}{{Martin} et~al.}{2018}]{Martin2018}
{Martin} G.,  {Kaviraj} S.,  {Devriendt} J.~E.~G.,  {Dubois} Y.,   {Pichon} C.,
   2018, \mn@doi [\mnras] {10.1093/mnras/sty1936}, \href
  {https://ui.adsabs.harvard.edu/abs/2018MNRAS.480.2266M} {480, 2266}

\bibitem[\protect\citeauthoryear{{Martini}, {Dicken}  \&
  {Storchi-Bergmann}}{{Martini} et~al.}{2013}]{Martini2013}
{Martini} P.,  {Dicken} D.,   {Storchi-Bergmann} T.,  2013, \mn@doi [\apj]
  {10.1088/0004-637X/766/2/121}, \href
  {https://ui.adsabs.harvard.edu/abs/2013ApJ...766..121M} {766, 121}

\bibitem[\protect\citeauthoryear{{McDermid} et~al.,}{{McDermid}
  et~al.}{2015}]{Mcdermid2015}
{McDermid} R.~M.,  et~al., 2015, \mn@doi [\mnras] {10.1093/mnras/stv105}, \href
  {https://ui.adsabs.harvard.edu/abs/2015MNRAS.448.3484M} {448, 3484}

\bibitem[\protect\citeauthoryear{{Micha{\l}owski} et~al.,}{{Micha{\l}owski}
  et~al.}{2019}]{Michalowski2019}
{Micha{\l}owski} M.~J.,  et~al., 2019, \mn@doi [\aap]
  {10.1051/0004-6361/201936055}, \href
  {https://ui.adsabs.harvard.edu/abs/2019A&A...632A..43M} {632, A43}

\bibitem[\protect\citeauthoryear{{Micha{\l}owski} et~al.,}{{Micha{\l}owski}
  et~al.}{2024}]{Michalowski2024}
{Micha{\l}owski} M.~J.,  et~al., 2024, \mn@doi [\apj]
  {10.3847/1538-4357/ad1b52}, \href
  {https://ui.adsabs.harvard.edu/abs/2024ApJ...964..129M} {964, 129}

\bibitem[\protect\citeauthoryear{{Moffett}, {Driver}, {Lange}, {Robotham},
  {Kelvin}  \& {GAMA Team}}{{Moffett} et~al.}{2016}]{Moffett2016}
{Moffett} A.~J.,  {Driver} S.~P.,  {Lange} R.,  {Robotham} A.,  {Kelvin} L.,
  {GAMA Team} 2016, in American Astronomical Society Meeting Abstracts \#227.
  p. 408.01

\bibitem[\protect\citeauthoryear{{Pernet}, {Boecker}  \&
  {Mart{\'\i}n-Navarro}}{{Pernet} et~al.}{2024}]{Pernet2024}
{Pernet} E.,  {Boecker} A.,   {Mart{\'\i}n-Navarro} I.,  2024, \mn@doi [\aap]
  {10.1051/0004-6361/202449308}, \href
  {https://ui.adsabs.harvard.edu/abs/2024A&A...687L..14P} {687, L14}

\bibitem[\protect\citeauthoryear{{Porter-Temple} et~al.,}{{Porter-Temple}
  et~al.}{2022}]{Porter-Temple2022}
{Porter-Temple} R.,  et~al., 2022, \mn@doi [\mnras] {10.1093/mnras/stac1936},
  \href {https://ui.adsabs.harvard.edu/abs/2022MNRAS.515.3875P} {515, 3875}

\bibitem[\protect\citeauthoryear{{Press}, {Teukolsky}, {Vetterling}  \&
  {Flannery}}{{Press} et~al.}{1992}]{Press1992}
{Press} W.~H.,  {Teukolsky} S.~A.,  {Vetterling} W.~T.,   {Flannery} B.~P.,
  1992, {Numerical recipes in FORTRAN. The art of scientific computing}.
{Cambridge University Press}

\bibitem[\protect\citeauthoryear{{Robotham} et~al.,}{{Robotham}
  et~al.}{2011}]{Robotham2011}
{Robotham} A.~S.~G.,  et~al., 2011, \mn@doi [\mnras]
  {10.1111/j.1365-2966.2011.19217.x}, \href
  {https://ui.adsabs.harvard.edu/abs/2011MNRAS.416.2640R} {416, 2640}

\bibitem[\protect\citeauthoryear{{Rowlands} et~al.,}{{Rowlands}
  et~al.}{2012}]{Rowlands2012}
{Rowlands} K.,  et~al., 2012, \mn@doi [\mnras]
  {10.1111/j.1365-2966.2011.19905.x}, \href
  {https://ui.adsabs.harvard.edu/abs/2012MNRAS.419.2545R} {419, 2545}

\bibitem[\protect\citeauthoryear{{Saintonge} et~al.,}{{Saintonge}
  et~al.}{2016}]{Saintonge2016}
{Saintonge} A.,  et~al., 2016, \mn@doi [\mnras] {10.1093/mnras/stw1715}, \href
  {https://ui.adsabs.harvard.edu/abs/2016MNRAS.462.1749S} {462, 1749}

\bibitem[\protect\citeauthoryear{{Sansom} et~al.,}{{Sansom}
  et~al.}{2019}]{Sansom2019}
{Sansom} A.~E.,  et~al., 2019, \mn@doi [\mnras] {10.1093/mnras/sty3021}, \href
  {https://ui.adsabs.harvard.edu/abs/2019MNRAS.482.4617S} {482, 4617}

\bibitem[\protect\citeauthoryear{{Schawinski} et~al.,}{{Schawinski}
  et~al.}{2007}]{Schawinski2007}
{Schawinski} K.,  et~al., 2007, \mn@doi [\apjs] {10.1086/516631}, \href
  {https://ui.adsabs.harvard.edu/abs/2007ApJS..173..512S} {173, 512}

\bibitem[\protect\citeauthoryear{{Schneider} \& {Maiolino}}{{Schneider} \&
  {Maiolino}}{2024}]{Schneider2024}
{Schneider} R.,  {Maiolino} R.,  2024, \mn@doi [\aapr]
  {10.1007/s00159-024-00151-2}, \href
  {https://ui.adsabs.harvard.edu/abs/2024A&ARv..32....2S} {32, 2}

\bibitem[\protect\citeauthoryear{{Serra} \& {Trager}}{{Serra} \&
  {Trager}}{2007}]{Serra2007}
{Serra} P.,  {Trager} S.~C.,  2007, \mn@doi [\mnras]
  {10.1111/j.1365-2966.2006.11188.x}, \href
  {https://ui.adsabs.harvard.edu/abs/2007MNRAS.374..769S} {374, 769}

\bibitem[\protect\citeauthoryear{{Simonian} \& {Martini}}{{Simonian} \&
  {Martini}}{2017}]{SimonianMartini2017}
{Simonian} G.~V.,  {Martini} P.,  2017, \mn@doi [\mnras]
  {10.1093/mnras/stw2623}, \href
  {https://ui.adsabs.harvard.edu/abs/2017MNRAS.464.3920S} {464, 3920}

\bibitem[\protect\citeauthoryear{{Smith} et~al.,}{{Smith}
  et~al.}{2012}]{Smith2012}
{Smith} M.~W.~L.,  et~al., 2012, \mn@doi [\apj] {10.1088/0004-637X/748/2/123},
  \href {https://ui.adsabs.harvard.edu/abs/2012ApJ...748..123S} {748, 123}

\bibitem[\protect\citeauthoryear{{Tody}}{{Tody}}{1986}]{Tody1986}
{Tody} D.,  1986, in {Crawford} D.~L.,  ed.,  Society of Photo-Optical
  Instrumentation Engineers (SPIE) Conference Series Vol. 627, Instrumentation
  in astronomy VI. p.~733, \mn@doi{10.1117/12.968154}

\bibitem[\protect\citeauthoryear{{Trussler}, {Maiolino}, {Maraston}, {Peng},
  {Thomas}, {Goddard}  \& {Lian}}{{Trussler} et~al.}{2021}]{Trussler2021}
{Trussler} J.,  {Maiolino} R.,  {Maraston} C.,  {Peng} Y.,  {Thomas} D.,
  {Goddard} D.,   {Lian} J.,  2021, \mn@doi [\mnras] {10.1093/mnras/staa3545},
  \href {https://ui.adsabs.harvard.edu/abs/2021MNRAS.500.4469T} {500, 4469}

\bibitem[\protect\citeauthoryear{{Vaghmare}, {Barway}, {V{\"a}is{\"a}nen},
  {Ramphul}, {Wadadekar}  \& {Kembhavi}}{{Vaghmare}
  et~al.}{2018}]{Vaghmare2018}
{Vaghmare} K.,  {Barway} S.,  {V{\"a}is{\"a}nen} P.,  {Ramphul} R.,
  {Wadadekar} Y.,   {Kembhavi} A.~K.,  2018, \mn@doi [\mnras]
  {10.1093/mnras/sty2217}, \href
  {https://ui.adsabs.harvard.edu/abs/2018MNRAS.480.4931V} {480, 4931}

\bibitem[\protect\citeauthoryear{{Valiante} et~al.,}{{Valiante}
  et~al.}{2016}]{Valiante2016}
{Valiante} E.,  et~al., 2016, \mn@doi [\mnras] {10.1093/mnras/stw1806}, \href
  {https://ui.adsabs.harvard.edu/abs/2016MNRAS.462.3146V} {462, 3146}

\bibitem[\protect\citeauthoryear{{Vazdekis}, {S{\'a}nchez-Bl{\'a}zquez},
  {Falc{\'o}n-Barroso}, {Cenarro}, {Beasley}, {Cardiel}, {Gorgas}  \&
  {Peletier}}{{Vazdekis} et~al.}{2010}]{Vazdekis2010}
{Vazdekis} A.,  {S{\'a}nchez-Bl{\'a}zquez} P.,  {Falc{\'o}n-Barroso} J.,
  {Cenarro} A.~J.,  {Beasley} M.~A.,  {Cardiel} N.,  {Gorgas} J.,   {Peletier}
  R.~F.,  2010, \mn@doi [\mnras] {10.1111/j.1365-2966.2010.16407.x}, \href
  {https://ui.adsabs.harvard.edu/abs/2010MNRAS.404.1639V} {404, 1639}

\bibitem[\protect\citeauthoryear{{Walcher}, {Coelho}, {Gallazzi}, {Bruzual},
  {Charlot}  \& {Chiappini}}{{Walcher} et~al.}{2015}]{Walcher2015}
{Walcher} C.~J.,  {Coelho} P.~R.~T.,  {Gallazzi} A.,  {Bruzual} G.,  {Charlot}
  S.,   {Chiappini} C.,  2015, \mn@doi [\aap] {10.1051/0004-6361/201525924},
  \href {https://ui.adsabs.harvard.edu/abs/2015A&A...582A..46W} {582, A46}

\bibitem[\protect\citeauthoryear{{Werle} et~al.,}{{Werle}
  et~al.}{2020}]{Werle2020}
{Werle} A.,  et~al., 2020, \mn@doi [\mnras] {10.1093/mnras/staa2217}, \href
  {https://ui.adsabs.harvard.edu/abs/2020MNRAS.497.3251W} {497, 3251}

\bibitem[\protect\citeauthoryear{{Woo}, {Walters}, {Archinuk}, {Faber},
  {Ellison}, {Teimoorinia}  \& {Iyer}}{{Woo} et~al.}{2024}]{Woo2024}
{Woo} J.,  {Walters} D.,  {Archinuk} F.,  {Faber} S.~M.,  {Ellison} S.~L.,
  {Teimoorinia} H.,   {Iyer} K.,  2024, \mn@doi [\mnras]
  {10.1093/mnras/stae1114}, \href
  {https://ui.adsabs.harvard.edu/abs/2024MNRAS.530.4260W} {530, 4260}

\bibitem[\protect\citeauthoryear{{Worthey}}{{Worthey}}{1994}]{Worthey1994B}
{Worthey} G.,  1994, \mn@doi [\apjs] {10.1086/192096}, \href
  {https://ui.adsabs.harvard.edu/abs/1994ApJS...95..107W} {95, 107}

\bibitem[\protect\citeauthoryear{{Worthey}, {Faber}, {Gonzalez}  \&
  {Burstein}}{{Worthey} et~al.}{1994}]{Worthey1994}
{Worthey} G.,  {Faber} S.~M.,  {Gonzalez} J.~J.,   {Burstein} D.,  1994,
  \mn@doi [\apjs] {10.1086/192087}, \href
  {https://ui.adsabs.harvard.edu/abs/1994ApJS...94..687W} {94, 687}

\bibitem[\protect\citeauthoryear{{Worthey}, {Tang}  \& {Serven}}{{Worthey}
  et~al.}{2014}]{Worthey2014}
{Worthey} G.,  {Tang} B.,   {Serven} J.,  2014, \mn@doi [\apj]
  {10.1088/0004-637X/783/1/20}, \href
  {https://ui.adsabs.harvard.edu/abs/2014ApJ...783...20W} {783, 20}

\bibitem[\protect\citeauthoryear{{Wright} et~al.,}{{Wright}
  et~al.}{2010}]{Wright2010}
{Wright} E.~L.,  et~al., 2010, \mn@doi [\aj] {10.1088/0004-6256/140/6/1868},
  \href {https://ui.adsabs.harvard.edu/abs/2010AJ....140.1868W} {140, 1868}

\bibitem[\protect\citeauthoryear{{York} et~al.,}{{York}
  et~al.}{2000}]{York2000}
{York} D.~G.,  et~al., 2000, \mn@doi [\aj] {10.1086/301513}, \href
  {https://ui.adsabs.harvard.edu/abs/2000AJ....120.1579Y} {120, 1579}

\bibitem[\protect\citeauthoryear{{Young} et~al.,}{{Young}
  et~al.}{2011}]{Young2011}
{Young} L.~M.,  et~al., 2011, \mn@doi [\mnras]
  {10.1111/j.1365-2966.2011.18561.x}, \href
  {https://ui.adsabs.harvard.edu/abs/2011MNRAS.414..940Y} {414, 940}

\bibitem[\protect\citeauthoryear{{da Cunha}, {Charlot}  \& {Elbaz}}{{da Cunha}
  et~al.}{2008}]{daCunha2008}
{da Cunha} E.,  {Charlot} S.,   {Elbaz} D.,  2008, \mn@doi [\mnras]
  {10.1111/j.1365-2966.2008.13535.x}, \href
  {https://ui.adsabs.harvard.edu/abs/2008MNRAS.388.1595D} {388, 1595}

\bibitem[\protect\citeauthoryear{{de Jong} et~al.,}{{de Jong}
  et~al.}{2013}]{deJong2013}
{de Jong} J.~T.~A.,  et~al., 2013, The Messenger, \href
  {https://ui.adsabs.harvard.edu/abs/2013Msngr.154...44D} {154, 44}

\bibitem[\protect\citeauthoryear{{van de Voort} et~al.,}{{van de Voort}
  et~al.}{2018}]{vandevoort2018}
{van de Voort} F.,  et~al., 2018, \mn@doi [\mnras] {10.1093/mnras/sty228},
  \href {https://ui.adsabs.harvard.edu/abs/2018MNRAS.476..122V} {476, 122}

\makeatother
\end{thebibliography}

\appendix
\section{Stellar Population Weights Plots}
\label{sec:ppxf_plots}  

In this Appendix we present output plots from pPXF luminosity weighted full spectrum fitting using MILES SSPs, made with 'baseFe' models from the MILES website\protect\footnotemark. These show a grid of [M/H]$_L$ vs. log$_{10}$ Age$_L$, with post-regularization template weight fraction at each grid point represented by colour, seen in the colour scale to the right side of each grid. See Fig.~\ref{fig:fig_example_ppxf} and \citet{Cappellari2023} for a full explanation of pPXF weights plots.

Fig.~\ref{fig:ppxf_plots} shows pPXF fits for our sample of dusty ETGs observed with SALT.
Fig.~\ref{fig:synthetic_spectra_plots} shows pPXF fits for various models constructed from MILES SSPs.

\newpage
\onecolumn
\begin{figure*}
    \centering
    \caption{Presented below is a [M/H]$_L$ vs. log$_{10}$ Age$_L$ grid for the central $R_e/8$ aperture of each target ETG, observed with SALT and fitted using MILES SSPs. Grids for GAMA227266, 422436 and 546040 
    show old, high metallicity stellar populations. Grids for GAMA99687, 136847, 227264, 560238 and 3576053 contain an old, high metallicity stellar population component, and an intermediate age, lower metallicity component. Grids for GAMA79849, 272990 and GAMA570227 show early continuous star formation and an intermediate age component. As discussed in Sections~\ref{subsec:analysis_population} and \ref{subsec:analysis_alpha_lick}, spectra for GAMA85416, 298980 and 569555 contain strong emission lines, which may have led to anomalous results from full spectrum fitting. GAMA65075 grid contains an old, uncertain metallicity stellar population component, a young, high metallicity component and a trace intermediate age components.}
    \begin{subfigure}{0.33\textwidth}
        \centering
        \caption*{GAMA65075}
        \includegraphics[width=\columnwidth]{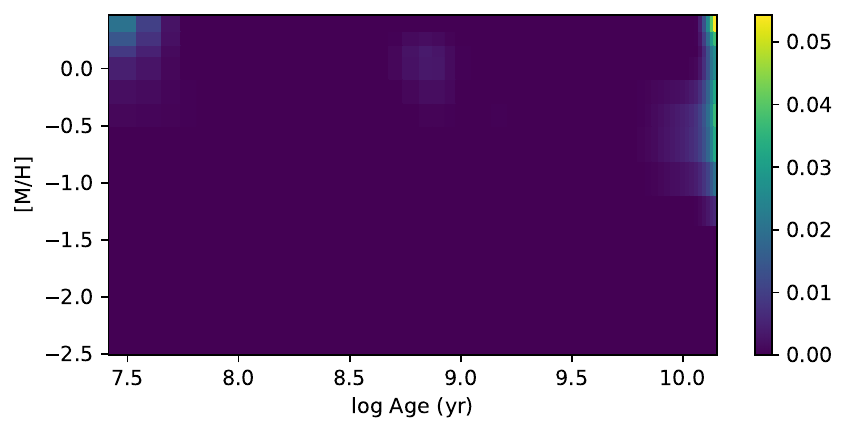}
    \end{subfigure}
    \hfill
    \begin{subfigure}{0.33\textwidth}
        \centering
        \caption*{GAMA79849}
        \includegraphics[width=\columnwidth]{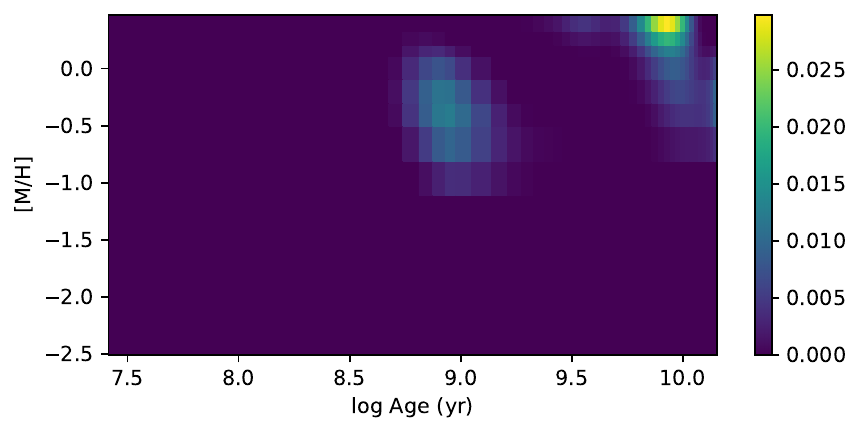}
    \end{subfigure}
    \hfill
    \begin{subfigure}{0.33\textwidth}
        \centering
        \caption*{GAMA85416}
        \includegraphics[width=\columnwidth]{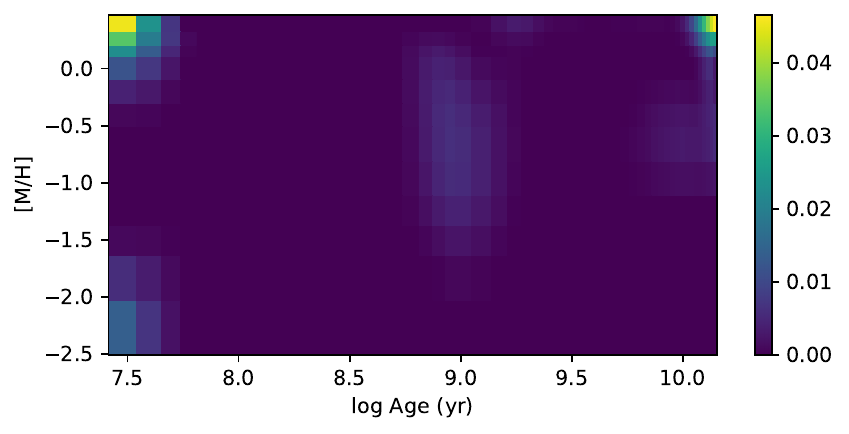}
    \end{subfigure}
    \newline
    \begin{subfigure}{0.33\textwidth}
        \centering
        \caption*{GAMA99687}
        \includegraphics[width=\columnwidth]{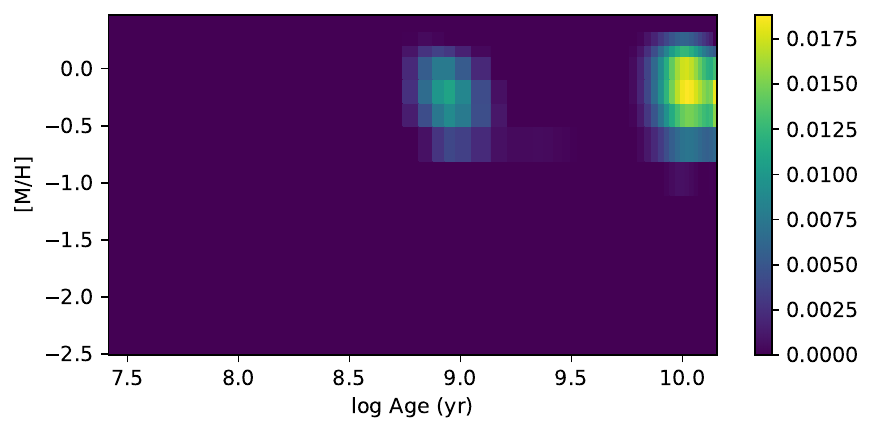}
    \end{subfigure}
    \hfill
    \begin{subfigure}{0.33\textwidth}
        \centering
        \caption*{GAMA136847}
        \includegraphics[width=\columnwidth]{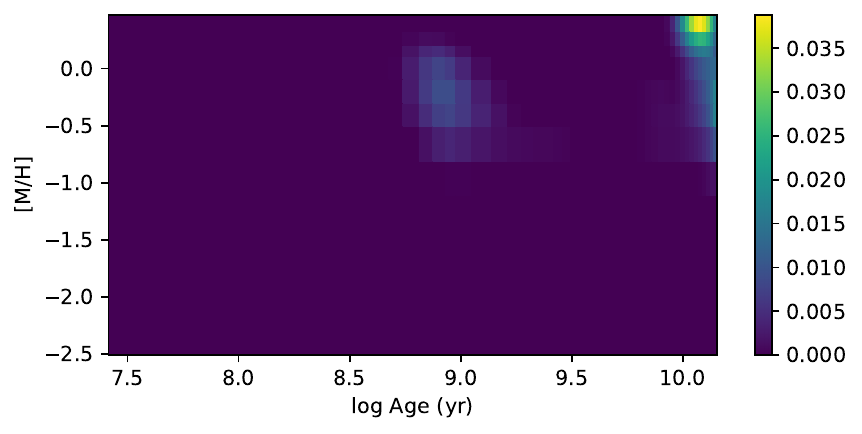}
    \end{subfigure}
    \hfill
    \begin{subfigure}{0.33\textwidth}
        \centering
        \caption*{GAMA227264}
        \includegraphics[width=\columnwidth]{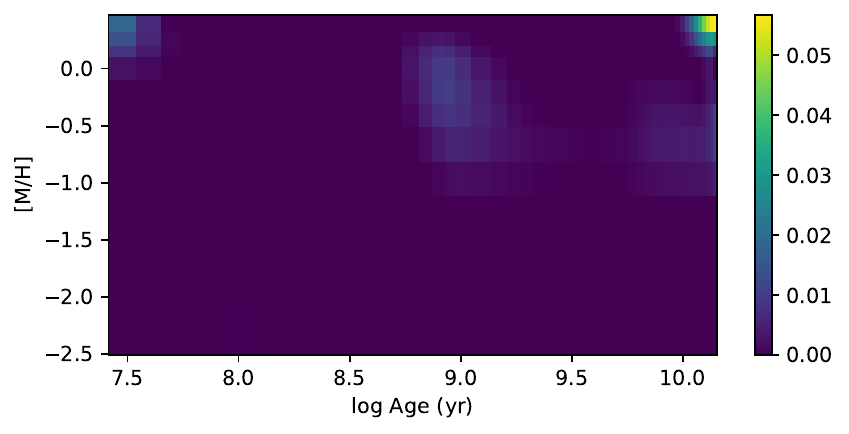}
    \end{subfigure}
    \newline
    \begin{subfigure}{0.33\textwidth}
        \centering
        \caption*{GAMA227266}
        \includegraphics[width=\columnwidth]{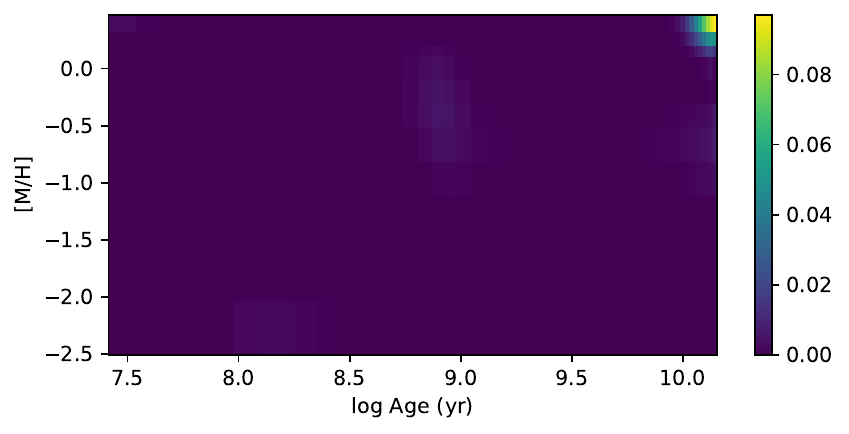}
    \end{subfigure}
    \hfill
    \begin{subfigure}{0.33\textwidth}
        \centering
        \caption*{GAMA272990}
        \includegraphics[width=\columnwidth]{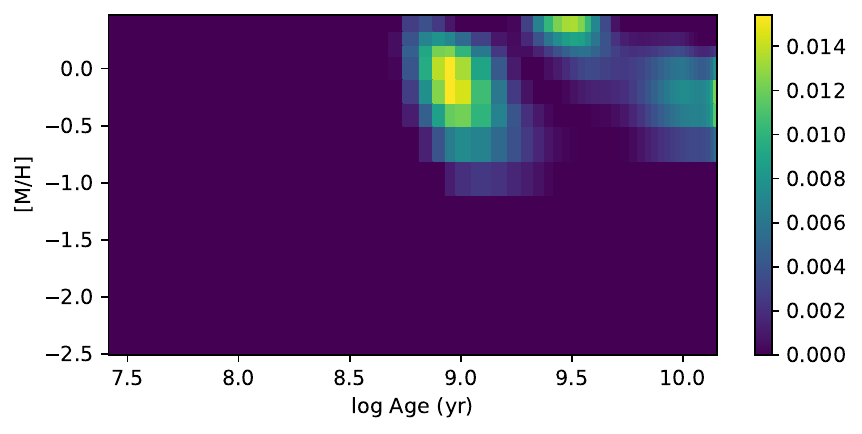}
    \end{subfigure}
    \hfill
    \begin{subfigure}{0.33\textwidth}
        \centering
        \caption*{GAMA298980}
        \includegraphics[width=\columnwidth]{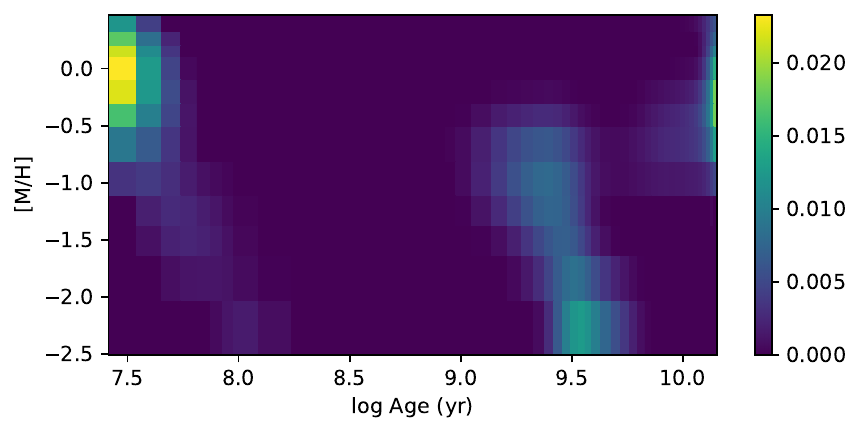}
    \end{subfigure}
    \newline
    \begin{subfigure}{0.33\textwidth}
        \centering
        \caption*{GAMA422436}
        \includegraphics[width=\columnwidth]{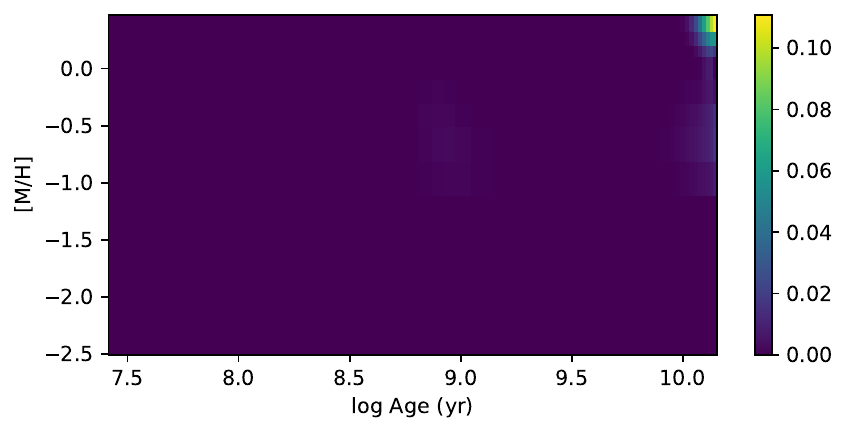}
    \end{subfigure}
    \hfill
    \begin{subfigure}{0.33\textwidth}
        \centering
        \caption*{GAMA546040}
        \includegraphics[width=\columnwidth]{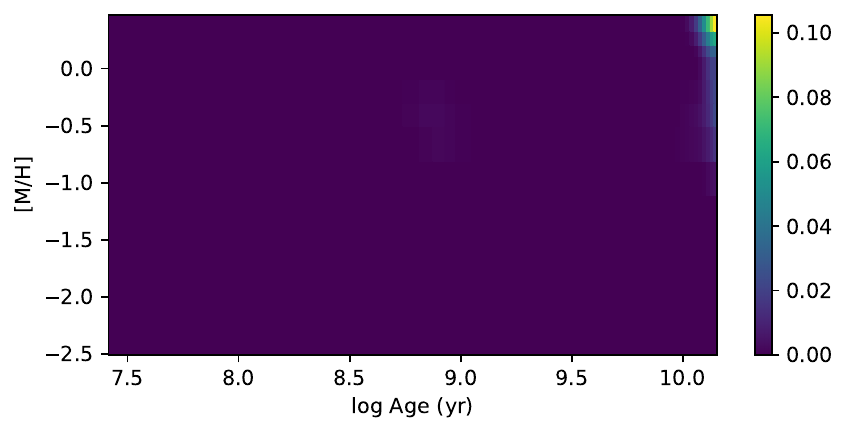}
    \end{subfigure}
    \hfill
    \begin{subfigure}{0.33\textwidth}
        \centering
        \caption*{GAMA560238}
        \includegraphics[width=\columnwidth]{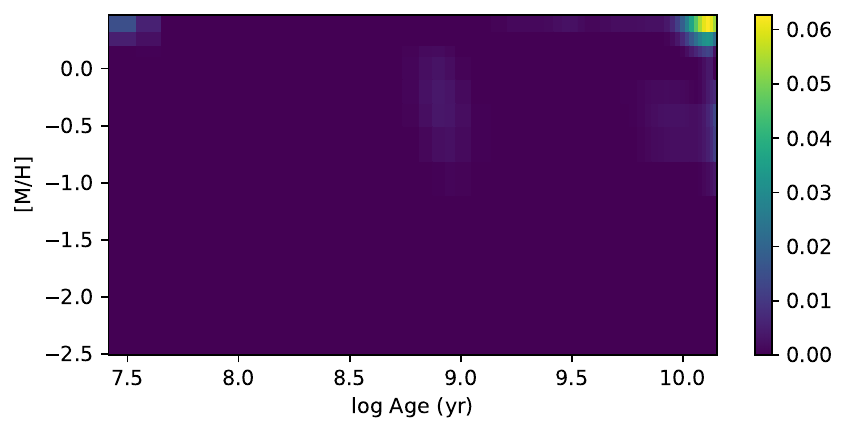}
    \end{subfigure}
    \newline
    \begin{subfigure}{0.33\textwidth}
        \centering
        \caption*{GAMA569555}
        \includegraphics[width=\columnwidth]{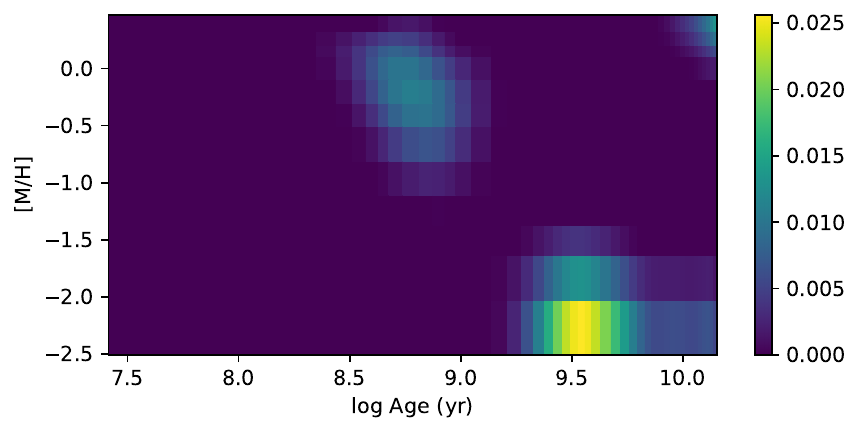}
    \end{subfigure}
    \hfill
    \begin{subfigure}{0.33\textwidth}
        \centering
        \caption*{GAMA570227}
        \includegraphics[width=\columnwidth]{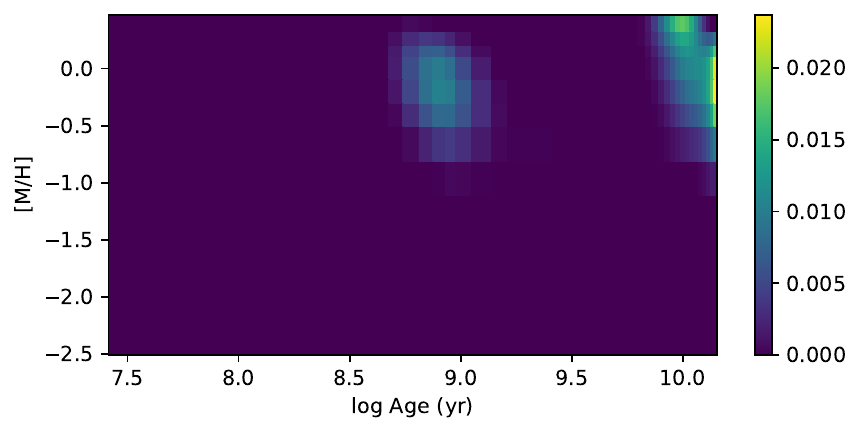}
    \end{subfigure}
    \hfill
    \begin{subfigure}{0.33\textwidth}
        \centering
        \caption*{GAMA3576053}
        \includegraphics[width=\columnwidth]{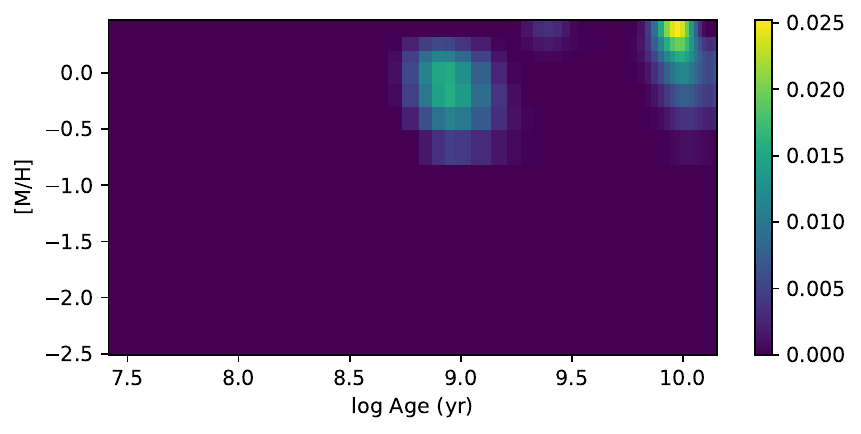}
    \end{subfigure}
\vskip\baselineskip
\label{fig:ppxf_plots}
\end{figure*}

\begin{figure*}
    \centering
    \caption{Presented below are [M/H]$_L$ vs. log$_{10}$ Age$_L$ grids from pPXF luminosity-weighted full spectrum fitting of synthetic test spectra made using MILES SSPs, as described in Section~\ref{subsec:analysis_population}. The top two grids show outputs from fitting synthetic SSPs with equally weighted components of 13, 12 and 11~Gyr, with [M/H]=+0.06 (metal rich) and [M/H]=-0.35 (metal poor) populations on the left and right sides respectively. The next two grids show old populations, with a 2\% mass-weighted, metal-rich, younger component of 1~Gyr. The third row shows old populations, with a 2\% mass-weighted, metal-poor, younger component of 1~Gyr. The fourth row shows a 1-Gyr populations only, with metal-rich (left) or metal-poor (right) SSP. The lowest row, on the left, shows the fit to a 1~Gyr SSP with intermediate metallicity of [M/H]=-0.25. The lowest row, on the right, shows the fit to a model with a spread of equally mass-weighted components from 13 to 1~Gyr, to simulate continuous star formation. White points show the values of the input SSPs for each model fitted. The layout of each grid is as described in Fig.~\ref{fig:ppxf_plots}.}
    \vspace{-1.5cm}   \includegraphics[width=1.0\columnwidth]{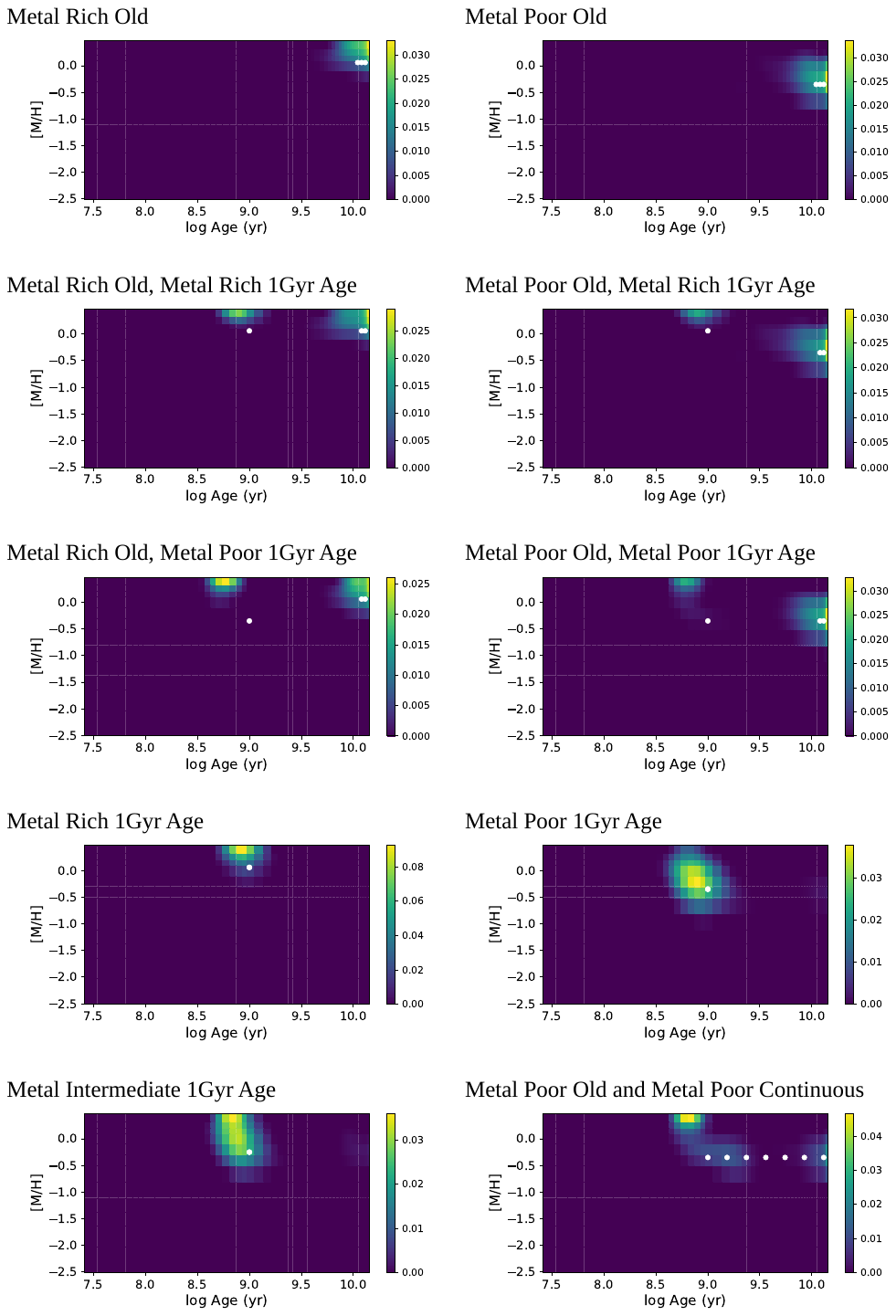}
\label{fig:synthetic_spectra_plots}
\end{figure*}

\bsp	
\label{lastpage}
\end{document}